%&Amstex
%
%
%
%        GROUP INVARIANT SOLUTIONS WITHOUT TRANSVERSALITY
%
%        Ian  Anderson,  Mark Fels,  Charles Torre
%
%        Final Version        March 8
%
%

%  Fonts for GROUP INVARIANT SOLN
%

% small hat letters

		\def\hatx{{\hat x}}

% small bar letters

% small tilde letters

\def\tilg{{\tilde g}}

\def\tilp{{\tilde p}}	 	
\def\tils{{\tilde s}}		
		\def\tilx{{\tilde x}}

% tilde greek letters
\def\tilomega{{\tilde \omega}}

% small bold letters

\def\bff{{\text{\bf f}}}

\def\bfu{{\text{\bf u}}}
\def\bfv{{\text{\bf v}}}		
	
\def\bfx{{\text{\bf x}}}
\def\bfy{{\text{\bf y}}}

%  small star letters

%lists

% Group action notation 

\def\bigG{\eufm G}

\def\tilbigG{\tilde{\bigG}}

\def\liebigG{\Cal G}

% Vertical Notation

% Horizontal Notation

\def\nothing#1{}

% Lie Group and Lie  algebra  notation

\def\LieSO{\text{\bf SO}}

\def\LieDiff{\text{\bf Diff}}

\def\lieh{\eufm h}

% Orbit, Stabilizer  and Fx Notation.

\def\Orb{\eusm O}

\def\Nor{\text{\rm Nor}}

%Maps 
\def\id{\text{id}}
\def\q{{\eufm q}}
\def\qM{{\eufm q_{\sssize M}}}

% Misc Math Symbols

\def\real{\text{\bf R}}

\def\vect#1{\frac{\partial \hfill}{\partial #1}}

\def\mychi{{\hbox{\raise 2 pt \hbox{$\chi$}}}}

% Bundle  Notation

\def\VertM{\Vert \kern-.2pt M}
\def\VertMpt#1{\Vert_{#1}\kern-.4pt M}

\def\VertE{\Vert  \kern-.3ptE}
\def\VertEpt#1{\Vert \kern-.3pt{}_{#1}\kern-.4pt E}

\def\TM{T\kern -1.2pt M}
\def\TMpt#1{T\kern -2pt{}_{#1} M}

\def\TbarM{T\kern -1.2pt \barM}
\def\TbarMpt#1{T\kern -2pt{}_{#1} \barM}

\def\TE{T\kern -1.2pt E}
\def\TEpt#1{T\kern -2pt{}_{#1} E}

\def\Vert{\text{\rm Vert}}

\def\Inv{{\text{\rm Inv}}}

% Bar Capital Letters

\def\barA{
\hbox{\kern 3 true pt
\vbox{\hrule width 4 true pt height .4 true pt \kern .9 true pt
\hbox{\kern -3 true pt $A$}}}}

\def\barB
{\hbox{\kern 2 true pt 
\vbox{\hrule width 7 true pt height .4 true pt \kern .9 true pt
\hbox{\kern -2 true pt $B$}}}}

\def\barE{
\hbox{\kern 2 true pt 
\vbox{\hrule width 7.3 true pt height .4 true pt \kern .9 true pt
\hbox{\kern -2 true pt $E$}}}}

\def\smallbarE{
\hbox{\kern 1.3 true pt 
\vbox{\hrule width 4.3 true pt height .4 true pt \kern .7 true pt
\hbox{\kern -1.3 true pt $\ssize E$}}}}

\def\barF{
\hbox{\kern 2 true pt 
\vbox{\hrule width 7.3 true pt height .4 true pt \kern .9 true pt
\hbox{\kern -2 true pt $F$}}}}

\def\barK{
\hbox{\kern 2.3 true pt 
\vbox{\hrule width 7.5true pt height .4 true pt \kern .9 true pt
\hbox{\kern -2.3 true pt $K$}}}}

\def\smallbarK{
\hbox{\kern 1.3 true pt 
\vbox{\hrule width 4.3 true pt height .4 true pt \kern .7 true pt
\hbox{\kern -1.3 true pt $\ssize K$}}}}

\def\barL{
\hbox{\kern 2 true pt 
\vbox{\hrule width 7.3 true pt height .4 true pt \kern .9 true pt
\hbox{\kern -2 true pt $L$}}}}

\def\barM{
\hbox{\kern 2.3 true pt 
\vbox{\hrule width 8.5  true pt height .3 true pt \kern .9 true pt
\hbox{\kern -2.3 true pt $M$}}}}

\def\smallbarM
{\hbox{\kern 1.7 true pt 
\vbox{\hrule width   6.6 true pt height .4 true pt \kern .7 true pt
\hbox{\kern -1.7 true pt $\ssize M$}}}}

\def\barR{
\hbox{\kern 2 pt
\vbox{\hrule width 7.5 true pt height .4 true pt \kern .9 true pt
\hbox{\kern -2 true pt $R$}}}}

\def\barU{
\hbox{ 
\vbox{\hrule width 7 true pt height .3 true pt \kern .9 true pt
\hbox{\kern-1 pt $U$}}}}

\def\barV{
\hbox{ 
\vbox{\hrule width 7 true pt height .4 true pt \kern .9 true pt
\hbox{\kern-1 pt $V$}}}}

\def\barX{
\hbox{\kern 2.3  true pt 
\vbox{\hrule width 7 true pt height .3 true pt \kern .9 true pt
\hbox{\kern -2.3 true pt $X$}}}}

\def\barY{
\hbox{ 
\vbox{\hrule width 7.5 true pt height .4 true pt \kern .9 true pt
\hbox{$Y$}}}}

\def\barZ{
\hbox{\kern 2.1  true pt 
\vbox{\hrule width 6.5 true pt height .4 true pt \kern .9 true pt
\hbox{\kern -2.1 true pt $Z$}}}}

\def\barsigma{
\hbox{\kern 1.2 true pt                       
\vbox{\hrule width 5.8 true pt height .4 true pt \kern 1.1 true pt 
\hbox{\kern -1.2 true pt $\sigma$}}}}                

\def\barpi{
\hbox{\kern .8 true pt                        
\vbox{\hrule width 5.5 true pt height .3  true pt \kern 1.1 true pt 
\hbox{\kern -.8 true pt $\pi$}}}}                

\def\barkappa{
\hbox{\kern .8 true pt                        
\vbox{\hrule width 5.5 true pt height .3  true pt \kern 1.1 true pt 
\hbox{\kern -.8 true pt $\kappa$}}}}

\def\barsmallpi{
\hbox{\kern .4 true pt                        
\vbox{\hrule width 4.5 true pt height .4 true pt \kern 1.1 true pt 
\hbox{\kern -.4 true pt $\ssize \pi$}}}}

\def\barphi{
\hbox{\kern 1.5 true pt                        
\vbox{\hrule width 4.5 true pt height .4 true pt \kern 1.1 true pt 
\hbox{\kern -1.5 true pt $\phi$}}}}

\def\barPhi{
\hbox{\kern 1.5 true pt                        
\vbox{\hrule width 4.5 true pt height .5 true pt \kern 1.1 true pt 
\hbox{\kern -1.5 true pt $\Phi$}}}}

\def\barmu{
\hbox{\kern 1.5 true pt                        
\vbox{\hrule width 5.5 true pt height .5 true pt \kern 1.1 true pt 
\hbox{\kern -1.5 true pt $\mu$}}}}

\def\barnu{
\hbox{\kern 1.5 true pt                        
\vbox{\hrule width 5.5 true pt height .5 true pt \kern 1.1 true pt 
\hbox{\kern -1.5 true pt $\nu$}}}}

\def\barDelta{%
\hbox{\kern 2  true pt%                       
\vbox{\hrule width 6 true pt height .4 true pt \kern .9 true pt% 
\hbox{\kern -2 true pt $\Delta$\kern -3 true pt}}\kern 3 true 
pt}}              

\def\barlambda{
\hbox{\kern 1 true pt                        
\vbox{\hrule width 4pt height .4 true pt \kern .9 true pt  
\hbox{\kern -1 true pt$\lambda$}}}}

\def\baromega{
\hbox{\kern .8 true 
pt                                                                     
\vbox{\hrule width 5.5 true pt height .4 true pt \kern 1 true pt  
\hbox{\kern -.8 true pt $\omega$}}}}

\def \barCalR{
\hbox{\kern  .2  true pt 
\vbox{\hrule width 7.5 true pt height .4 true pt \kern .9 true pt
\hbox{\kern -1 true pt $\Cal R$}}}}

%
%Cap tilde letters
%

\def\tilA{
\hbox{\kern 3 true pt
\vbox{\hrule width 4 true pt height .4 true pt \kern .9 true pt
\hbox{\kern -3 true pt $A$}}}}

\def\tilB
{\hbox{\kern 2 true pt 
\vbox{\hrule width 7 true pt height .4 true pt \kern .9 true pt
\hbox{\kern -2 true pt $B$}}}}

\def\tilE{\widetilde E}

\def\tilF{
\hbox{\kern 2 true pt 
\vbox{\hrule width 7.3 true pt height .4 true pt \kern .9 true pt
\hbox{\kern -2 true pt $F$}}}}

\def\tilK{
\hbox{\kern 2.3 true pt 
\vbox{\hrule width 7.5true pt height .4 true pt \kern .9 true pt
\hbox{\kern -2.3 true pt $K$}}}}

\def\smalltilK{
\hbox{\kern 1.3 true pt 
\vbox{\hrule width 4.3 true pt height .4 true pt \kern .7 true pt
\hbox{\kern -1.3 true pt $\ssize K$}}}}

\def\tilL{
\hbox{\kern 2 true pt 
\vbox{\hrule width 7.3 true pt height .4 true pt \kern .9 true pt
\hbox{\kern -2 true pt $L$}}}}

\def\tilM{{\hbox{\kern 3pt $\widetilde{\NoBlackBoxes\hbox to 10pt {\kern 
-3 pt $M$}}$}}}

\def\tilM{\widetilde M}

\def\tilR{
\hbox{\kern 2 pt
\vbox{\hrule width 7.5 true pt height .4 true pt \kern .9 true pt
\hbox{\kern -2 true pt $R$}}}}

\def\tilU{\widetilde U}

\def\tilW{{\hbox{\kern 3pt $\widetilde  {\NoBlackBoxes\hbox to 10pt { 
\kern -3 pt $W$}}$} }}

\def\tilZ{
\hbox{\kern 2.1  true pt 
\vbox{\hrule width 6.5 true pt height .4 true pt \kern .9 true pt
\hbox{\kern -2.1 true pt $Z$}}}}

\def\tilsigma{
\hbox{\kern 1.2 true pt                       
\vbox{\hrule width 5.8 true pt height .4 true pt \kern 1.1 true pt 
\hbox{\kern -1.2 true pt $\sigma$}}}}                

\def\tilpi{\tilde \pi}

\def\tilkappa{\tilde \kappa}

\def\tilphi{
\hbox{\kern 1.5 true pt                        
\vbox{\hrule width 4.5 true pt height .4 true pt \kern 1.1 true pt 
\hbox{\kern -1.5 true pt $\phi$}}}}

\def\tilPhi{
\hbox{\kern 1.5 true pt                        
\vbox{\hrule width 4.5 true pt height .5 true pt \kern 1.1 true pt 
\hbox{\kern -1.5 true pt $\Phi$}}}}

\def\tilmu{
\hbox{\kern 1.5 true pt                        
\vbox{\hrule width 5.5 true pt height .5 true pt \kern 1.1 true pt 
\hbox{\kern -1.5 true pt $\mu$}}}}

\def\tilnu{
\hbox{\kern 1.5 true pt                        
\vbox{\hrule width 5.5 true pt height .5 true pt \kern 1.1 true pt 
\hbox{\kern -1.5 true pt $\nu$}}}}

\def\tilDelta{\tilde \Delta}

\def\tillambda{
\hbox{\kern 1 true pt                        
\vbox{\hrule width 4pt height .4 true pt \kern .9 true pt  
\hbox{\kern -1 true pt$\lambda$}}}}

\def\tilomega{
\hbox{\kern .8 true 
pt                                                                     
\vbox{\hrule width 5.5 true pt height .4 true pt \kern 1 true pt  
\hbox{\kern -.8 true pt $\omega$}}}}

\def \tilCalR{
\hbox{\kern  .2  true pt 
\vbox{\hrule width 7.5 true pt height .4 true pt \kern .9 true pt
\hbox{\kern -1 true pt $\Cal R$}}}}

%
%  hat letters
%

\def\hatA{
\hbox{\kern 3 true pt
\vbox{\hrule width 4 true pt height .4 true pt \kern .9 true pt
\hbox{\kern -3 true pt $A$}}}}

\def\hatB
{\hbox{\kern 2 true pt 
\vbox{\hrule width 7 true pt height .4 true pt \kern .9 true pt
\hbox{\kern -2 true pt $B$}}}}

\def\hatF{
\hbox{\kern 2 true pt 
\vbox{\hrule width 7.3 true pt height .4 true pt \kern .9 true pt
\hbox{\kern -2 true pt $F$}}}}

\def\hatK{
\hbox{\kern 2.3 true pt 
\vbox{\hrule width 7.5true pt height .4 true pt \kern .9 true pt
\hbox{\kern -2.3 true pt $K$}}}}

\def\smallhatK{
\hbox{\kern 1.3 true pt 
\vbox{\hrule width 4.3 true pt height .4 true pt \kern .7 true pt
\hbox{\kern -1.3 true pt $\ssize K$}}}}

\def\hatL{
\hbox{\kern 2 true pt 
\vbox{\hrule width 7.3 true pt height .4 true pt \kern .9 true pt
\hbox{\kern -2 true pt $L$}}}}

\def\hatR{
\hbox{\kern 2 pt
\vbox{\hrule width 7.5 true pt height .4 true pt \kern .9 true pt
\hbox{\kern -2 true pt $R$}}}}

\def\hatZ{
\hbox{\kern 2.1  true pt 
\vbox{\hrule width 6.5 true pt height .4 true pt \kern .9 true pt
\hbox{\kern -2.1 true pt $Z$}}}}

\def\hatsigma{
\hbox{\kern 1.2 true pt                       
\vbox{\hrule width 5.8 true pt height .4 true pt \kern 1.1 true pt 
\hbox{\kern -1.2 true pt $\sigma$}}}}                

\def\hatpi{
\hbox{\kern .8 true pt                        
\vbox{\hrule width 5.5 true pt height .3  true pt \kern 1.1 true pt 
\hbox{\kern -.8 true pt $\pi$}}}}                

\def\hatkappa{
\hbox{\kern .8 true pt                        
\vbox{\hrule width 5.5 true pt height .3  true pt \kern 1.1 true pt 
\hbox{\kern -.8 true pt $\kappa$}}}}

\def\hatsmallpi{
\hbox{\kern .4 true pt                        
\vbox{\hrule width 4.5 true pt height .4 true pt \kern 1.1 true pt 
\hbox{\kern -.4 true pt $\ssize \pi$}}}}

\def\hatphi{
\hbox{\kern 1.5 true pt                        
\vbox{\hrule width 4.5 true pt height .4 true pt \kern 1.1 true pt 
\hbox{\kern -1.5 true pt $\phi$}}}}

\def\hatPhi{
\hbox{\kern 1.5 true pt                        
\vbox{\hrule width 4.5 true pt height .5 true pt \kern 1.1 true pt 
\hbox{\kern -1.5 true pt $\Phi$}}}}

\def\hatmu{
\hbox{\kern 1.5 true pt                        
\vbox{\hrule width 5.5 true pt height .5 true pt \kern 1.1 true pt 
\hbox{\kern -1.5 true pt $\mu$}}}}

\def\hatnu{
\hbox{\kern 1.5 true pt                        
\vbox{\hrule width 5.5 true pt height .5 true pt \kern 1.1 true pt 
\hbox{\kern -1.5 true pt $\nu$}}}}                

%
%
%              
%
%  This is a package of macros for AMS-TEX developed by Ian M. Anderson
%
%  The principle components of this package are  (i) an automatic 
%  equation  and statement numbering program, (ii) an automatic 
%  bibliography generation routine.
%
%
%  REVISED December 1994
%
%

%
%  These are the setup macros to determine page size.

\def\State#1 {\statementtag#1 }

\catcode`\^^J=10
\magnification=\mag
\documentstyle{amsppt}
\pagewidth{6.00 truein}
\pageheight{8.00 truein}
\hcorrection{.4 truein}
\vcorrection{.25 truein}

\font\deffont=cmbxti10

\topskip= 28pt
\baselineskip     = 16 true pt  
%\smallskipamount  = 7 true pt  plus 2pt minus 2pt
%\medskipamount    = 12 true pt  plus 4pt minus 4pt
%\bigskipamount    = 21 true pt  plus 6pt minus 6pt 
%abovedisplayskip = 21 true pt  plus 6pt minus 6pt
%\belowdisplayskip = 21 true pt  plus 6pt minus 6pt

%\nologo
\TagsOnRight

%
% These  are new document-enddocument macros. They open-close files for 
% equation-labeling and bib. making

\def\ikedocument{
\ifbibmakemode\immediate\openout1= \jobname.bib\fi
\ifmultisection\immediate\openout2=\jobname.eqn\fi}

%
% These macros override amsppt output routine

\headline={
\ifnum\pageno=\lastsectpageno\hfil \else
{\ifodd\pageno\rightheadline \else\leftheadline\fi}\fi
}

\footline={\hss\the\pageno\hss}
\def\pagecontents{\ifvoid\topins\else\unvbox\topins\fi
\dimen0=\dp255 \unvbox255
\ifvoid\footins\else\vskip\skip\footins\footnoterule\unvbox\footins\fi}

% This are true-false toggles used in the equation numbering routines
%
% Set \proofmodetrue to get the internal equation numbers printed in the
% right-hand margin
%
% Set \multisectiontrue if the paper contains several chapters and 
% equation numbers  are to be passed forward
% 
% Set \bibmakemodetrue to generate the citations needed for bibiography
%
% Set \bibcitemodetrue to obtain citation numbers in the text
% 

\newif\ifproofmode    \proofmodefalse
\newif\ifbibmakemode  \bibmakemodefalse
\newif\ifbibcitemode  \bibcitemodefalse
\newif\ifmultisection \multisectionfalse
\newif\ifmulteq        \multeqtrue

%
% These are macros for initializing sections.

\newcount\lastsectpageno
\lastsectpageno=0
\define\sectpagenocs#1{sect\expandafter\romannumeral#1pageno}

\define\initializesection#1{\sectionnumber=#1%
\equationnumber=1 \statementnumber=0\exercisenumber=0
\ifmultisection
\global\advance\lastsectpageno by 1
\pageno=\lastsectpageno
\immediate\write2{\noexpand\expandafter\def%
\noexpand\csname\sectpagenocs{\the\sectionnumber}\noexpand\endcsname
{\the\pageno}}
\fi}

\define\forwardsectpageno{\immediate\write2{\lastsectpageno=\the\pageno}}

\define\tocpageno#1{
\expandafter\csname\sectpagenocs{#1}\expandafter\endcsname}

%
%These macros are for equation numbering 

\newcount\equationnumber
\newcount\eqnolet
\newcount\sectionnumber
\newcount\statementnumber
\newcount\exercisenumber
\def\strutdepth{\dp\strutbox}

\def\margintagleft#1{\strut\vadjust{\kern-\strutdepth
{\vtop to \strutdepth{\baselineskip\strutdepth\vss\llap{\sevenrm
#1\quad}\null}}}}

\def\ifundefined#1{\expandafter\ifx\csname#1\endcsname\relax}%

\def\eqcs#1{s\expandafter\romannumeral\the\sectionnumber% 
eq\romannumeral#1}

\def\writeeqcs#1{\ifmultisection%
\immediate\write2{\noexpand\expandafter\def% 
\noexpand\csname\eqcs#1\noexpand\endcsname{\the\equationnumber}}\fi%
}

\def\makeeqcs#1{\expandafter\xdef\csname\eqcs#1\endcsname%
{\the\equationnumber}%
}

\def\equationtag#1#2{
\ifundefined{\eqcs#1}
\else
\message{Tag \the\sectionnumber.#1.#2 already exists}
\fi%
\tag"(\the\sectionnumber.\the\equationnumber#2)%
\ifproofmode\rlap{\quad\sevenrm#1}\fi%
\writeeqcs{#1}\makeeqcs{#1}%
\global\eqnolet=\equationnumber%
\global\advance\equationnumber by 1%
\global\def\eqrefnumber{#1}"%
}

\def\Tag#1 {\equationtag{#1}{\empty}}

\define\Tagletter#1 {\tag\the\sectionnumber.\the\eqnolet#1}

\define\equationlabel#1#2#3{%
{\xdef\cmmm{\csname
 s\romannumeral#1eq\romannumeral#2\endcsname}\expandafter\ifx
 \csname s\romannumeral#1eq\romannumeral#2\endcsname
\relax\message{Equation #1.#2.#3 not defined ...}\fi(#1.\cmmm#3)}}

\define\eq#1{\equationlabel{\the\sectionnumber}{#1}{}}

%
%These macros are for numbering statements

\define\statementtag#1 {%
\expandafter\ifx\csname %
s\expandafter\romannumeral\the\sectionnumber %
stat\romannumeral#1\endcsname\relax\else%
\message{Statementtag \the\sectionnumber.#1 already exists ...}\fi%
\ifproofmode%
\margintagleft{#1}\fi\global\advance\statementnumber by 1%
\expandafter\xdef\csname %
s\expandafter\romannumeral\the\sectionnumber %
 stat\romannumeral#1\endcsname{\the\statementnumber}\ifmultisection %
\immediate\write2{\noexpand\expandafter\def\noexpand\csname %
s\expandafter\romannumeral\the\sectionnumber%
stat\romannumeral#1%
\noexpand\endcsname{\the\statementnumber}}%
\fi\the\sectionnumber.\the\statementnumber}

\define\statementlabel#1#2{\xdef\cmmm{\csname 
s\romannumeral#1stat\romannumeral#2\endcsname}\expandafter\ifx
 \csname 
s\romannumeral#1stat\romannumeral#2\endcsname\relax\message{Statement
 #1.#2 not defined ...}\fi#1.\cmmm}

\define\statement#1{\statementlabel{\the\sectionnumber}{#1}}
\define\st#1{\statementlabel{\the\sectionnumber}{#1}}

%These macros are for numbering  exercises

\define\exercisetag#1 {%
\expandafter\ifx\csname %
s\expandafter\romannumeral\the\sectionnumber %
exer\romannumeral#1\endcsname\relax\else%
\message{exercisetag \the\sectionnumber.#1 already exists ...}\fi%
\ifproofmode%
\margintagleft{#1}\fi\global\advance\exercisenumber by 1%
\expandafter\xdef\csname %
s\expandafter\romannumeral\the\sectionnumber %
 exer\romannumeral#1\endcsname{\the\exercisenumber}\ifmultisection %
\immediate\write2{\noexpand\expandafter\def\noexpand\csname %
s\expandafter\romannumeral\the\sectionnumber%
exer\romannumeral#1%
\noexpand\endcsname{\the\exercisenumber}}%
\fi{\bf\the\sectionnumber.\the\exercisenumber.\quad }}

\define\exerciselabel#1#2{\xdef\cmmm{\csname 
s\romannumeral#1exer\romannumeral#2\endcsname}\expandafter\ifx
 \csname 
s\romannumeral#1stat\romannumeral#2\endcsname\relax\message{exercise
 #1.#2 not defined ...}\fi#1.\cmmm}
\define\Exer#1 {\noindent\exercisetag{#1} }

\def\ex#1{\exerciselabel{\the\sectionnumber}{#1}}
\def\Sol#1{\noindent {\bf \exerciselabel{\the\sectionnumber}{#1}\quad}}

%
% These are bibliography macros
%
% With bibmakemode true, citations are listed in jobname.bib
%
% Use the DOS sort program to sort jobname.bib 
%
% Create the  References
%
% With bibcitemode true, the citation numbers are printed

\def\bibyear#1:#2 {\edef\reftag{#1\romannumeral#2}}

\font\slr=cmsl10

\def\cite#1{\catcode`-=11\ifbibmakemode\immediate\write1{#1}[{\bf00}]%
\fi%
\ifbibcitemode%
\bibyear#1  % 
\expandafter\ifx\csname\reftag bibno\endcsname\relax{\message{#1 not in 
bibfile}[{\bf00}]}%
\else%
[{\slr \csname\reftag bibno\endcsname}\hbox{\kern 1pt}]%
\fi\fi
}
   
\def\referencetag#1#2#3#4#5#6#7#8 
{\edef\reftag{#1#2#3#4#5\romannumeral#6#7#8}
\ifbibmakemode
\immediate\write3{\noexpand\def\csname \reftag bibno\endcsname
{\the\refnumb}}
\fi}

\def\referencetag#1:#2 
{\edef\reftag{#1\romannumeral#2}
\ifbibmakemode
\immediate\write3{\noexpand\def\csname \reftag bibno\endcsname
{\the\refnumb}}
\fi}

\newcount\refnumb
\refnumb=0
\def\InitializeRef{
\ifbibmakemode\immediate\openout3=\jobname.ref\fi
\NoBlackBoxes
\medskip}

%
%This gives the date

\define\today{\ifcase\month\or January \or February \or March
\or April  \or May \or June \or July \or August \or September 
\or October \or November \or December \fi\space
\oldnos{\number\day}, \oldnos{\number\year}}

%\hook is the inner evaluation macro
\def\hook{\mathbin{\raise2.5pt\hbox{\hbox{{\vbox{\hrule height.4pt 
width6pt depth0pt}}}\vrule height3pt width.4pt depth0pt}\,}}

\def\real{\text{\bf R}}

\def\pr{\operatorname{pr}}

\def\Div{\operatorname{Div}}

%
% These are various format macros

\define\Ex#1 {\par\medpagebreak\noindent{\bf Example #1.} }
\define\endEx{\qed\par\medpagebreak}
\define\Rem#1 {\par\medpagebreak\noindent{\bf Remark #1.\ } \ }

%
% These macros adjust spaces between equations.

\define\squash#1#2#3{\vspace{-#1\jot}\intertext{#2}\vspace{-#3\jot}}
\define\eqtext#1#2#3{{\vskip -#1\jot}\noindent#2{\vskip -#3\jot}}

%
% Note macros

\newcount\endnoteno
\newcount \nb
\endnoteno=1
\nb=1
\def\NB{\ifproofmode${}^{\the\nb}$ \global\advance\nb by 1\fi}
\def\myitem{\item{\bf[\the\endnoteno]}\global\advance \endnoteno by 1} 

%
%   QED 

\def\qed{\hfill\hbox{\vrule width 4pt height 6pt depth 1.5 pt}}
\def\eqqed{\tag"\hbox{\vrule width 4pt height 6pt depth 1.5 pt}"}

%
%Add to the basic hyphenation table
\hyphenation{dif-fer-en-tial di-men-sion-al Helm-holtz
     Cum-mings  Czech-o-sla-vak }

%
% Defaults

\multisectiontrue
\proofmodetrue
%\bibmakemodetrue

\pageno=1

\def\hatDelta{%
\hbox{\kern 2  true pt%                       
\vbox{\hrule width 6 true pt height .4 true pt \kern .9 true pt% 
\hbox{\kern -2 true pt $\Delta$\kern -3 true pt}}\kern 3 true 
pt}}              

\def\hatlambda{
\hbox{\kern 1 true pt                        
\vbox{\hrule width 4pt height .4 true pt \kern .9 true pt  
\hbox{\kern -1 true pt$\lambda$}}}}

\def\hatomega{
\hbox{\kern .8 true 
pt                                                                     
\vbox{\hrule width 5.5 true pt height .4 true pt \kern 1 true pt  
\hbox{\kern -.8 true pt $\omega$}}}}

\def \hatCalR{
\hbox{\kern  .2  true pt 
\vbox{\hrule width 7.5 true pt height .4 true pt \kern .9 true pt
\hbox{\kern -1 true pt $\Cal R$}}}}                

%
% References
%

\catcode`-=11

\def \abraham-marsdenmcmlxxviiiabibno {1}
\def \anderson-felsmcmxcviiabibno {2}
\def \anderson-felsmcmxcixabibno {3}
\def \anderson-fels-torremmabibno {4}
\def \arms-gotay-jenningsmcmxcabibno {5}
\def \beckers-harnad-perroud-winternitzmcmlxxviiiabibno {6}
\def \beckers-harnad-jasselettemcmlxxixabibno {7}

\def \bluman-kumeimcmlxxxixabibno {10}
\def \bruning-heintzemcmlxxixabibno {11}
\def \bondi-pirani-robinsonmcmlixabibno {12}
\def \coquereaux-jadczykmcmlxxxviiiabibno {13}
\def \david-kamran-levi-winternitzmcmlxxxviabibno {14}
\def \eells-rattomcmxciiiabibno {15}
\def \fels-olvermcmxcviiabibno {16}

\def \fushchich-shtelen-slavutskymcmlxxviabibno {18}
\def \gaeta-morandomcmxcviiabibno {19}
\def \gotay-bosmcmlxxxviabibno {20}
\def \grundland-winternitz-zakrzewskimcmxcviabibno {21}
\def \harnad-schnider-vinetmcmlxxixabibno {22}

\def \jackiw-rebbimcmlxxviabibno {24}
\def \kovalyov-legare-gagnonmcmxciiiabibno {25}

\def \legare-harnadmcmlxxxivabibno {27}

\def \rogers-shadwickmcmlxxxixabibno {34}

\def \winternitz-grundland-tuszynskimcmlxxxviiabibno {42}

\catcode`-=12

\bibcitemodetrue
\proofmodefalse

\def\bfDelta{\bold\Delta}
\def\bfA{\text{\bf A}}

\loadeusm
\loadeufm
\define\kappaG{\kappa_{\sssize G }}
\define\kappaGamma{\kappa_{\sssize\Gamma}}
\define\tilkappaG{\tilkappa_{\sssize G } }
\def\kappaGpt#1{\kappa_{\sssize G,#1 } }
\def\kappaGammapt#1{\kappa_{\sssize\Gamma,#1}}

\ikedocument

\def\rightheadline{}

\ikedocument

\vskip 2 in
\heading   
GROUP INVARIANT SOLUTIONS  WITHOUT TRANSVERSALITY 
\endheading

\nopagenumbers

\footnote""{\today}
\footnote""{ Research supported  by NSF grants DMS--9403788 and 
PHY--9732636}
\vskip 40pt
\centerline{
\vbox{\hsize 130pt 
\centerline{  {\smc Ian M. Anderson}}
\centerline{Department of Mathematics}
\centerline{Utah State University}
\centerline{Logan, Utah 84322}}\qquad
\vbox{\hsize 130pt
\centerline{ { \smc Mark E. Fels}}
\centerline{Department of Mathematics}
\centerline{Utah State University}
\centerline{Logan, Utah 84322}}\qquad
\vbox{\hsize 130pt
\centerline{ { \smc Charles  G.  Torre}}
\centerline{Department of Physics}
\centerline{Utah State University}
\centerline{Logan, Utah 84322}
}
}
\vskip40pt

\subheading{Abstract}   We present a  generalization of Lie's method for 
finding the group invariant solutions to a
system of partial differential equations. Our generalization relaxes  the  
standard  transversality assumption and   encompasses
the  common situation where the  reduced differential equations for the  
group invariant  solutions  involve both fewer 
dependent and  independent variables.    The  theoretical basis for our 
method is   provided   by a general  existence theorem 
for  the   invariant sections, both local and global,
of    a  bundle  on which a finite dimensional Lie group    acts.   
A  simple and natural extension of  our characterization of invariant 
sections leads to an  intrinsic  characterization of the
reduced equations  for the group invariant solutions for a system of 
differential equations.  The  characterization  of both  the   invariant 
sections and
the  reduced equations  are summarized 
schematically by  the kinematic and dynamic reduction diagrams and are 
illustrated by a number of examples from fluid mechanics,  harmonic maps, 
and general relativity.   This work    also provides the  theoretical 
foundations  for  a further detailed study of   the  reduced equations
for group invariant solutions.  

\subheading{Keywords} Lie symmetry reduction,  group invariant solutions,  
kinematic reduction diagram,  dynamic reduction diagram. 
\vfill

\newpage

\initializesection{1}
\def\rightheadline{\smc \hfil  Group Invariant Solutions without 
Transversality \hfil\folio}
\def\leftheadline{ \smc \hfil   Group Invariant Solutions without 
Transversality \hfil\folio}

\pageno=1
\line{}

\subheading{1. Introduction  } Lie's   method of symmetry reduction   for  
finding the group invariant  solutions to
partial differential equations is widely recognized as one of  the most 
general and effective  methods for  obtaining
exact solutions  of non-linear  partial differential equations.  In recent 
years Lie's method    has been  described in a number of excellent texts 
and  survey articles (see, for example, Bluman and Kumei 
\cite{bluman-kumei:1989a},  Olver \cite{olver:1993a}, Stephani 
\cite{stephani:1989a}, 
Winternitz \cite{winternitz:1990a})  and  has been systematically applied  
to  differential equations  arising  in a broad spectrum of
disciplines (see, for example,  Ibragimov  \cite{ibragimov:1995a} or 
Rogers and Shadwick \cite {rogers-shadwick:1989a}). 
It  came, therefore, as quite a surprise to the  present	authors that 
Lie's method,  as it is  conventionally  described,  does not provide  an 
appropriate 
theoretical   framework for  the derivation  of such  celebrated  
invariant solutions as the Schwarzschild solution
of the vacuum  Einstein equations, the  instanton and monopole  solutions 
in  Yang-Mills theory  or the Veronese map
for the harmonic map equations. The  primary objectives  of this paper 
are   to  focus  attention  on this deficiency 
in the literature on Lie's method, to describe the  elementary steps 
needed to correct this problem, and  to give a precise
formulation  of the reduced   differential  equations for the group 
invariant solutions   which arise from   this generalization of Lie's 
method. 

A second impetus for the present  article   is to provide   the   
foundations   for a  systematic  study of 
 the interplay  between the formal   geometric  properties of   a system 
of differential equations,  such as the conservation laws, symmetries, 
Hamiltonian structures, variational principles,  local solvability, formal 
integrability  and  so on,  and   those  same properties of   the reduced 
equations for the group invariant solutions.  Two problems merit special 
attention. First, 
one can interpret the  principle of  symmetric  criticality  
\cite{palais:1979a}, \cite{palais:1985a}  as the problem of determining 
those  group actions  for which
 the  reduced  equations  of  a system of Euler-Lagrange equations are  
derivable from a
canonically  defined  Lagrangian.  Our previous work   
\cite{anderson-fels:1997a} on this problem, and  the  closely related   
problem of  reduction of conservation laws,  was  cast entirely within  
the  context of transverse group actions. Therefore, 
 in order to extend our results to include the reductions that one  
encounters in field theory  and differential geometry,  
one    needs the  more general  description of Lie symmetry reduction 
obtained  here.   Secondly,  there do  not
appear to be     any general  theorems   in the literature  which insure 
the local existence  of group invariant solutions to differential 
equations;   however,
as one step in this  direction the  results  presented here    can  be 
used    to  determine  when a  system of  differential 
equations  of Cauchy-Kovalevskaya  type    remain   of   
Cauchy-Kovalevskaya type under reduction \cite{anderson-fels-torre:2000a}.

We  begin   by    quickly reviewing   the  salient steps of   Lie's  
method 
and  then  comparing Lie's method  with the standard derivation of  the 
Schwarzschild  solution  of  
the vacuum Einstein equations.  This  will clearly demonstrate the 
difficulties  with the classical Lie  approach.
In  section  3  we  describe,   in  detail,  a   general  method for   
characterizing the  group  invariant  sections of a given bundle. 
In section 4 
the  reduced equations  for  the group invariant solutions are 
constructed  in the case where reduction in  both the number of 
independent and dependent variables
can occur.   We  define the  residual   symmetry  group  of the  reduced 
equations in section 5.
In section 6  we   illustrate, at some length,  these results with  a 
variety of examples.   In  the  appendix we  briefly outline    some of the
technical  issues  underlying the    general 
theory of Lie symmetry  reduction  for the group invariant solutions of 
differential equations.

\sectionnumber=2
\equationnumber=1
\statementnumber=0

\subheading{2.  Lie's Method for Group Invariant Solutions}
Consider a system of   second-order partial
differential equations
$$
	\Delta_\beta(x^i, u^\alpha, u^\alpha_i, u^\alpha_{ij}) = 0
\Tag101
$$
for the $m$   unknown functions $u^\alpha$, $\alpha=1$,\dots, $m$,  as 
functions of the $n$ independent variables  $x^i$, $i=1$,\dots, $n$.
 As usual, $u^\alpha_i$ and $u^\alpha_{ij}$ denote the first and second 
order partial 
derivatives of the functions $u^\alpha$. We have assumed that the 
equations  \eq{101} are second-order
and that the number of equations coincides with the number of  unknown 
functions strictly for the 
sake of simplicity.  A fundamental
feature  of Lie's  entire approach to symmetry reduction of differential 
equations, and one that  contributes 
greatly to its  broad applicability,  is  that the Lie algebra of 
infinitesimal  symmetries of a  system of  differential  equations can be 
systematically   and readily  determined. We are not  so much concerned 
with this aspect of Lie's work and accordingly assume that  the symmetry 
algebra of   \eq{101} is  given. 
Now  let  $\Gamma$ be   a finite dimensional  Lie  subalgebra  of   the 
symmetry algebra of \eq{101}, generated   by
vector fields   
$$
V_a = \xi^i_a(x^j)\vect{x^i} + \eta^\alpha_a(x^j, u^\beta)\vect{u^\alpha},
\Tag102
$$
where $a=1$,\dots, $p$.  A map $s\:\real^n \to \real^m$ given by  
$u^\alpha = s^\alpha(x^i)$ is said to be invariant under the Lie algebra  
$\Gamma$  if  the
graph  is invariant under the  local
flows of the vector fields \eq{102}. One finds this to be the case if and 
only  if the functions $s^\alpha(x^i)$ satisfy the 
{\deffont infinitesimal invariance   equations}
$$
	\xi^i_a(x^j) \frac{\partial s^\alpha}{\partial x^i} = \eta^\alpha_a( x^j, 
s^\beta(x^j))
\Tag103
$$
for all $a=1$, 2,\dots, $p$.
 The method of Lie symmetry  reduction consists of explicitly solving the 
infinitesimal 
invariance equations \eq{103} and substituting the  solutions of \eq{103} 
into \eq{101}  to   derive the
reduced equations for  the  $\Gamma$ invariant solutions.

In order to solve \eq{103} it is  customarily  assumed   (see, for 
example,   Olver \cite{olver:1993a}, Ovsiannikov  
\cite{ovsiannikov:1982a}, 
 or  Winternitz \cite{winternitz:1990a}) that  the rank of  the matrix
$\bigl[ \xi^i_a(x^j )\bigr] $ is  constant, say $q$,  and  that  the Lie 
algebra of vector fields satisfies the {\deffont local
transversality condition} 
$$
	\text{rank} [ \xi^i_a(x^j) \bigr] = 
            \text{rank} [\xi^i_a(x^j), \eta^\alpha_a(x^j, u^\alpha)]. 
\Tag104
$$
Granted \eq{104},  it then follows that  there  exist   local coordinates 
$$
	\tilx^r =\tilx^r(x^j), \quad \hatx^k=\hatx^k(x^j)
\quad\text{and}\quad
            v^\alpha= v^\alpha(x^j,u^\beta),
\Tag105
$$
on the space of  independent and  dependent variables, 
where $r=1$,\dots, $n-q$, $k=1$,\dots, $q$, and $\alpha=1$,\dots, $m$,  
such that, 
in these new coordinates, 
the vector fields $V_a$ take the form
$$
	V_a = \sum^q_{l=1}\hat \xi^l_a(\tilx^r, \hatx^k)\vect{\hatx^l}.
\Tag108
$$
The coordinate functions $\tilx^r$ and $v^\alpha$ are   the  
infinitesimal  invariants  for the Lie algebra of vector fields $\Gamma$.
In these coordinates the infinitesimal invariance  equations \eq{103} for 
$v^\alpha=v^\alpha( \tilx^r, \hatx^k)$ 
can be  explicitly integrated to  give $v^\alpha=v^\alpha(\tilx^r)$, where 
the $v^\alpha(\tilx^r)$
are arbitrary smooth functions.
One  now inverts the   relations   \eq{105}  to find  that  the explicit 
solutions to \eq{103} are given by
$$
s^\alpha(\tilx^r,\hatx^k)= u^\alpha(\tilx^r, \hatx^k, v^\alpha(\tilx^r)).
\Tag106
$$
Finally one substitutes \eq{106}  into   the differential equations 
\eq{101} to 
arrive at the reduced system of differential equations
$$
	\tilDelta_\beta(\tilx^r, v^\alpha,  v^\alpha_r,  v^\alpha_{rs}) = 0.
\Tag107
$$
Every solution of \eq{107} therefore  determines, by \eq{106},   a 
solution of  \eq{101}  which also satisfies the invariance condition 
\eq{103}.
 In many applications
of Lie reduction one picks the  Lie algebra of vector fields \eq{102} so 
that   $q=n-1$ in which
case there is only   one  independent  invariant  $\tilx$  on $M$ and 
\eq{107}  is a system of ordinary differential 
equations.  

For the  vacuum  Einstein equations   the independent variables $x^i$, 
$i=0$, \dots, 3, are the  local coordinates on a 
4 dimensional  spacetime, the  dependent variables are the 10  components 
$g_{ij}$ 
of the spacetime metric  and the  differential equations \eq{101} are given
by the vanishing of the  Einstein  tensor $ G^{ij} = 0$.  In the case of 
the  spherically symmetric, stationary  solutions to the 
vacuum Einstein equations the relevant   infinitesimal symmetry generators 
on spacetime  are
$\dsize  V_0 	= \vect{x^0}$,
$$
                  V_1  =    x^3 \vect{x^2}-x^2\vect{x^3},\qquad 
                  V_2  = -x^3 \vect{x^1} +x^1 \vect{x^3} 
\quad\text{and}\quad
                  V_3 = x^2 \vect{x^1} - x^1 \vect{x^2}   
$$
and the  symmetry conditions, as represented by the Killing equations 
$\Cal L_{V_a} g_{ij} = 0$,
lead to the familiar  ansatz  (in spherical coordinates)
$$
	ds^2  = A(r) dt^2  + B(r) dt dr + C(r) dr^2 + D(r) ( d\phi^2 + 
\sin(\phi)^2 d\theta^2).
\Tag118
$$
The  substitution of \eq{118} into the field equations leads to a  system 
of  ODE  whose  
general solution   leads to   the Schwarzschild  solution to the vacuum 
Einstein field equations.

What happens if we attempt to derive  the Schwarzschild  solution using 
the  classical
Lie ansatz \eq{106}?  To begin, it is  necessary to lift the vector fields 
$V_a$  to the space of  independent 
and dependent variables  in order to account for  the  induced  action of 
the  infinitesimal spacetime    transformations
on the  components of  the metric. These  lifted vector fields are  
$\widehat V_0 = V_0$ and
$$
	\widehat V_k= V_k  -  2 \frac{\partial V^l_k}{\partial x^i} g_{lj}\,  
\frac{\partial \hfill}{\partial g_{ij}}.
\Tag120
$$
In  terms of  these lifted vector fields, the   infinitesimal  invariance  
equations 
\eq{103 }  then coincide exactly with  the  Killing equations. 
However,  \eq{106}  cannot possibly coincide  with \eq{118} since   the 
latter contains only 4 arbitrary functions
$A(r)$, $B(r$), $C(r)$, $D(r)$ whereas  \eq{106} would  imply that the 
general  stationary  rotationally invariant  metric
 depends upon  10 arbitrary  functions of $r$. This discrepancy is  easily 
accounted for --- in this example
$$
\text{rank}\bigl[ V_0, V_1, V_2, V_3\bigr] =3
\quad\text{while}\quad
\text{rank}\bigl[\hat  V_0,  \hat V_1, \hat V_2,  \hat V_3\bigl] =4,
$$
and hence {\it  the   local transversality condition  \eq{104} does not 
hold}.  Indeed, whenever  the  local transversality condition
fails, the  general solution to the infinitesimal  invariance   equation 
will  depend  upon    fewer arbitrary functions than the original number 
of 
dependent  variables. The   reduced differential equations   will    be  a 
system of   equations
with  both fewer  independent and dependent variables.

We remark that  in many  of the exhaustive classifications of invariant 
solutions using Lie reduction   either the number of   
independent variables is 2  and hence, typically, the number of vector 
fields $V_a$ is   one,  or   there is just a single  dependent variable 
and  
\eq{101} is a scalar partial differential equation. In either  
circumstance  the   local transversality condition is normally 
satisfied and the  ansatz  \eq{106}  gives the  correct  solution to the  
infinitesimal invariance equation \eq{103}. However,
once the number of  independent and dependent variables exceed these  
minimal thresholds,  as is the case in  most physical
field theories,   the  local transversality condition  is  likely to 
fail.  
\bigskip

\sectionnumber=3
\equationnumber=1
\statementnumber=0

\subheading{3.  An Existence Theorem for Invariant Sections}  Let  $M$ be 
an $n$-dimensional  manifold
and $\pi\:E\to M$ a bundle over $M$.  
In our applications to Lie symmetry reduction  the manifold $M$ serves as 
the space of  independent variables and
the bundle $E$  plays the role  of  the total space  of  independent and 
dependent variables. 
We  refer to points of $M$ with  local coordinates  $(x^i)$ and  to points 
of  $E$ with local coordinates 
$(x^i, u^\alpha)$,   for which the  projection map $\pi$ is given  by  
$\pi(x^i,u^\alpha) = (x^i)$. 
In many applications $E$  either  is  a
trivial bundle $E= M\times N$,  a vector  bundle over $M$, or a  fiber 
bundle  over $M$  with  finite dimensional structure group.
However, for the   purposes of this paper  one need only suppose that  
$\pi$ is  a smooth
submersion.
We let $E_x=\pi^{-1}(x)$ denote the fiber of  $E$ over the point $x\in M$. 

Now let $G$ be a finite  dimensional  Lie group which acts  smoothly on   
$E$.  We  assume that $G$  acts projectably on $E$ in  the sense that  the 
action  of each  element of $G$ is a fiber 
preserving transformation on $E$ ---  if $p,q$ lie in a common fiber, then 
so do $g\cdot p$ and $g \cdot q$.
Consequently, there is a  smooth induced action of $G$ on $M$.   The  
action of $G$ on the   space of sections of $E$   is then given  by  $$
(g\cdot s)(x) =   g \cdot s(g^{-1} \cdot x).
\Tag204
$$
for each  smooth section $s\:M\to E$. 

A section $s$ is  invariant if $g\cdot s  = s$ for all $g\in G$. More 
generally, we have the following definition.

\proclaim{Definition \State201 } Let $G$ be a smooth  projectable group 
action on the bundle
$\pi\:E\to M$ and let $U \subset M$ be open. Then a smooth section $s\:U 
\to E$ is 
$G$ {\deffont  invariant},  if for all  $x\in U$ and $g\in G$ such that 
$g\cdot x\in U$,
$$
s(g\cdot x) = g\cdot s(x).
\Tag201
$$
\endproclaim

Let $\Gamma$ be the Lie algebra  of vector fields on $E$ which  are the 
infinitesimal
generators for the action of $G$  on $E$. Since the action of $G$ is 
assumed projectable,  
any  basis  $V_a$, $a =1$,\dots, $p$ assumes the local coordinate form 
\equationlabel{2}{102}{}. 
If  $g_t$ is a one-parameter  subgroup of $G$   with associated 
infinitesimal generator $V_a$ on $E$,  then  by 
differentiating  the  invariance condition $s(g_t\cdot x) = g_t \cdot 
s(x)$  one finds that the 
component functions $s^\alpha(x^i)$ satisfy
the   infinitesimal invariance condition \equationlabel{2}{103}{}.
 If $s$ is  globally defined on all of $M$
and if $G$ is connected,  then   the infinitesimal   invariance criterion 
\equationlabel{2}{103}{} implies  \eq{201}.
This may not be  true if  $G$ is not connected or if  $s$ is only  defined 
on a  proper open  subset of $M$.
  
For    the purposes  of finding  group invariant solutions of differential 
equations,   we shall take the group $G$ to be a symmetry group of the
given system of differential equations. The task at hand is to 
explicitly    identify   the  space  of $G$ invariant
sections of $E$  with  sections of an auxiliary  bundle 
$\pi_{\tilkappaG}\:\tilkappaG(E) \to \tilM$ and  to  construct
 the   differential  equations  for the $G$  invariant sections  as a 
reduced   system of differential equations on the sections of  
$\pi_{\tilkappaG}\:\tilkappaG(E) \to \tilM$.

Our characterization of the $G$ invariant sections of $E$ is based upon 
the following  key  observation.   Suppose that $p\in E$ and  that there 
is a $G$ invariant section  $s:U\to E$ with $s(x) = p$, where $x\in U$.  
Let 
$G_x= \{\,g\in G \,|\, g\cdot x =x \,\}$ be the  {\deffont isotropy 
subgroup of $G$ at $x$}. Then,  for every  $g\in G_x$,   we compute
$$
g\cdot  p = g \cdot s(x) = s( g \cdot x) = s(x) = p.
\Tag205
$$
This  equation  shows that  the isotropy subgroup $G_x$ constrains the 
admissible values   that  an invariant section can  assume at the point 
$x$.   Accordingly,  we  define the {\deffont  kinematic bundle  
$\kappaG(E)$  for the action of $G$ on $E$ } by
$$
\gather
 	\kappaG(E)  = \bigcup_{x\in M}\kappaGpt{x}(E) 
\\
\squash{5}{where}{5}
              \kappaGpt{x}(E) =\bigl\{\, p\in E_x \, |\, g \cdot p = p 
\quad\text{for all}\quad g \in G_x\, \bigr\}.
\Tag230
\endgather
$$
It is easy to  check that  $\kappaG(E)$  is a $G$ invariant subset of  $E$ 
and therefore the action of  $G$ restricts to
an  action on  $\kappaG(E)$. 

Let 
$\tilM= M/G$ and  $\tilkappaG(E) =\kappaG(E)/G$   be the  quotient spaces 
for the actions
of $G$ on $M$ and  $\kappaG(E)$. We 
 define the {\deffont kinematic  reduction  diagram   for the action of 
$G$ on $E$ }to be the commutative diagram 
$$\CD 
	\tilkappaG(E)	            @<\q_{\kappaG}  <<	 
\kappaG(E)                    @>\iota>> E     \\
            @V \pi_{\tilkappaG} 
VV                                           @V\pi   
VV                      @VV\pi V   \\
            \tilM     @< \qM<<   M           @>\id >>     M. 
\endCD
\Tag206
$$
In this diagram $\iota$ is   the inclusion map  of  the kinematic bundle  
$\kappaG(E)$  into $E$,  $\id\:M\to M$ is the identity map, 
the maps $\qM$ and $\q_{\kappaG}$  are the  projection  maps to the   
quotient spaces and $\pi_{\tilde \kappaG}$
is the  surjective  map  induced by   $\pi$. The next lemma  summarizes 
two of the key properties of  the
kinematic reduction diagram.

 \proclaim{Lemma  \State206 }  Let $G$ act  projectably on $E$.
\smallskip
\noindent
{\bf[i]}   Let $p\in \kappaG(E)$ and $g\in G$. If $\pi(g \cdot p )  = 
\pi(p)$, then  $g\cdot p = p$.
\smallskip
\noindent
{\bf[ii]}  If $\tilp\in \tilkappaG(E)$   and $x\in M$  satisfy $ 
\pi_{\tilde \kappa}(\tilp) = \qM(x)$, 
then   there is a unique point  $p\in \kappaG(E)$ such that 
$\q_{\kappaG}(p) = \tilp$ and $\pi(p) = x$.   
\endproclaim

\demo{Proof} {\bf[i]}  Let  $x= \pi(p)$. If $\pi(g \cdot p )  = \pi(p)$, 
then $g\cdot x = x$ and  therefore, since
$p\in \kappaGpt{x}(E)$, we conclude that $g\cdot p  = p$.

\smallskip
\noindent
{\bf[ii]}  Since $\q_{\kappaG}\:\kappaG(E) \to \tilkappaG(E)$ is 
surjective, there  is a point  $p_0 \in \kappaG(E)$
which projects to $\tilp$.  Let $x_0 = \pi(p_0)$. Then $\qM(x_0) = \qM(x)$ 
and  hence, by definition of the
quotient map $\qM$,  there is a $g\in G$ 
such that $g \cdot x_0 = x$.  The  point $p=g \cdot p_0$  projects  under 
$\q_{\kappaG}$ to $\tilp$ and to
$x$ under $\pi$ so that the  existence of the  point $p$ is established.
Suppose $p_1$ 	and $p_2$  are two points   in   $\kappaG(E)$ which 
project  to $\tilp$ and $x$  under $\q_{\kappaG}$ and
$\pi$ respectively. Then  $p_1$ and $p_2$ belong to the   same fiber 
$\kappaGpt{x}(E)$ and  are  related by  a group
element $g\in G$, that is, $g \cdot p_1  = p_2$.  Since $\pi(p_1) = 
\pi(p_2)$,  it follows that $\pi(g\cdot p_1) = \pi(p_1)$.
Since $p_1\in  \kappaGpt{x}(E)$, we  infer from {\bf[i]}  that $g \cdot 
p_1= p_1$ and therefore $p_1 = p_2.$ 
\qed
\enddemo 

This   simple lemma
immediately implies that every  local section  $\tils:\tilU\to 
\tilkappaG(E)$, where
$\tilU$ is an open subset of $\tilM$,  uniquely determines   a   
$G$-invariant section $s\:U \to \kappaG(E)$, where $U = \q_{\sssize 
M}^{-1}(\tilU)$, 
such that
$$
	\q_{\kappaG}(s(x)) = \tils(\qM(x)).
\Tag207
$$
To insure that  this correspondence  between  the $G$ invariant sections 
of $E$ and the sections of
$\tilkappa_G(E)$  extends to a correspondence between smooth sections it 
suffices to  insure 
that  $\pi_{\tilkappaG}\:\tilkappaG(E) \to\tilM $ is a smooth bundle.

\proclaim{Theorem \State202 }{\smc (Existence Theorem for $G$ invariant 
sections)} Suppose that $E$  admits  a   
kinematic  reduction   diagram  \eq{206} such that  $\kappaG(E)$ is an  
imbedded subbundle of $E$,  the quotient spaces
$\tilM$ and $\tilkappaG(E)$ are smooth manifolds,  and 
$\pi_{\kappaG}\:\tilkappaG(E) \to \tilM$ is a bundle.

Let $\tilU$ be any open set  in $\tilM$  and  let $U= \q_{\ssize 
M}^{-1}(\tilU)$. Then  \eq{207} defines a   
one-to-one correspondence between the  $G$ invariant  smooth sections  
$s\:U\to E$  and the smooth  sections 
$\tils\:\tilU \to \tilkappaG(E)$.
\endproclaim

We can describe the  kinematic reduction  diagram in  local coordinates as 
follows. Since $\pi_{\tilkappaG}\:$ $\tilkappaG(E) \to \tilM$ 
is a  bundle we  begin with  local coordinates  $\pi_{\tilkappaG}\: 
(\tilx^r, v^a) \to (\tilx^r)$ for $\tilkappaG(E)$,  where 
$r=1,\ldots, \dim\tilM$ and $a$ ranges from 1 to the fiber dimension of 
$\tilkappaG(E)$.   
Since
$\qM\:M\to \tilM$ is a submersion, we can use the  coordinates $\tilx^r$ 
as  part of a   local coordinate system $(\tilx^r, \hatx^k)$
on $M$.  Here $k=1,\ldots, \dim M - \dim \tilM$  and,  for  fixed values 
of $\tilx^r$,  the points  $(\tilx^r, \hatx^k)$ all lie
 on a common  $G$  orbit.  As a consequence of  Lemma \st{206}{\bf[ii]} 
one can prove that  $\q_{\kappaG}$  restricts to a  diffeomorphism
between the fibers  of $\kappaG(E)$ and $\tilkappaG(E)$ and  hence one can 
use  
$(\tilx^r, \hatx^k, v^a) $ as a system of  local coordinates on 
$\kappaG(E)$. Finally, 
let $(\tilx^r, \hatx^k, u^\alpha) \to (\tilx^r,\hatx^k)$ be a
system of local coordinates on $E$.  Since $\kappaG(E)$ is an imbedded 
sub-bundle of $E$,   the  inclusion map $\iota\:\kappaG(E)\to E$   assumes 
the form 
$$
	\iota(\tilx^r,\hatx^k, v^a)
              	=(\tilx^r, \hatx^k,  \iota^\alpha(\tilx^r,\hatx^k, v^a)),
\Tag208
$$
where the  rank of the  Jacobian matrix $\dsize \left[ \frac{\partial 
\iota^\alpha}{\partial v^a} \right] $ is maximal.
In these  coordinates the  kinematic $G$ reduction diagram  \eq{206}  
becomes
$$
\CD 
	(\tilx^r, v^a)	            @<\q_{\kappaG}  <<	  (\tilx^r, \hatx^k, 
v^a)                @>\iota>> 
            ( \tilx^r, \hatx^k, \iota^\alpha(\tilx^r, \hatx^k, v^a))   \\
            @V \pi_{\tilkappaG} VV                              @V\pi  
VV                                           @V\pi VV   \\
            (\tilx^r)     @< \qM<<            (\tilx^r, \hatx^k)    @>\id 
>>    (\tilx^r, \hatx^k) .   
\endCD
\Tag209
$$
These coordinates are  readily constructed  in  most applications.
If $v^a = \tils^a(\tilx^r)$ is a local section  of $\tilkappaG(E)$, then 
the corresponding $G$ invariant section of 
$E$ is given by   
$$
	s^\alpha(\tilx^r, \hatx^k) = \iota^\alpha(\tilx^r, \hatx^k, 
\tils^a(\tilx^r)).
\Tag210
$$
Notice that when $\iota$ is the identity map,  \eq{210} reduces to 
\equationlabel{2}{106}{}.
{\it  The   formula  \eq{210} is the full and  proper generalization  of 
the classical Lie  prescription \equationlabel{2}{106}{}  for 
infinitesimally  invariant
sections  of  transverse actions.}

In general   the fiber  dimension of $\kappaG(E)$ will be  less than that  
of $E$,  while the fiber dimension of
$\tilkappaG(E)$ is always the same as that of $\kappaG(E)$.   Thus, in our 
description of  the 
$G$ invariant sections of $E$,   {\deffont fiber  reduction},  or 
reduction in the number of dependent variables,
occurs in the right square  of the  diagram \eq{206}  while {\deffont  
base reduction}, or reduction in the
number of independent variables,  occurs  in the left  square of 
\eq{206}.  

We  now consider the case of an infinitesimal  group action   on $E$,  
defined directly  by  a $p$-dimensional  Lie algebra $\Gamma$ of 
vector fields  \equationlabel{2}{102}{}. These vector fields   need not  
be the infinitesimal generators 
of a  global  action of a Lie group $G$ on $E$.
If  the   rank  of the  coefficient matrix $[\xi^i_a(x^j)]$ is $q$,  then  
there are  locally defined functions $\phi^a_\epsilon(x^j)$,  
where $\epsilon = 1$,\dots, $p-q$, such that  
$$
	\sum^p_{a=1}\phi^a_\epsilon(x^j) \xi^i_a(x^j) =0.
$$  
Consequently, if we  multiply
the   infinitesimal  invariance equation \equationlabel{2}{103}{}  by the  
functions $\phi^a_\epsilon(x^j)$ and sum on $a=1,\dots,p$, we find that
the  invariant sections $s^\alpha(x^j)$ are constrained  by the algebraic 
equations
$$
	\sum^p_{a=1}\phi^a_\epsilon(x^j) \eta_a^\alpha (x^j, s^\beta(x^j)) = 0.
\Tag212
$$
These conditions are the infinitesimal counterparts to  equations 
\eq{205}  and accordingly  we define the   
the {\deffont infinitesimal  kinematic bundle} $\kappaGamma(E) = 
\bigcup_{x\in M} \kappaGammapt{x}(E)$,
 where 
$$
\align
	\kappaGammapt{x}(E) 
&             =\bigl\{\, (x^j,u^\beta) \in E_x \,| 
\,\sum^p_{a=1}\phi^a_\epsilon(x^j) \eta_a^\alpha (x^j, u^\beta) = 0 \, 
\bigr\}
\\
\vspace{2\jot}
&             = \bigl\{\, p \in E_x \,| \, Z(p) = 0 \quad\text{for all 
$Z\in \Gamma$ such that  
	$\pi_*(Z(p))= 0$ }\,  \bigr\}. 
\Tag211
\endalign
$$

In  most applications  the algebraic  conditions defining $\kappaGamma(E)$ 
are easily  solved.
 The  Lie algebra  of vector fields $\Gamma$
restricts to a Lie algebra of  vector fields on $\kappaGamma(E)$ which  
now satisfies the infinitesimal transversality 
condition \equationlabel{2}{104}{}.   One then arrives  at \eq{209}  as  a 
local coordinate  description of the infinitesimal  kinematic
diagram for  $\Gamma$, where  the coordinates $(\tilx^r, v^a)$ are  now 
the  infinitesimal  invariants for the action of
$\Gamma$ on $\kappaGamma(E)$.

It is not difficult to  show that $\kappaGpt{x}(E) \subset 
\kappaGammapt{x}(E)$, with equality holding  whenever the  
isotropy group $G_x$ is connected.

In the case where  $E$ is  a vector bundle, the infinitesimal kinematic 
bundle  appears in  Fels and   Olver \cite{fels-olver:1997a}.
For applications  of   the kinematic bundle  to the classification of  
invariant tensors and spinors see
 \cite{beckers-harnad-perroud-winternitz:1978a} and   
\cite{beckers-harnad-jasselette:1979a}.  

\bigskip

\def\D{\Cal D}
\def\tilD{\tilde \Cal D}
\def\bff{{\text{\bf f}}}
\def\tilbff{\tilde \bff}

\sectionnumber=4
\equationnumber=1
\statementnumber=0

\subheading{4. Reduced  Differential Equations for Group Invariant 
Solutions}  Let $G$ be a Lie group acting projectably
on the bundle  $\pi\:E\to M$ and let $\Delta=0$ be a system of  $G$ 
invariant differential equations  for the sections of $E$.  
In order to describe  geometrically the reduced equations $\tilDelta = 0$  
for the  $G$ invariant solutions
to $\Delta = 0$ we  first formalize   the definition of a  system of 
differential equations.  

To this end, let $\pi^k\:J^k(E) \to M$ be the $k$-th order jet bundle of  
$\pi\:E\to M$. A point  $\sigma =j^k(s)(x)$ in $J^k(E)$ represents
the  values of a local section $s$ and all its derivatives to  order $k$ 
at the point $x\in M$.  Since $G$ acts naturally on the space of 
sections of $E$ by \equationlabel{3}{204}{},  the action of $G$ on $E$  
can be lifted  
(or prolonged)  to an  action on $J^k(E)$ by setting 
$$ 
	g\cdot \sigma  = j^k(g\cdot s)( g \cdot  x), 	\quad\text{where $\sigma  = 
j^k(s)(x)$}
$$ 
Now let $\pi\:\D \to J^k(E)$ be a vector
bundle over $J^k(E)$ and suppose that  the Lie group acts projectably on 
$\D$ in  a manner which covers the  action of
$G$ on $J^k(E)$.  A {\deffont  differential operator}  is a   section  
$\Delta :J^k(E) \to \D$.     The  differential operator  $\Delta$  
is $G$ invariant    if  it is  invariant  in the sense of Definition 
\statementlabel{3}{201}, that is,
$$
	g \cdot \Delta(\sigma ) = \Delta (g \cdot  \sigma)
$$
for all $ g\in G$ and all points $\sigma  \in J^k(E)$. A   section $s$ of 
$E$ defined on an open  set $U\subset M$ is a solution to the differential 
equations $\Delta = 0$  if $\Delta(j^k(s)(x)) = 0$  for all  $x\in  U$.

Typically,  the  bundle $\Cal D\to J^k(E)$ is  defined  as the    
pullback  bundle of a  vector  bundle  $V$ (on which $G$ acts)  over $E$ 
or  $M$
by the projections $\pi^k\:J^k(E) \to E$ or  $\pi^k_M\:J^k(E) \to M$
and the  action of $G$ on $\Cal D$ is  the  action  jointly induced from  
$J^k(E)$ and $V$.

Our goal now is to construct  a  bundle $\tilD\to J^k(\tilkappaG(E))$ and  
a differential  operator $\tilDelta\:J^k(\tilkappaG(E))$  $\to\tilD$
such that  the correspondence \equationlabel{3}{207}{}   restricts to a  
1-1 correspondence between the  $G$ invariant solutions of $\Delta=0$ and 
the
solutions of $\tilDelta = 0$.

One  might   anticipate  that the  required  bundle $\tilD\to 
J^k(\tilkappaG(E))$ can be  constructed  by a direct application of
kinematic  reduction    to $\D \to J^k(E)$.   However,  one can readily 
check that  the quotient space  of $J^k(E)$ by the
prolonged action of $G$    does {\it not}  in general   coincide with   
the jet space  $J^k(\tilkappaG(E))$ so that the kinematic  reduction 
diagram  for
the action of $G$ on $\D$ will not lead to a bundle over 
$J^k(\tilkappaG(E))$.  For example, if $G$ is  the  group
acting on $M\times\real \to M$  by rotations in the  base $M=\real^2 
-\{(0,0)\}$, then   $J^2(E)/G$ is a 7 dimensional manifold
whereas $J^2(\tilkappaG(E))$ is 4 dimensional.     
This difficulty  is easily  circumvented by introducing the  {\deffont  
bundle of invariant $k$-jets}
$$
\align
	\Inv^k(E)=\{\, \sigma\in 
&	J^k(E) \, |\, \sigma =j^k(s)(x_0),
 \\
\vspace{2\jot}
 &	\text{where s is a $G$ invariant section defined in a neighborhood of 
	$x_0$}\,\}.
\Tag238
\endalign
$$
This  bundle  is  studied  in Olver  \cite{olver:1993a} although the 
importance of 
 these   invariant jet spaces  to the   general theory of symmetry 
reduction of differential equations  is not   as widely acknowledged 
in the literature as it should be.  

The quotient space  $\Inv^k(E)/G$  coincides  with the  jet space 
$J^k(\tilkappaG(E))$. 
We let $\D_{\sssize \Inv} \to \Inv^k(E)$  be the restriction of  $\D $ to 
the bundle of  invariant $k$-jets  and to this  we now apply
our  reduction  procedure to arrive at the {\deffont  dynamic reduction 
diagram}
$$
\CD
\tilkappaG(\D_{\sssize \Inv}) @<\q<<    \kappaG(\D_{\sssize \Inv}) 
@>\iota>> \D_{\sssize \Inv}  @>\iota_{\Inv}>> \D
\\
@V\tilde \pi VV                 @V\pi VV                               
@V\pi VV                @VV \pi V
\\
J^k(\tilkappaG(E)) @<\q_{\sssize \Inv}<<  \Inv^k(E) @>\id>>\Inv^k(E)    @> 
\iota^k>> J^k(E) .
\endCD
\Tag305
$$
Theorem  \statementlabel{3}{202} insures that there is a one-to-one 
correspondence
between the   $G$ invariant sections of $\D_{\sssize \Inv} \to \Inv^k(E)$ 
and the  sections of  $\tilkappaG(\D_{\sssize \Inv}) \to 
J^k(\tilkappaG(E))$.
Any $G$ invariant differential operator $\Delta\:J^k(E) \to \D$  restricts 
to a  $G$ invariant differential operator
$\Delta_{\sssize \Inv}\:\Inv^k(E)\to \D_{\sssize \Inv}$ and thus   
determines a  differential operator
$\tilDelta\: J^k(\tilkappaG(E)) \to \tilkappaG(\D) $. This is the reduced  
differential operator  whose solutions  describe the
$G$ invariant solutions  for  the original operator $\Delta$.  

To describe  diagram  \eq{305} in local  coordinates,  we begin with the  
coordinate  description  \equationlabel{3}{209}{}
of the kinematic reduction diagram and we let   
$$
	(\tilx^r, \hatx^k, u^\alpha, u^\alpha_r, u^\alpha_k, u^\alpha_{rs}, 
u^\alpha_{rk}, u^\alpha_{kl},\ldots)
$$
 denote the
standard jet coordinates on $J^k(E)$.  
 Since the  invariant sections   are  parameterized by  functions $v^a= 
v^a(\tilx^r)$,  coordinates for  $\Inv^k(E)$ are 
$$
	(\tilx^r, \hatx^k, v^a, v^a_r, v^a_{rs},\dots)
$$
In accordance with \equationlabel{3}{210}{},  the inclusion map 
$$
	\iota\:\Inv^k(E) \to J^k(E)
$$
 is  given by  
$$
	\iota(\tilx^r, \hatx^k, v^a, v^a_r, v^a_{rs},\ldots) =
	(\tilx^r, \hatx^k,  u^\alpha, u^\alpha_r, u^\alpha_k, u^\alpha_{rs}, 
u^\alpha_{rk}, u^\alpha_{kl}, \ldots),
\Tag308
$$
where  by  a  formal application of the chain rule,
$$
\aligned
	u^\alpha
&=         \iota^\alpha(\tilx^r, \hatx^i, v^a),
\qquad
               u^\alpha_r
=         \frac{\partial \iota^\alpha}{\partial\tilx^r} + 
\frac{\partial\iota^\alpha}{\partial v^a} v^a_r, 
\qquad
               u^\alpha_k
=           \frac{\partial \iota^\alpha}{\partial \hatx^k},
\\
\vspace{2\jot}
	u^\alpha_{rs}
&            =  \frac{\partial^2 \iota^\alpha}{\partial\tilx^r\partial 
\tilx^s} +
                    \frac{\partial^2\iota^\alpha}{\partial v^a\partial 
\tilx^s} v^a_r  +                     
                     \frac{\partial^2\iota^\alpha}{\partial v^a\partial 
\tilx^r} v^a_s  +                      
\frac{\partial^2\iota^\alpha}{\partial v^a \partial v^b} v^a_r 
v^b_s                                      
+\frac{\partial\iota^\alpha}{\partial v^a} v^a_{rs}, 
\endaligned
$$
and so on.  The  quotient map 
$$
 	\q_{\sssize \Inv}\:\Inv^k(E)\to  J^k(\tilkappaG(E))
$$
is  given simply by
$$
	\q_{\sssize \Inv}(\tilx^r, \hatx^k, v^a, v^a_r, v^a_{rs},\ldots)= 
(\tilx^r, v^a, v^a_r, v^a_{rs},\ldots).
$$

Next  let $\bff{\,}^A$  be a  local  frame  field for the vector bundle 
$\D$.   The  differential operator  $\Delta\:J^k(E) \to \D$ can be written 
 in terms of the  standard coordinates on $J^k(E)$ and in this local frame 
by
$$
 	\Delta= 
	\Delta_A(\tilx^r, \hatx^k,  u^\alpha, u^\alpha_r, u^\alpha_k, 
u^\alpha_{rs}, u^\alpha_{rk}, u^\alpha_{kl}, \ldots)\, \bff{\,}^A.
\Tag311
$$
The  restriction of $\Delta$ to $\Inv^k(E)$ defines  the section  
$\Delta_{\Inv}\:\Inv^k(E) \to \D_{\sssize \Inv}$
by
$$
	\Delta_{\Inv} = \Delta_{\Inv,A}(\tilx^r, \hatx^k , v^a, 	v^a_r, 
v^a_{rs},\ldots) \, \bff{\,}^A,
\Tag312
$$ 
where the component functions $\Delta_{\Inv,A}(\tilx^r, \hatx^k, v^a, 
v^a_r, v^a_{rs},\ldots)$ are defined as the
composition of the maps    \eq{308} and  the component maps $\Delta_A$.   
Since $\Delta$  is a $G$ invariant differential operator,
$\Delta_{\Inv}$  is a $G$ invariant differential operator and  hence  
necessarily factors through the 
kinematic  bundle $\kappaG(\Cal D_{\Inv})$,
$$
	\Delta_{\Inv}\:\Inv^k(E) \to \kappaG(\D_{\sssize \Inv}).
$$
Our general existence theory for invariant sections  implies that we  can 
also find a locally defined,  $G$ invariant
frame  $\bff{\,}^Q_{\Inv}$  for $\kappaG(\D_{\sssize  \Inv})$. The 
inclusion
map  $\kappaG(\D_{\sssize \Inv}) \to \D_{\sssize \Inv}$  is represented  
by writing  each  vector $\bff{\,}^Q_{\sssize \Inv}$ as
a linear combination of the  vectors  $\bff{\,}^A$,
$$ 
\bff{\,}^Q_{\sssize \Inv} = M^Q_A \bff{\,}^A,
$$
where the coefficients $M^Q_A$ are  functions on $\Inv^k(E)$. 
The invariant operator  $\Delta_{\Inv}$
can be  expressed as
$$
	\Delta_{\Inv}=
               \Delta_{\Inv,Q}( \tilx^r, \hatx^k,  v^a, v^a_r, 
v^a_{rs},\ldots) \, \bff{\,}^Q_{\Inv}.
$$

Finally,  
the   $G$ invariant  frame $\bff{\,}^Q_{\Inv}$   determines   a frame $ 
\tilbff{\,}^Q$ on  $\tilkappaG(E)$,
the invariance  of  $\Delta$  implies that the  component functions  
$\Delta_{\Inv,Q}$ are  necessarily independent
of  the parametric  variables $\hatx^k$,  that is, 
$$
\tilDelta_Q( \tilx^r,  v^a, v^a_r, v^a_{rs},\ldots)= \Delta_{\Inv, Q}( 
\tilx^r, \hatx^k, \ v^a, v^a_r, v^a_{rs},\ldots)
$$
and  the reduced  differential operator  is
$$
\tilDelta = \tilDelta_Q( \tilx^r, v^a, v^a_r, v^a_{rs},\ldots) 
\,\tilbff{\,}^Q.
$$

At first sight,  this general framework  may  appear to be  rather 
cumbersome and overly complicated. 
However, as we shall see  in examples,  every  square  in the
dynamic reduction  diagram  \eq{305}  actually corresponds  to  the 
individual steps that one performs in practice.
\bigskip

\sectionnumber=5
\equationnumber=1
\statementnumber=0

\subheading{5. The Automorphism Group of the Kinematic Bundle} Let 
$\bigG$  be the   full   group of
 projectable symmetries on $E$  for    a given system of differential 
equations on $J^k(E)$ and let $G\subset \bigG$
be  a fixed  subgroup  for  which the group invariant  solutions are  
sought.  It is   commonly noted (again, within the
context of  reduction with transversality)  that $\text{Nor}(G,\bigG)$,   
the normalizer of   $G$ in $\bigG$, 
preserves the space of  invariant sections and    that   
$\text{Nor}(G,\bigG)/G$ 
is a  symmetry  group of the reduced equations. However, because this is 
a  purely algebraic  construction which  does not 
take into account the action of $\bigG$ on $E$, this construction may not 
yield the  largest  possible residual symmetry group  
or may   result in a residual  group which  does  not act effectively on 
$\tilkappaG(E)$. 
These difficulties are easily  resolved. We let $\Orb_p(G)$ denote orbit 
of $G$ through a point $p\in E$.

\proclaim{Definition \State501 } Let $\bigG$ be a  group of  
fiber-preserving transformations acting on 
$\pi\:E\to M$ and let $G$ be a subgroup of $\bigG$.  Assume that  $E$   
admits    a kinematic reduction
diagram \equationlabel{3}{206}{} for the action of $G$ on $E$.  
\smallskip
\noindent
{\bf[i]}    The    {\deffont automorphism group}  $\tilbigG$  for the 
kinematic  bundle  
$\pi\:	 \kappaG(E)  \to M$ 
is  the subgroup of $\bigG$ which stabilizes the set of  all  the $G$  
orbits in $\kappaG(E)$, that is,   
$$
\align
	\tilbigG
&         =\bigl\{a\in\bigG \,| \, 	a \cdot \Orb_p(G) = \Orb_{a\cdot p}(G) 
           \text{\ and\  }  a^{-1} \cdot \Orb_p(G) = \Orb_{a^{-1}\cdot 
p}(G)\text{\ for all\ } p\in \kappaG(E) \,\bigr\}.       
\Tag532
\endalign
$$
\smallskip
\noindent
{\bf[ii]} The {\deffont  global isotropy subgroup}  of  $\bigG$,   as it 
acts on the space of 
$G$ orbits of $\kappaG(E)$,  is 
$$
               \tilbigG{}^*=\big\{\, a\in  \bigG\, | \, a \cdot \Orb_p(G) 
= \Orb_p(G) 
                  \quad \text{for all}\quad p\in\kappaG(E)\,\big\}.
\Tag534
$$
\smallskip
\noindent
{\bf[iii]} The {\deffont residual symmetry group}   is   
$\tilbigG_{\text{eff}} = \tilbigG/\tilbigG^*$.
\endproclaim
 
 The  key  property of  
$\tilbigG{}^*$ is that  it is the largest subgroup of $\bigG$  with  
exactly the same reduction diagram  and  invariant sections as $G$.
This is an important  interpretation of the group $\tilbigG{}^*$ --- from 
the
viewpoint of  kinematic reduction,  one  should   generally  replace  the  
group $G$   
by the  group $\tilbigG{}^*$.  For computational  purposes,  it is often
advantageous to  use the fact  that
$\tilbigG{}^*$  fixes every  $G$ invariant section  of $E$.
It is not  difficult to check that  $\Nor(\tilbigG{}^*, \bigG)=\tilbigG$, 
that
the quotient  group $\tilbigG_{\text{eff}} = \tilbigG/\tilbigG{}^*$ acts 
effectively and  projectably
on the reduced   bundle $ \tilkappaG(E) \to \tilM$ and that, if $\bigG$ is 
a symmetry group of of a 
differential  operator  $\Delta$, then  $\tilbigG_{\text{eff}}$ is always 
a  symmetry  group 
of the reduced  differential operator  $\tilDelta$.

Similarly, if $\liebigG$ is a Lie algebra of projectable  vector fields  
on $E$  and $\Gamma\subset \liebigG$,
we   define the    { \deffont infinitesimal automorphism  algebra} of 
$\kappaGamma(E)$ as the  Lie subalgebra of vector fields   given by
$$
	\tilde \liebigG
         =\bigl\{\, Y\in \liebigG \, | \, [\,Z,\,Y\,]_p \in 
\text{span}(\Gamma) (p) 	\quad\text{for all $p\in \kappa_{\lower 1.5pt 
\hbox{$\ssize \Gamma$}}(E)$ and all $Z\in \Gamma$}\,\bigr\},
\Tag540
$$
and the  associated  isotropy subalgebra  for $\kappa_{\lower 1.5pt \hbox 
{$\ssize \Gamma$}}(E)$
$$
            \tilde \liebigG^*
        = \bigr\{ Y\in  \liebigG \, | \,  Y_p\in \text{span}(\Gamma)(p) 
\quad\text{for all 	$p\in \kappa_{\lower 1.5pt \hbox{$\ssize\Gamma$}}(E)$} 
\, \bigl\}. 
\Tag541
\endalign
$$
When $\bigG$ is a finite dimensional Lie group and  $\liebigG = 
\Gamma(\bigG)$, then  it is readily checked   that  $\tilde\liebigG= 
\Gamma(\tilbigG)$ and
$ \tilde\liebigG{}^*= \Gamma(\tilbigG{}^*)$.

Since the automorphism group $\tilbigG$  acts on the  $k$-jets  of 
invariant sections $\Inv^k_G(E)$, 
 this  group also plays an important role in dynamic reduction.   
Specifically, let us   suppose that  $\bigG$ acts on the  vector bundle 
$\Cal D \to J^k(E)$
and that  $\Delta\:J^k(E) \to \Cal D$  is a  $\bigG$ invariant section.  
Then $\Delta_{\sssize \Inv}\: \Inv^k(E) \to 
\Cal D_{\sssize \Inv}$  is always invariant  under  the action of 
$\tilbigG$ and accordingly   the operator $\Delta_{\sssize \Inv}$
always  factors through the   kinematic bundle for the action of 
$\tilbigG$  on $\Cal D_{\sssize \Inv}$,  where for 
$\sigma \in \Inv^k(E)$,
$$
	\kappa_{\tilbigG, \sigma}(\Cal D_{\sssize \Inv}) =
                \bigl \{ \,\Delta\in \Cal D_{{\sssize \Inv},\sigma} \, | 
\, g\cdot \Delta = \Delta \quad \text{for all}\quad
                g\in \tilbigG_\sigma \bigr \}.
$$
We note that
$$
\kappa_{\sssize \tilbigG}(\Cal D_{\sssize \Inv}) \subset  \kappaG(\Cal 
D_{\sssize \Inv})
$$
and consequently one can refine the  dynamic  reduction  diagram from 
\equationlabel{4}{305}{} to
$$
\CD
\tilde\kappa_{\sssize \tilbigG}(\D_{\sssize \Inv})@<\q<<    
\kappa_{\sssize \tilbigG}(\D_{\sssize \Inv}) @>\iota>> \D_{\sssize \Inv} 
 @>\iota _{\Inv}>> \D
\\
@V\pi VV                 @V\pi VV                               @V\pi 
VV                @VV \pi V
\\
J^k(\tilkappaG(E)) @<\q_{\sssize \Inv}<<  \Inv^k(E) @>\id >>\Inv^k(E)    
@>\iota^k >> J^k(E) ,
\endCD
$$
where the  quotient maps to the left are still  by the action of $G$.  

Given the actions of $G $ on $\pi\: E\to M $ and also $\bigG$ on $\Cal D 
\to J^k(E)$,  it  sometimes
happens that  
$$
	\kappa_{\tilbigG, \sigma}(\Cal D_{\sssize \Inv}) = 0.
\Tag525
$$
{\it  In this case  every  $G$ invariant section of $E$  is automatically  
a solution to $\Delta = 0$ for  every $\bigG$ 
invariant  operator  $\Delta \:J^k(E) \to \Cal D$} --- such  sections are 
called  {\deffont universal solutions.}
Previous  work on this subject  (see Bleecker  \cite{bleecker:1979a},  
\cite{bleecker:1979b},  Gaeta and Morando  \cite{gaeta-morando:1997a})  
have emphasized  a variational
approach which, from the viewpoint of the dynamic reduction diagram and 
the automorphism group of the kinematic 
reduction diagram, may not always  be necessary.

\sectionnumber=6
\equationnumber=1
\statementnumber=0

\subheading{6. Examples}  In this section we    find  the kinematic and 
dynamic reduction  diagrams for  the group  invariant
solutions  for some  well-known  systems of  differential equations in 
applied mathematics, differential geometry,  and  mathematical physics.  
We  begin  by deriving the  rotationally invariant solutions of the  Euler 
equations for incompressible fluid flow. As   noted
by Olver  \cite{olver:1993a} (p. 199),  these  solutions cannot be    
obtained by the classical Lie  ansatz.   The   general  theory of symmetry 
reduction  without transversality leads to some interesting  new   
classification  problems  for   group invariant solutions   which we   
briefly illustrate  by presenting  another  reduction of  the  Euler 
equations.  

In our second set of  examples we   consider  reductions of the harmonic 
map equations. 
We show    the   classic Veronese map from $S^2\to S^4$  is  an example 
of    a universal   solution.  
 In Example  5.4  we  consider another symmetry reduction of  the  
harmonic 
map equation which  nicely  illustrates   the construction of the 
reduced   kinematic  space   for  quotient  manifolds $\tilM$ with 
boundary.  

In our third set of examples, the 
Schwarzschild  and plane  wave  solutions  of the  vacuum Einstein 
equations  are  re-examined  in  the  context  of symmetry reduction
without  transversality. 
We   
demonstrate      the importance  of the automorphism group  in 
understanding     the geometric  properties of the   kinematic  bundle  
and, as well,    qualitative  features  of  the reduced equations.  

Finally, some   elementary examples from mechanics are   used  to 
demonstrate the basic  differences  between  symmetry reduction  for group 
invariant solutions  and  symplectic reduction of Hamiltonian systems.  

Although space does not  permit us to do so,  the  kinematic  and  dynamic 
reduction  diagrams are  also nicely illustrated by  symmetry  reduction 
of the Yang-Mills equations  as
found, for example, in  \cite{harnad-schnider-vinet:1979a},  
\cite{kovalyov-legare-gagnon:1993a},  \cite{legare-harnad:1984a}. In 
particular,  it is  interesting to note that  the   invariance   
properties of    the classical instanton  solution to the  Yang-Mills 
equations  (Jackiw and Rebbi 
 \cite{jackiw-rebbi:1976a})
 imply that  it  is  a  universal solution in the sense of  equation
\equationlabel{5}{525}{}.

\heading
Euler  Equations for Incompressible Fluid Flow 
\endheading

The Euler equations are a   system of 4 first order equations in  4  
independent and  dependent variables. 
The  underlying  bundle $E$ for these  equations 
is  the trivial  bundle  $\real^4 \times \real^4  \to \real^4$ with 
coordinates $(t, \bfx, \bfu,p)\ \to (t, \bfx)$, 
where $\bfx=(x^1,x^2,x^3)$ and $\bfu=(u^1,u^2,u^3)$  and the equations 
are 
$$
 	\bfu _t + \bfu \cdot \nabla \bfu 
            = -\nabla p \quad\text{and}\quad 
                \nabla \cdot  \bfu  = 0.
\Tag401
$$
The full symmetry group  $\bigG$ of the Euler equations is well-known 
(see, for example,  \cite{ibragimov:1995a},  \cite{olver:1993a},  
\cite{rogers-shadwick:1989a}) 

\Ex{\State401 .  \ \smc Rotationally Invariant Solutions of the Euler 
Equations}  The symmetry group of the
 Euler equations  contains  the  group  $G= \LieSO(3)$, acting  on $E$, by
$$
	R \cdot (t, \bfx, \bfu,p) = (t, R\cdot\bfx,  R\cdot \bfu, p) = (t, 
R^i_jx^j, R^i_j u^j,p),
\Tag402
$$
for $R=(R^i_j)\in \LieSO(3)$.   To insure that the action of $G$ on the 
base  $\real^4$ is regular we restrict to 
the open set $M\subset \real^4$ where $||\bfx|| \neq 0$. 
The infinitesimal generators for  this action  are 
$$
	V_k=  \varepsilon_{kij}x^i\vect{x^j} + \varepsilon_{kij}u^i\vect{u^j}. 
\Tag403
$$

We first  construct the  kinematic reduction  diagram for  this action.  
For a  given   point   $x_0=(t_0,\bfx_0) \in M$,  the   isotropy  subgroup
$G_{x_0}$   for the action of $G$ on $M$ is the  subgroup  
$\LieSO(2)_{\bfx_0}\subset \LieSO(3)$ which fixes the vector  $\bfx_0$ in 
$\real^3$.
 Since the   only vectors invariant under   all rotations  about a  given 
axis of rotation  are vectors  along the axis of rotation, we deduce
that for $x_0 \in M$,
$$
\align
	\kappaGpt{x_0}(E) 
&             =  \{\,(t_0,\bfx_0,\bfu,p)\, |\, R\cdot \bfu= 
\bfu\quad\text{for all\quad $R\in \LieSO(2)_{\bfx_0}$ }\,\}  
\\
\vspace{2\jot}
&             =\{\,( t_0,\bfx_0,\bfu,p)\, |\, \bfu = A\bfx_0\quad  
\text{for some\quad  $A\in \real$}\,\}.
\endalign            
$$

The same  conclusion  can be obtained by  infinitesimal considerations. 
Indeed,  
 the infinitesimal isotropy vector  field at $x_0$ for the action on $M$ 
is 
$$
	Z= x^k_0   \varepsilon_{kij}x^i\vect{x^j}
$$
and therefore,  if  $(t,\bfx,\bfu,p)\in \kappaGpt{x}(E) $, we must have, 
by \equationlabel{3}{211}{},
$$
	x^k\varepsilon_{kij}u^i\vect{u^j} =0.
$$
This implies that  $\bfx \times \bfu =0$ and so $\bfu$  is parallel to 
$\bfx$.  

Either way, we conclude that 
$\kappaG(E)$ is   a  two dimensional trivial  bundle  $(t,\bfx,A,B) \to 
(t, \bfx)$,  where the inclusion map
$\iota\:\kappaG(E) \to E$ is 
$$
	\iota(t,\bfx,A, B) = (t, \bfx,\bfu,p),\quad\text{ where \quad $\bfu = A 
\bfx$\quad  and  \quad $p =B$}.
$$  
The  invariants for the 
action of $G$  on $M$ are  $t$ and $r= \sqrt{x^2+ y^2 +z^2}$  so that the 
kinematic reduction diagram for the action of 
$\LieSO(3)$ on $E$ is 
$$
\CD
  	(t, r, A,  B)  @<\q_{\kappaG }<< (t,\bfx, A ,B)     @> \iota >>  
(t,\bfx,\bfu,p) 
\\
              @V \pi_{\tilde \kappaG} VV              @V\pi VV      @ VV  
\pi V
 \\
              (t,r)           @<\q_M < <                
(t,\bfx)            @>> \id >  (t,\bfx). 
\endCD
\Tag404
$$
In accordance with equation  \equationlabel{3}{210}{},  each  section  
$A=A(t,r)$ and $B=B(t,r)$ of $\tilkappaG(E)$    determines the  
rotationally 
invariant section  
$$
	\bfu = A(r,t) \, \bfx \quad\text{and}\quad   p = B(r,t)
\Tag405
$$
of $E$. 

The  computation of the reduced  equations for the rotationally invariant 
solutions to the Euler equations  now  proceeds  as follows. 
From  \eq{405} we compute
$$
	u^i_t = A_tx^i, \quad 
	u^i_j = A\delta^i_j + A_r \frac{x^ix_j}{r}\quad\text{and}\quad  
	p_i =  B_r  \frac{x^i}{r}
\Tag406
$$
so that  the Euler equations  \eq{401}  become
$$
	 A_tx^i  + Ax^j\bigl(A\delta^i_j + A_r \frac{x^ix_j}{r} \bigr)=- B_r  
\frac{x^i}{r} 
\quad\text{and}\quad 
                3A + r A_r = 0  
\Tag407
$$
which  simplify to the  differential equations  
$$
	A_t + A(A + r A_r)  = - \frac{B_r}{r}   \quad\text{and}\quad   3 A + r 
A_r = 0 
\Tag408
$$
on $J^1(\tilkappaG(E))$.  These equations are readily integrated to give 
$$
A= \frac{a}{r^3} \quad\text{and}\quad B=  \frac{\dot a }{r}  - \frac{ a^2 
}{2 r^4} + b  
\Tag410
$$
for arbitrary functions  $a(t)$ and $b(t)$   and  the  rotationally 
invariant solutions to the 
Euler equations  are 
$$
\bfu =   \, \frac{a}{r^3} \bfx \quad\text{and}\quad   p =\frac{\dot  a 
}{r}  - \frac{ a^2  }{2 r^4} + b. 
$$

We note that for the  Lie algebra of vector fields \eq{403}, the matrix 
on the right side of \equationlabel{2}{104}{}, namely
$$ 
\bmatrix
0           &   -x^3        &     x^2     &         0     &       -u^3    
&    u^2   \\
x^3       &   0              &   -x^1     &      u^3    &         0       
&    -u^1   \\
-x^2     &  x^1           &    0         &      -u^2   &         u^1   &   
0            
\endbmatrix
$$
has full rank 3 whereas the matrix on the left side of 
\equationlabel{2}{104}{}, 
consisting of the first three columns of \equationlabel{2}{104}{},  has 
rank 2.
The local transversality condition \equationlabel{2}{104}{} fails and   
the solution \eq{410}  to the Euler equations {\it cannot\/}  be obtained 
using the classical Lie  prescription.

To  describe  the derivation of   the reduced equations   in the  
context   of invariant differential operators and the dynamic reduction 
diagram 
we  introduce  the bundle  $ \D =J^1(E) \times \real^3 \times \real$  
with  sections $\dsize \vect{u^i}\otimes \, dt $ and $dt$ 
and   define the   differential operator $\Delta$ on $\D$ by
$$
	\Delta =  [u^i_t  + u^ku^i_k  + \delta^{ij}  p_j]  \,  \vect{u^i}\otimes 
\, dt   +   [u^i_i ]\, dt. 
\Tag447
$$
This operator is  invariant under the full symmetry group of the Euler 
equations.
The induced action of $G=\LieSO(3)$  on $J^1(E)$ is  given by 
$$
R\cdot (t, x^i, u^i, p, u^i_j, p_j) =  (t, R^i_rx^r, R^i_r u^r,  p, 
R^i_rR^s_ju^r_s, R^r_s p_r)
\quad\text{where $R\in \LieSO(3).$}
$$
Coordinates  for the   bundle of invariant  jets $\Inv^1(E)$ are  
$(t,x^i,A, A_t,A_r, B, B_t,B_r)$ and  \eq{406} defines
the inclusion map $\iota:\Inv^1(E) \to J^1(E)$.  A basis for the   $G$ 
invariant sections of $\D_{\Inv}\to \Inv^1(E)$ is  given 
by
$$
\bff{\,}^1= x^i \vect{u^i} \otimes dt \quad\text{and}\quad\bff{\,}^2 = dt.
$$ 
Let $\tilde \bff{\,}^1$ and $\tilde \bff{\,}^2$ be the corresponding 
sections of $\tilkappaG(\D_{\sssize \Inv})$. 

We are now ready to work  though the dynamic reduction diagram  
\equationlabel{4}{305}{}, starting with  the  
Euler operator  as  a section $\Delta \: J^1(E) \to \D$. Restricted to the 
invariant jet  bundle  $\Inv^1(E)$, 
$\Delta$  becomes
$$
	\Delta_{\sssize\Inv} = 
               [ A_tx^i  + Ax^j \bigl(A\delta^i_j + A_r 
\frac{x^ix_j}{r}\bigr)+ B_r  \frac{x^i}{r}  ] \, \vect{u^i}\otimes \, dt  +
               [\delta^j_i( A \delta^i_j + A_r\frac{ x^i x_j}{r}) ]\, dt. 
$$ 
Restricting   $\Delta$ to $\Inv^1(E)$  is precisely the  first step one  
takes   in  practice in computing the reduced equations and corresponds
to the right most square  in the  dynamic reduction diagram.

Next, because $\Delta_{\sssize \Inv}$ is $G$ invariant  it  is    
necessarily a linear combination of the  two invariant sections 
$\bff{\,}^1$ and $\bff{\,}^2$  and therefore  factors though the  
kinematic bundle $\kappaG(\D_{\Inv})$.  
This  means we can  write $\Delta_\Inv$  as a section of 
$\kappaG(\D_{\sssize \Inv})$, namely,
$$
	 \Delta_{\sssize\Inv}=
               [  A_t   + A \bigr(A   + A_r  \frac{x^jx_j}{r}\bigl) + 
\frac{1}{r}B_r    ] \, \bff{\,}^1 + 
               [3  A  + A_r  (\delta_i^j \frac{x^i x_j}{r})]\, \bff{\,}^2
$$ 
This  corresponds to the  center  commutative  square in the dynamic 
reduction diagram \equationlabel{4}{305}{} and coincides with the fact 
that  
equation \eq{407} contained a  common factor  $x^i$ --  a common factor 
which  insures that the  time evolution equation for
$\bfu$  reduces to a single  time evolution  equation  for $A$.

Finally, as a   $G$  invariant section of  $\kappaG(\Cal D_{\Inv})$, a 
bundle on which $G$  always acts transversally, we are assured that
$\Delta_\Inv$  descends to a differential operator   $\tilDelta$ on the   
bundle $J^1(\tilkappaG(E) )$. In this example
this implies that the independent variables $(t, x^i)$   appear  only 
though the  invariants for the action of $G$ on $M$, in this case  $t$ and 
$r$, and
so 
$$
	\tilDelta= [A_t +A( A + r A_r)  +\frac{B_r}{r}]\,\tilde\bff{\,}^1 +    [3 
A + r A_r]\, \tilde \bff{\,}^2  . 
\eqqed
$$
\medskip

\Ex{\State402  .\ \smc  A new reduction of the Euler equations} It is 
possible   to give a  complete classification of all possible symmetry 
reductions of the Euler equations 
\eq{401} to a  system of ordinary differential equations  in   three or  
fewer  dependent  variables  \cite{fels:1999a}.  A  number of  authors  
have  obtained    complete    lists of   reductions of  various 
differential  equations
(see, for example,   
\cite{david-kamran-levi-winternitz:1986a},\cite{fushchich-shtelen-slavutsky:1976a},  
\cite{grundland-winternitz-zakrzewski:1996a},  
\cite{winternitz-grundland-tuszynski:1987a})    but   this  particular 
classification of  reductions of the Euler equations may be  the first 
such 
classification of   group invariant  solutions  which explicitly  
requires   non-trivial  isotropy  in  the  group action on the space of 
independent  variables.
There are  too  many  cases to list  the results of  this  classification  
here,    but we do present  one more  reduction of  the Euler equations, 
one which   does  not  seem to appear elsewhere in the literature. 

For this example  it will be convenient to write $\bfx=(x,y,z)$ and 
$\bfu=(u,v,w)$. 
The infinitesimal  generators for  the  group action are   
$
\Gamma=\{ \,V_0, V_1, V_2 = V_{x,\alpha} +  V_{y,\beta},
V_3 = V_{y,\alpha} - V_{x,\beta}\,\},
$
where 
$$
\alignat2
	V_0 &= x\partial_x + y\partial_y + z\partial_z +  u\partial_u  + 
v\partial_v + w \partial_w + 2 p \partial_p, 
&\qquad
	V_1 & = y\partial_x - x \partial_y  + v\partial_u - u\partial_v,
\\
\vspace{2\jot}
V_{x,\alpha} &= \alpha \partial_x + \dot \alpha\partial_u -  x\ddot 
\alpha  \partial_p, \qquad\text{and}
&\qquad
V_{y,\beta} &= \beta \partial_y +  \dot \beta \partial_v  -y \ddot \beta   
\partial_p,
\endalignat
$$
 Here  $\alpha = \alpha(t)$ and $\beta = \beta(t)$ are such that 
 $ \alpha\ddot \beta  -  \ddot \alpha \beta=0$, or equivalently, 
$$
\alpha \dot \beta - \beta \dot \alpha = c =\text{constant}.
\Tag425
$$
This   condition   insures that $[V_2,V_3] = 0$ so that $\Gamma$  is 
indeed a  finite dimensional
Lie algebra of  vector fields.   In order that  $\Gamma$ have constant 
rank on  the base space,
we assume that  $xy \alpha \neq 0$ or $ yz\beta \neq 0$.

The horizontal components of $V_2$ and $V_3$  are given by
$$
\bmatrix V_2^M \\ V_3^M \endbmatrix = \bmatrix  \alpha  &\beta \\ - \beta 
& \alpha  \endbmatrix \bmatrix \partial_x \\ \partial_y \endbmatrix,
\quad
\text{so that}  
\quad
\bmatrix \partial_x \\ \partial_y \endbmatrix = \frac{1}{\delta} \bmatrix  
\alpha  &-\beta \\  \beta & \alpha  \endbmatrix 
\bmatrix V_2^M \\ V_3^M \endbmatrix,
$$
where $\delta = \alpha^2 + \beta^2$, and therefore 
at the point  $(t_0,\bfx_0)$,     the  horizontal  components  of the 
vector field
$$
Z= V_1 -   y_0 \frac{\alpha(t_0) V_2 -\beta(t_0) V_3}{\delta(t_0)}  +  
x_0\frac{\beta(t_0) V_2 + \alpha(t_0) V_3}{\delta(t_0)}  
\Tag437
$$
 vanish.   The isotropy condition \equationlabel{3}{211}{}  defining the 
fiber of the  kinematic  bundle $\kappaGammapt{x}(E)$  leads, from the
coefficients of $\partial_u$, $\partial_v$ and $\partial_p$,    to the 
relations
$$
v  =   \frac{y \alpha  - x \beta}{\delta}\dot \alpha   + \frac{ x\alpha + 
y\beta}{\delta}{\dot \beta} 
\quad\text{and}\quad
u=  \frac{x \alpha + y \beta}{\delta} \dot \alpha + \frac{x\beta - 
y\alpha}{\delta}{\dot \beta}.
\Tag427
\endalign
$$
We therefore conclude that  the  kinematic  bundle  has fiber dimension  
2  with  fiber coordinates  $w$ and $p$.  However,   these
coordinates are not invariant  under the   action of  $\Gamma$  on 
$\kappaGamma(E)$ and cannot be used  in 
the local coordinate  description  \equationlabel{3}{209}{} of the  
kinematic  reduction diagram.  Restricted to 
$\kappaGamma(E)$, the  vector fields $V_i$  become
$$
\alignat2 
V'_0 &= x\partial_x + y\partial_y + z\partial_z   + w \partial _w + 2 p 
\partial_p, 
&\quad
V'_1 & = y\partial_x - x \partial_y,  
\\
\vspace{2\jot}
 V'_2 &= \alpha \partial_x  +\beta\partial_y   - (\ddot \alpha  x  + 
\ddot\beta y ) \partial_p,\quad\text{and}\quad
&\quad
 V'_3 &= -\beta \partial_x + \alpha \partial_y  -(-\ddot \beta x    
+\ddot  \alpha y )\partial_p.     
\endalignat
$$

Note  that  these restricted  vector fields  now  satisfy the  
infinitesimal  transversality condition \equationlabel{2}{104}{}.  
Invariants for this
action are $t$,
$$
A =\frac{w}{z}\quad\text{and}\quad B=  \bigl(2p +   \frac{\alpha \ddot 
\alpha + \beta \ddot  \beta}{\delta}(x^2 +y^2)\bigr)/z^2.
\Tag433
$$
To verify that $B$  satisfies  $V'_2(B) =  V'_3(B) = 0$ one must use   $ 
\alpha \ddot \beta  =\ddot \alpha \beta$. 
The   kinematic   reduction diagram for the action of $\Gamma$  on $E$ is 
therefore
$$
\CD
(t, A,B)   @<\q_{\kappaGamma}  << (t, \bfx, A, B) @>\iota >> (t,\bfx, 
\bfu, p) \\  
@V\tilde \pi VV                                                           
@V\pi VV                               @VV \pi V \\
   (t)        @    << \qM<                              
(t,\bfx)            @>> \id > (t,\bfx),  
\endCD
$$
where the  inclusion map  $\iota$  is  defined  by \eq{427} and the 
solutions to \eq{433} for  $w$ and $p$.
The  general invariant section is then, on putting  $\sigma= \ln \delta$,  
$$
\alignat2 
	u &= x  \frac{\dot\sigma}{2} - y \frac{c}{\delta}, &\qquad
               v& =  x\frac{c}{\delta}  +     y\frac{\dot \sigma}{2}, 
\\
\vspace{2\jot}
               w& = zA(t) , &\qquad  p &=  - \frac{\alpha \ddot \alpha + 
\beta \ddot \beta}{2\delta}\,(x^2 +y^2)   + \frac{z^2}{2} B(t).
\endalignat
$$ 
Note that the  $u$ and $v$  components are   uniquely   determined  from 
the  isotropy    conditions \eq{437} and \eq{427}   and  that the  
arbitrary functions $A(t)$ and $B(t)$
defining these invariant sections appear only in  the  $w$ and $p$  
components.

\def\bfA{\text{\bf A}}

We now turn to the dynamic reduction diagram.  Since  we  are  treating 
the  Euler equations as the  section \eq{447}  of the
tensor  bundle  $\D$  we  can anticipate the    form of $\Delta_{\sssize 
\Inv}$ by  computing the  $\Gamma$ invariant tensors  of the form
$$
T=  P\,\partial_u\otimes  d\,t +  Q\,\partial_v  \otimes d\,t   + 
R\,\partial_w\otimes d \,t  + S  d\,t .
\Tag428
$$
The  isotropy condition  $\Cal L_Z T=0$ at  $x_0$, where $Z$ is defined by 
\eq{437},  shows immediately that   $P= Q=0$  from which it 
follows that
$$
\bff{\,}^1=  z\vect{w}\otimes \, dt  \quad\text{and}\quad  \bff{\,}^2 =   
d\,t 
$$ 
are    a basis for the  $\Gamma$  invariant   fields of the type 
\eq{428}.   This  calculation  shows that the $\partial _u\otimes dt$ and
$\partial_v \otimes dt$ components of the reduced Euler equations  must 
vanish  identically and, consistent with this conclusion,
  one readily computes
$$
\align
	\Delta_{\sssize \Inv} 
&= 	\bigl[\frac{ \ddot {\sigma} }{2} + ( \frac{ \dot \sigma}{2})^2 - 
\bigl( \frac{c}{\delta}\bigr)^2-
               \frac{ (\alpha \ddot \alpha + \beta \ddot 
\beta)}{\delta}\bigr]\,[ x \vect{u} \otimes  dt +  y \vect{v} \otimes dt]
             + \bigl( \dot A + A^2    +  B \bigr)  \bff{\,}^1 +  
\bigl(\dot \sigma + A \bigr) \bff{\,}^2 
\\
\vspace{2\jot}
&	=
              \bigl( \dot A + A^2    +  B \bigr)  \bff{\,}^1  +  
\bigl(\dot \sigma + A \bigr) \bff{\,}^2 .
\endalign
$$
Thus,  the reduced differential equations are
$$
\dot A + A^2   + B = 0  \quad \text{and}\quad   \dot  \sigma + A = 0
$$   
which determine    $A$ and $B$ algebraically. 
In conclusion,  for  each  choice of $ \alpha$ and $\beta$ there is  
precisely  {\it one \/}  
$\Gamma$ invariant  solution to the Euler equations given by
$$
\gather
	u  = x \frac{ \alpha \dot \alpha + \beta \dot \beta}{\alpha^2+\beta^2} - 
y \frac{ \alpha \dot \beta - \dot \alpha \beta }{\alpha^2+\beta^2},
\qquad
	v = x \frac{ \alpha \dot \beta - \dot \alpha \beta  }{\alpha^2+\beta^2} +
	y \frac{ \alpha \dot \alpha + \beta \dot \beta}{\alpha^2+\beta^2},
\qquad
	w  = -2z \frac{ \alpha \dot \alpha + \beta \dot \beta}{\alpha^2+\beta^2}, 
\\
\vspace{2\jot}
	p =  -\frac{1}{2} (x^2 +y^2 ) \frac{ \alpha \ddot \alpha + \beta \ddot 
\beta}{\alpha^2+\beta^2} + 
      	z^2\biggl( \frac{ \alpha \ddot \alpha + \beta \ddot 
\beta}{\alpha^2+\beta^2}
	+  \bigl(\frac{ \alpha \dot \beta  - \beta \dot 
\alpha}{\alpha^2+\beta^2}\bigr)^2 
	-3  \bigl(\frac{ \alpha \dot \alpha + \beta \dot \beta}{\alpha^2+\beta^2} 
\bigr)^2  \biggr).
                       \eqqed
\endgather
$$

\medskip

\def\D{\Cal  D}

\heading Harmonic Maps  \endheading

For our next   examples  we  look  at two    well-known reductions of  the 
harmonic map equation  for maps between spheres. 
For these examples the bundle $E$ is  $S^n\times S^m \to S^n$ which we  
realize as  a subset of  $\real^{n+1}\times \real^{m+1}$ 
by
$$
	E = \bigl\{ \, (\bfx, \bfu) \in \real^{n+1}\times \real^{m+1} \, | \, 
\bfx\cdot  \bfx = \bfu \cdot \bfu = 1 \, \bigr\}.
$$
Let $G$  be  a Lie  subgroup of   $\LieSO(n+1)$,  let  $\rho\:G\to 
\LieSO(m+1)$ be a  Lie group  homomorphism
and define the action of $G$ on $E$  by
$$
R \cdot  (\bfx, \bfu)  = ( R\cdot  \bfx, \rho(R) \cdot \bfu) 
\quad\text{for}\quad  R\in G.
$$
The  kinematic bundle for the $G$ invariant  sections of  $E$  has fiber
$$
	\kappaGpt{\bfx}(E) = 
                \bigl\{(\bfx, \bfu) \in E  \, |\,  \rho(R) \cdot \bfu  = 
\bfu \quad
                \text{for all $R\in G$ such that $R \cdot \bfx= \bfx$}\, 
\bigr\}.
$$
We  identify the   jet space  $J^2(E)$   with a  submanifold  of  
$J^2(\real^{n+1},\real^{m+1})$ by 
$$
	J^2(E) = \{\,(\bfx, \bfu,  \partial_i\bfu,  \partial_{ij}\bfu) \in 
J^2(\real^{n+1},\real^{m+1}) \, | 
	\,  \bfx \cdot  \bfx = 1,   \bfu \cdot  \bfu =1,   \bfu \cdot \partial_i  
\bfu  = 0,  \bfu \cdot  \partial_{ij} \bfu    +
               \partial_i \bfu  \cdot 	\partial_j  \bfu = 0\,   \}. 
$$
Since the  harmonic map operator  (or tension field)  is  a  tangent 
vector   to the  target  sphere $S^m$ at each point
 $\sigma\in  J^2(E)$, we   let
$$
\Cal D = \{ ( \sigma, \bfDelta) \in J^2(E) \times \real^{m+1} \,|\,   \bfu 
\cdot \bfDelta = 0 \}.
\Tag429
$$
 By  combining  Proposition    I.1.17 (p.19)  and   Lemma VII.1.2 (p.129)  
in  Eells and Ratto   \cite{eells-ratto:1993a}, 
 it follows that   one can   write the  harmonic map operator 
$\Delta\:J^2(E) \to \Cal D$ as the map
$$  
	\Delta(\sigma) =
 	\bigl[\Delta^{\real^{n+1}}u^\alpha    + x^i x^j 	u^\alpha_{ij}  +n 
x^iu^\alpha_i -  \lambda u^\alpha \bigr ]\, 	\vect{u^\alpha},
\Tag434
$$
where  
$$
	\lambda =  \delta_{\alpha\beta}[\delta^{ij}u^\alpha_i u^\beta_j -  x^i 
x^j u^\alpha _i u^\beta_j]  \quad\text{and}\quad 
               \Delta^{\real^{n+1}}u^\alpha = -  \delta^{ij}u^\alpha_{ij}.
$$
This operator is invariant under the induced  action of   $\bigG=  
\LieSO(n+1) \times \LieSO(m+1)$ on $E$.

\Ex{ \State412 .  \  \smc  Harmonic maps from $S^2$ to $S^4$} For our 
first example we take $E= S^2\times S^4  \to S^2$ and we look for harmonic 
maps which are invariant  under the  standard action of  $\LieSO(3)$ 
acting  on $S^2$.   It can be proved that, up to conjugation,  there   are 
three  distinct  group homomorphisms
$\rho\:\LieSO(3) \to  \LieSO(5)$,  which  lead to  the following   three 
possibilities for the   infinitesimal  generators  of $\LieSO(5) $  acting 
on $E$.
$$
\gather
\text{Case I}\quad 
\cases 
V_1=  z \partial_y -  y\partial_z, \\ 
\vspace{2\jot}
V_2=   x\partial_z -  z\partial_x,  \\ 
\vspace{2\jot}
V_3=  y \partial_x - x\partial_y  . \\ 
\vspace{2\jot}
\endcases
\qquad\qquad
\text {Case II}\quad 
\cases 
V_1=  z\partial_y -  y \partial_z    -     u^2\partial_{u^3}+ 
u^3\partial_{u^2},  
\\  
\vspace{2\jot}
V_2=  x \partial_z  - z\partial_x   -    u^3\partial_{u^1}+ 
u^1\partial_{u^3},     
 \\ 
\vspace{2\jot}
 V_3=  y\partial_x  - x\partial_y  -     u^1\partial_{u^2} + 
u^2\partial_{u^1}. 
\\ 
\endcases
\\
\vspace{4\jot}
\text{Case III} \quad \cases
 V_1=       z\partial_y -    y\partial_z
  	+u^2\partial_{u^1} - u^1\partial_{u^2} +(u^4 -\sqrt{3}u^5) 
\partial_{u^3} -u^3\partial_{u^4} +	\sqrt{3}u^3\partial_{u^5},
 \\
\vspace{2\jot}
  V_2=   x\partial_z - z\partial_x   
	-u^3\partial_{u^1} +(u^4 +\sqrt{3}u^5)\partial_{u^2} +u^1\partial_{u^3} 
-u^2\partial_{u^4} 	-\sqrt{3}u^2\partial_{u^5},
\\
\vspace{2\jot}
  V_3=  y\partial_x  - x \partial_y
	-2u^4\partial_{u^1} + u^3\partial_{u^2}-u^2\partial_{u^3} 
+2u^1\partial_{u^4}. 
\endcases
 \endgather
$$
In Case I,  the map   $\rho$ is the constant  map, and  in  Case II,  
$\rho$ is the standard inclusion of $\LieSO(3)$ into $\LieSO(5)$.  The  
origin of the
map  $\rho$  in Case III will be discussed shortly.

Since $\LieSO(3)$ acts transitively on $S^2$, the orbit manifold $\tilM$ 
consists of a single point,   the 
space of invariant  sections is  a  finite dimensional manifold,  and the 
reduced   differential  equations are  algebraic equations. 
The kinematic bundles   $\kappaG(E)$ are determined  in each case  from  
the  isotropy  constraint  
$$
	xV_1 + y V_2 + zV_3= 0.
$$
In   Case I   the  action is transverse,  the isotropy constraint is 
vacuous  and   the  kinematic bundle  is 
$\kappa_G(E) =  S^2\times S^4$. The invariant sections
are  given  by
$$
	\Phi_{\sssize I}(x,y,z) = (A,B,C,D,E),  
$$
where $A$,\dots, $E$ are constants and $A^2 + B^2 +C^2 +D^2 + E^2 = 1$.   
In Case II  the kinematic bundle is $S^2\times S^2$ and the invariant 
sections are 
$$
\Phi_{\sssize II}(x,y,z)=(Ax, Ay,Az, B,C),
$$ 
where $A$,  $B$,   $C$ are constants such that  $A^2 +B^2 + C^2 = 1$. We 
take $A\neq 0$, since otherwise  $\Phi_{\sssize II}$ becomes
a special case of $\Phi_{\sssize I}$.  In Case III,  $\kappaG(E)= S^2 
\times \{\,\pm1\,\}$ and the invariant sections are
$$
\Phi_{\sssize III}(x,y,z) = A
\sqrt{3}\bigr(xy, xz,  yz, \frac12(x^2 -y^2) , \frac{\sqrt{3}}{6}(x^2 +y^2 
- 2z^2)\bigl) ,
$$
where $A = \pm 1$.

Direct substitution    into \eq{434} easily shows that  the maps 
$\Phi_{\sssize I}$ and $\Phi_{\sssize III }$ automatically  satisfy
the harmonic  map equation.  The map $\Phi_{\sssize II}$ is  harmonic if 
and only if $B=C=0$ in which
case $\Phi_{\sssize II}$ is  either the  identity map or the antipodal map 
on $S^2$  followed  by the standard
inclusion into $S^4$. Despite the simplicity of these conclusions, it is  
nevertheless instructive to look
at the  corresponding dynamic  reduction diagrams.

In Case I, the invariant sections are constant and so 
 $$
\align
\Inv^2(E)&= \{(\bfx, \bfA) \in \real^3\times\real^5\, | \, \bfx\cdot \bfx 
= \bfA\cdot \bfA =1, \}
\\
\squash{3}{and}{3}
\Cal D_{\Inv}&=\{(\sigma, \bfDelta) \in  \Inv^2(E) \times  \real^5 \, |\, 
\bfA \cdot  \bfDelta  = 0 \}.
\endalign
$$
The automorphism  group  for the kinematic bundle in this case is 
$\tilbigG = \LieSO(3) \times \LieSO(5)$ which acts on 
$\Cal D_{\Inv}$ by
$$
	(R,S) \cdot  (\bfx, \bfA, \bfDelta) =
 	(R \cdot  \bfx, S \cdot \bfA,  S \cdot \bfDelta)\quad\text{for $R\in 
\LieSO(3)$ and
	$S\in \LieSO(5)$}.
$$
The isotropy constraint  for  $\kappa_{\ssize \tilbigG}(\Cal D_{\sssize 
\Inv})$    forces $\bfDelta$ to be a multiple
of  $\bfA$.  Hence, by the tangency condition  $\bfA\cdot \bfDelta=0$,  we 
have  $ \bfDelta = 0$ and 
$
\kappa_{\ssize \tilbigG, \sigma }(\Cal D_{\Inv}) = 0.
$
This  shows   that the map $\Phi_{\sssize I}$ is harmonic  by symmetry 
considerations alone and moreover that it is 
a  universal  solution  for  any operator   $\Delta\: J^k(S^2\times S^4) 
\to \Cal D$ with 
$\LieSO(3) \times \LieSO(5)$ symmetry.

In Case II, the harmonic map equations force $B= C= 0$ so that   the maps 
$\Phi_{\sssize II}$ are not
universal.  Interestingly however,  the  standard and antipodal 
inclusions  $S^2 \to S^4$ have  a larger symmetry 
group, namely $\LieSO(3) \times \LieSO(2) \subset \bigG$ and it is easily  
seen, using these larger symmetry 
groups,    that the   standard and antipodal inclusions   are  universal.  
It   is a  common phenomenon  that  
 the group invariant solutions to a system  of differential equations  
possess  a  larger  symmetry  group than the   original  group  used  in 
their   construction.  

In Case III  one finds  immediately that  $\kappaGpt{\sigma}(\Cal 
D_{\Inv}) = 0$ and    $\Phi_{\sssize III}$ 
is universal, again for  any operator   $\Delta\: J^k(S^2\times S^4) \to 
\Cal D$ with 
$\LieSO(3) \times \LieSO(5)$ symmetry.

The map $\Phi_{\sssize III}$ is  the classic Veronese map.   The  
symmetry  group  defining it  is based on a standard irreducible  
representation  of  $\LieSO(3)$ which  readily generalizes to  give 
harmonic maps  between various spheres of higher dimension. 
Specifically, starting with the standard action of
$\LieSO(n) $ on $V= \real^n$,  consider the induced action on 
$\text{Sym}^k_{\text{tr}}(V)$, the space of rank $k$  symmetric, 
trace-free tensors
or,  equivalently, on the space  $W=\Cal H^k(V)$  of   harmonic 
polynomials of  degree $k$  on $V$. The  standard metric  on $W$ is 
invariant under this action of $\LieSO(n)$ and in this way one obtains a 
Lie group  monomorphism  $\rho\:\LieSO(n) \to \LieSO(N)$,
where $N = \dim(W)=\binom{n+k-1}{k} - 1$.  For example, the  polynomials 
$$
u^1=xy,\quad, u^2=xz,\quad u^3=yz,\quad u^4= 1/2(x^2-y^2),\quad  
u^5=\sqrt{3}/6( x^2 +y^2 - 2z^2 ) 
$$
form an orthogonal basis for $\Cal H^2(\real^3)$ and the action of 
$\LieSO(3)$ on this space  determines the action of
$\LieSO(3)$  on $\real^3\times\real^5$ in Case III. For further  examples  
see  Eells and Ratto   \cite{eells-ratto:1993a} and Toth  
\cite{toth:1990a}. \endEx 

\Ex{\State420 . \  \smc  Harmonic Maps from $S^n$ to $S^n$}
A basic  result of Smith   \cite{smith:1975a}  states that each  element 
of  $\pi_n(S^n)  =\text{\bf  Z}$  can be  represented by a harmonic map 
(with respect to the standard metric)    provided $n\leq 7 $ or  $n = 
9$.   This  result,  which can be  established by  symmetry reduction of 
the 
harmonic map equation (see    Eells and Ratto   \cite{eells-ratto:1993a}  
and  Urakawa   \cite{urakawa:1993a}),   illustrates  a number
of interesting  features. First,  we  see that  much of   the general 
theory  which we have outlined could  be extended to the case
where $\tilM$ is a manifold with  boundary  and where  the   fibers of   
$\kappaG(E)$ change   topological type on the boundary.
 Secondly, we  find that the  invariant sections for  the   standard 
action of $G = \LieSO(n-1)\times \LieSO(2) \subset \LieSO(n+1)$ on $S^n$
are  slightly more  general than those  considered   in  
\cite{eells-ratto:1993a}  and   \cite{urakawa:1993a}. However,  a simple 
analysis
of the reduced equations,  based upon Noether's theorem, shows that the  
only solutions to the   reduced equations  are essentially those 
provided by the ansatz  used by  Eells and Ratto and  Urakawa.

If   $(R,S) \in G = \LieSO(n-1)\times \LieSO(2) \subset \LieSO(n+1)$ and 
$$
	(\bfx,\bfy,\bfu, \bfv)\in E\subset(\real^{n-1}\times \real^2)\times 
(\real^{n-1}\times \real^2),
$$
where $||\bfx||^2 +||\bfy^2|| =1$ and $||\bfu||^2 + ||\bfv||^2 =1$, 
then  the action of $G$ on $E=S^n \times S^n$ is given by
$$
	(R,S)(\bfx, \bfy, \bfu,\bfv) = 
	\bigl( 
	\bmatrix  R  &  0 \\   0 &  S \endbmatrix   \bmatrix   \bfx \\  \bfy  
\endbmatrix \, , \,  
               \bmatrix  R  &   0 \\  0 &  S  \endbmatrix  \bmatrix   
\bfu  \\  \bfv  \endbmatrix
                \bigr).
$$
The invariants for the action of $G$  on  the base  $\real^{n+1}$ are  $r= 
||\bfx||$ and  $s=|| \bfy||$ which, 
for points $(\bfx, \bfy)\in S^n$, are
related by  $r^2 +s^2 = 1$, where $r\geq 0$ and  $s\geq 0$.   
The quotient manifold  $\tilM= S^n/G$ is  therefore  diffeomorphic to the 
closed interval $[0, \pi/2]$.  

To describe the kinematic bundle   $\kappaG(E)$ we must consider  
separately  those points in $M$  for  which  (i) $s=0$, 
(ii)   $s\ne 0 $ and $r\ne 0$ and (iii) $r= 0$,  corresponding the  
left-hand boundary point, the interior points and the  right-hand boundary 
points of $\tilM$. 
For  $(\bfx, 0) \in S^n$, the isotropy subalgebra is  
$\LieSO(n-1)_{\bfx}\times \LieSO(2)$ and the  fiber of the  kinematic 
bundle consists 
of a  pair of points  
$$
	\kappaGpt{(\bfx,0)}(E) =  
               \{\, (\bfx,0, \bfu,\bfv) \, | \,  \bfu = \pm \bfx, 
\quad\text{and}\quad  \bfv= 0\,\}.
$$
For  points $(\bfx,\bfy) \in S^n$ with $r\neq0$  and $s\neq 0$  the 
isotropy group is $\LieSO(n-1)_{\bfx} \times \{\,I\,\}$
and  the fiber of the kinematic bundle is the ellipsoid of revolution
$$
	\kappaGpt{(\bfx,\bfy)}(E) =  
               \{\, (\bfx,\bfy, \bfu,\bfv) \, | \,  \bfu =   A\bfx, 
\quad\text{where}\quad  r^2 A^2 +  ||\bfv||^2 =1 \}.
$$
Invariant  coordinates on $\kappaGpt{(\bfx,\bfy)}(E)$ are  
$\dsize A =   \frac{\bfx\cdot \bfu}{r^2}$, $\dsize  B=    \frac{\bfy \cdot 
\bfv}{s^2}$  and 
$C =  \dsize    \frac{\bfy^{\perp} \cdot \bfv}{s^2}$ , where $\bfy^{\perp} 
= (0,-y^2, y^1)$, 
subject to  
$$ 
	 r^2A^2 +  s^2(B^2 +C^2) = 1.
\Tag436
$$ 
The inclusion map  from $\kappaGpt{(\bfx,\bfy)}(E)$ to  $E_{(\bfx,\bfy)}$ 
is
$$
\bfu = A\bfx \quad\text{and}\quad \bfv= B \bfy +  C\bfy^{\perp}.
$$ 
 At the points $(0,\bfy)$, 
 the isotropy  subalgebra  is $\LieSO(n) \times \{\,I\,\}$ and the  fiber 
of  the kinematic bundle is
the circle 
$$
                \kappaGpt{(0,\bfy)}(E)= 
                \{\,(0,\bfy,\bfu,\bfv)\,|\,\bfu=0	
                 \quad\text{and}\quad ||\bfv|| =1\,\}.
$$
%The   quotient space $\tilkappaG(E)$ is  shown in  Figure 1.  
The  $G$  invariant sections  are therefore described, as maps  
$\Phi\:\real^{n+1} \to \real^{n+1}$,  by
$$
	\Phi(\bfx, \bfy) =   A(t)\bfx +   B(t) \bfy +  C(t)\bfy^{\perp},     
\Tag422
$$ 
where $t$ is  the smooth function of $(\bfx, \bfy)$  defined by  
$\dsize
\cos(t) =  \frac{r}{r^2 + s^2}
$
 and
$\dsize 
\sin(t) = \frac{s}{r^2 + s^2},
$ 
and where
$
 \cos^2(t)  A^2(t) +  \sin^2(t) (B^2(t)  +C^2(t))      = 1.
$
The isotropy conditions at  the   boundary of $\tilM$  imply  that  the 
functions  $A$, $B$ and $C$ are subject to the  boundary conditions
$$
	A(0)  = \pm 1, \quad B(0) = 0, \quad C(0) = 0 ,
               \quad\text{and} \quad 
               A(\frac{\pi}{2}) = 0,\quad  B(\frac{\pi}{2})^2 	+ 
C(\frac{\pi}{2})^2 =1 .
\Tag421
$$
The  invariant sections considered  in  \cite{eells-ratto:1993a}  and  
\cite{urakawa:1993a} correspond to $C(t)= 0$. 
Note that the  space of     invariant sections \eq{422} is  preserved by  
rotations in the  $\bfv$ plane, that is,  rotations
in the  $BC$ plane  and  therefore $ \LieSO(2) \subset 
\tilbigG_{\text{eff}}$.

% \line{}
% \bigskip
% \vskip 4cm
% \midinsert
% \captionwidth{16cm}
% \hskip20pt{\special{eps:kbundle.eps x=12cm y=4cm}}
% \botcaption {Figure 1}  The  reduced kinematic bundle for  
% $\LieSO(n-1)\times \LieSO(2)$ invariant maps $s\:S^n \to S^n$
% \endcaption
% \endinsert
% \bigskip
By computing  $ \kappa_G(\Cal D_{\sssize \Inv})$  we  deduce that  the   
restricted   harmonic operator   $\Delta_{\sssize \Inv}$ is of the form 
$$
\Delta_{\sssize \Inv} =  
	\Delta_{\sssize A}\,  \bigl( \bfx   \cdot \vect{{\bfu}}\bigr)  + 
	\Delta_{\sssize B }\,  \bigl( \bfy \cdot \vect{{\bfv}}  \bigr)  + 
	\Delta_{\sssize C}\,   \bigl(\bfy^{\perp} \cdot \vect{{\bfv 
^\perp}}\bigr),
$$
where the     tangency condition \eq{429} reduces to
$$
r^2   A \Delta_{\sssize A}  + s^2 B\Delta_{\sssize B}  + s^2 C 
\Delta_{\sssize C} = 0.
$$
A  series of    straightforward  calculations, using  \eq{434},  now
 shows that the coefficients  of  the reduced  operator $\tilde \Delta$ 
are  
$$
\aligned
	\tilDelta_{\sssize A}
&
	=   -\ddot  A +  \bigl( n\frac{\sin(t)}{\cos(t)} - 
\frac{\cos(t)}{\sin(t)} \bigr) \dot  A   +
	n A  - \lambda A, 
\\
\vspace{2\jot}
	\tilDelta_{\sssize B}
&
	= -\ddot  B  +  \bigl( (n-2)\frac{\sin(t)}{\cos(t)} - 
3\frac{\cos(t)}{\sin(t)}\bigr ) \dot B  +
 	n B -\lambda B, 
\\
\vspace{2 \jot}
	\tilDelta_{\sssize C}
&
	=-\ddot  C +  \bigl( (n-2)\frac{\sin(t)}{\cos(t)} - 
3\frac{\cos(t)}{\sin(t)}\bigr ) \dot C  +  
	n C  - \lambda C,
\endaligned
\Tag423
$$
where 
$$
\align
	\lambda& =   \cos^2(t)\dot A^2  + \sin^2(t)\bigl( \dot B^2 + \dot C^2 
\bigr) +  
 	2\cos(t)\sin(t)\bigl (- A\dot A + B \dot B + C\dot C\bigr)  \\
\vspace{2\jot}
&              + (n-1) A^2  + 2 (B^2 + C^2) -\cos^2(t)  A^2 -  \sin^2(t) 
(B^2  +C^2).
	\endalign
$$

To analyze these equations, we  first invoke the principle of symmetric  
criticality and the  formulas in  \cite{anderson-fels:1997a}  for the 
reduced Lagrangian to conclude
that  these  equations are the  Euler-Lagrange equations for the  reduced 
Lagrangian
$$
	\tilde L = 
	\frac{1}{2}\cos(t)^{n-2}\sin(t)   \lambda\, dt
$$
subject,  of course, to the  constraint    \eq{436}.  From  knowledge of 
the automorphism group of the kinematic bundle
we  know that   this  Lagrangian  is  invariant under  rotations in   the  
$BC$  plane   and this leads to the  first integral
$$
	J= \cos(t)^{n-2} \sin(t) ^3(  B \dot  C - C  \dot B)
$$
for \eq{423}.
By the  boundary conditions   \eq{421},   $J$ must  vanish identically. 
Thus   $C(t) =  \mu B(t)$, for some constant $\mu$
and therefore  a rotation  in the  $\bfv ,\bfv^{\perp}$  plane  will 
rotate the 
 general invariant section  \eq{422} into   the section with $C(t) = 0$.   
We then have  $r^2 A^2 + s^2B^2 = 1$ and the change of
variables 
$$
	A(t) = \frac{\cos(\phi(t))}{\cos(t)}  \quad\text{and}  \quad  B(t) = 
\frac{\sin(\phi(t))}{ \sin(t)}
$$   
converts the   reduced operator  \eq{423} into the form found
in    \cite{eells-ratto:1993a} or  \cite{urakawa:1993a}.
\endEx

\heading  
	General Relativity
\endheading

We  now turn to  some examples of  Lie  symmetry reduction in general 
relativity which we again examine from the viewpoint
of the kinematic and dynamic reduction diagrams.  To study reductions of 
the Einstein  field equations,  we  take the bundle $E$ to  be  the  
bundle  $Q(M)$ of 
quadratic forms, with Lorentz signature, on  a  4-dimensional manifold 
$M$.  
A section of $E$ then corresponds to a  choice of  Lorentz metric  on 
$M$.  
We  view the  Einstein tensor  
$$
	\Delta =    G^{ij}(g_{hk}\,, g_{hk,l}\,, g_{hk,lm}) \vect{x^i} \otimes 
\vect{x^j}
$$
formally as  a section of $\Cal D \to J^2(E)$,  where $\Cal D$ is pullback 
of $V=\text{Sym}^2(T(M))$ to the  bundle of
2-jets $J^2(E)$.
The operator $\Delta$ is invariant under the Lie pseudo-group $ \bigG$ of 
all local diffeomorphisms of $M$.

Let $\Div_g $ be the covariant divergence operator (defined  by the metric 
connection for $g$)  acting on (1,1) tensors,
$$
\Div_g(S)  =  \nabla_i S^i_j \, dx^j .
$$
The contracted Bianchi identity is  $\Div_g \Delta^{\flat} = 0 $,  where 
$\Delta^{\flat}$ is  the operator obtained from $\Delta$
by lowering an index with the metric.

The  first  point we  wish to underscore  with the following  examples  is 
that  the kinematic reduction diagram   gives a  remarkably
 efficient  means of  solving the   Killing  equations  for the  
determination of the  invariant  metrics. Secondly,  we show that  
discrete symmetries,
which  will not  change the  dimension of the  reduced  spacetime  
$\tilM$,   can lead to isotropy constraints which  reduce the  fiber  
dimension of the 
kinematic bundle. Thirdly,
for $G$ invariant metrics,  the divergence operator  $\Div_g$  is  a $G$  
invariant operator to which the dynamical   reduction  procedure can
be applied to obtain  the reduction of the  contracted  Bianchi identities 
for the reduced equations. Throughout,  we emphasize the importance of 
the   residual symmetry   group
in analyzing  the  reductions of the 
field equations.

Finally, we remark  that our  conclusions in these examples are not 
restricted to the Einstein equations but in fact hold  for 
any  generally covariant  metric field  theories  derivable from  a 
variational principle.

\Ex{\State403 .\  \smc  Spherically Symmetric and   Stationary, 
Spherically Symmetric Reductions } We begin by looking at spherically 
symmetric solutions on the  four dimensional  manifold  $M= \real \times 
(\real^3 -\{\,0\,\})$,
with coordinates $(x^i) =(t,x,y,z)$ for  $i =0,1,2,3$.   
Although this is  a very well-understood example,  it is nevertheless 
instructive to consider it within the general  theory of  Lie symmetry 
reduction of differential
equations.   The  infinitesimal  generators  for $G=\LieSO(3)$ are   given 
by   \equationlabel{2}{120}{} and, just as in Example \st{401},
we  find that the  infinitesimal isotropy constraint   defining   
$\kappaGpt{x}(E)=\kappaGammapt{x}(E)$ is
$$
\varepsilon_{0kij} x^kg_{li}\vect{g_{lj}}= 0,
$$
or,  in terms of matrices, 
$$
	  ga  + a^{t}g = 0,    
\Tag412
$$      
where
$$
	a= \bmatrix 0&0&0 &0 \\ 0 & 0 & z& -y\\ 0 &  -z &0 &   x \\ 0 & y & -x & 
0 \endbmatrix
$$
and $g=[\,g_{ij}\,]$. These linear equations are easily solved  to  give 
$$
g	
=
A \bmatrix   
     1                   &    0                  &          0        
&         0           \\ 
     0                   &    0           &          0        &         
0            \\
     0                   &     0                 &    0     &            
0            \\
     0                   &     0                 &         0           
&     0        
\endbmatrix
+ 
B
\bmatrix
0                   &      x            &      y     &    z   \\
x                   &      0             &     0     &   0    \\
y                    &     0            &      0     &   0    \\
z                    &     0            &     0      &   0    \\
\endbmatrix
+
C
\bmatrix
 0                      &        0                &        0           
&     0               \\ 
 0                      &        x^2            &      xy           &     
xz              \\
0                       &        xy               &      y^2         &    
yz               \\
0                       &        xz               &      yz           
&     z^2  
\endbmatrix
+ 
D
\bmatrix
0                     &      0                  &         0            
&       0         \\
0                     &      1                  &         0             
&      0          \\
0                     &      0                  &         1             
&      0         \\
0                     &      0                  &         0              
&      1         
\endbmatrix
.
\Tag414
$$
The  fiber of the kinematic  bundle  $\kappaGpt{x}(E)$  is therefore 
parameterized by  four   variables
$A$,  $B$, $C$, $D$.
Since these  variables  are invariants for the action of $G$ restricted to 
$\kappaG(E)$ and   since the   invariants  for the action of $\LieSO(3)$ 
on  $M$ are $t$ and $r$,  
 the  kinematic reduction  diagram  for the  action of $\LieSO(3)$ on 
the   bundle of Lorentz metrics is
$$
\CD
(t,r, A,B,C,D)  @ <\q_{\kappaG}<<   (x^i,A,B, C,D) @ >\iota> > (x^i,  
g_{ij}) \\
@VVV                                             
@VVV                                                                                               
@VVV \\
(t,r) @< \q_M <<                           (x^i)        @ >\id >  
>                                                                            
(x^i),
\endCD
\Tag416
$$
where the inclusion map $\iota$ is given by \eq{414}.

Consequently,   the most general  rotationally invariant metric on $M$  is
$$
	ds^2
           = A(t,r)dt^2 +     2B(t,r) dt (x\,dx +y\,dy +z\,dz) 
+C(t,r)(x\,dx +y\,dy +z\,dz) ^2 + D(t,r) (dx^2+ dy^2 +dz^2).
$$
In  standard spherical coordinates $x=r \cos\theta \sin \phi$, $y= 
r\sin\theta\sin \phi $,  $z =\cos\phi$ this  takes 
the  familiar form (on re-defining  the coefficients $B$, $C$ and $D$) 
$$
ds^2
           = A(t,r)dt^2 +     B(t,r) dt  dr +C(t,r)dr^2 + D(t,r)  
d\,\Omega^2
\Tag445
$$
where $d\,\Omega^2 =  d\phi^2  + \sin^2 \phi \,d\theta^2$.

If we   enlarge the symmetry group to  include time  translations  $\dsize 
V_0= \vect{t}$, then  the kinematic  reduction diagram becomes
$$
\CD
(r, A,B,C,D)  @          <\q_{\kappaG}<<   (x^i,A, B,C,D) @ >\iota> > 
(x^i,  g_{ij}) \\
@VVV                                             
@VVV                                                                                               
@VVV \\
(r) @< \q_M <<                           (x^i)         @>\id>  
>                                                                            
(x^i).
\endCD
\Tag430
$$

At  first glance  there appears to be  little difference between the   two 
diagrams  \eq{416}  and  \eq{430}, but a  computation of the
automorphism  groups  reveals a  dramatic  difference  in the  geometry of 
the reduced bundles  $\tilkappaG(E)$ in \eq{416} and \eq{430}.  
This difference  is  best explained  in terms  of   general  results on 
Kaluza-Klein reductions of metric  theories as in, for example, 
Coquereaux  and Jadczyk  \cite{coquereaux-jadczyk:1988a}.
 From  our perspective, these authors  show that when the action of $G$ on 
$M$ is simple in the sense that the isotropy groups $G_x$
can all be conjugated in $G$ to a fixed  isotropy group $G_{x_0}$,  then  
the   reduced bundle 
$\tilkappaG(E)$ is  a  product   of  three  bundles over  $\tilM$,
$$
\tilkappaG(E) = Q(\tilM) \oplus  A(\tilM) \oplus Q_{\Inv}(K).
\Tag457
$$
Here 
\smallskip
\noindent
 {\bf[i]} $Q(\tilM)$ is the bundle of metrics on $\tilM$.
\smallskip
\noindent {\bf[ii]} $A(\tilM) = \Lambda^1(\tilM) \otimes (P\times_H 
\lieh)$ ,  where   $P$ is the principal $H$  bundle defined as
the  set of  points in $M$ with  isotropy group $G_{x_0}$and  $H = 
\text{Nor}(G_{x_0},G)/G_{x_0}$.
\smallskip
\noindent
 {\bf[iii]} $Q_{\Inv}(K)$ is the trivial  bundle whose fiber consist of 
the $G$ invariant metrics on
the homogeneous space $K =G/G_{x_0}$.

For \eq{416} one computes
the  residual symmetry  group $\tilbigG_{\text{eff}}$ to  be the 
diffeomorphism group   of $\tilM =\real\times \real^+$  and one finds 
that   the   coefficients  $A$, $B$, $C$   transform  as the  components 
of  a
 metric  on $\tilM$ and  that  $D$ is a scalar field  (which one 
identifies as  a map into the space  of  \LieSO(3) invariant metrics on 
$S^2$). 
Thus, for \eq{416},  we  find  that
$$
\tilkappaG(E) =  Q(\tilM) \oplus \Re ,
$$
where $\Re$ is a trivial line bundle over $\tilM$.  
By contrast,  for  the diagram  \eq{430}  the automorphism group  
$\tilbigG$   acts  on $M$ by 
$$
	r\to f(r) \quad \text{and}\quad  t \to \epsilon t  + g(r),
\Tag444 
$$
where $f\in \LieDiff(\real^+)$,  $g\in C^\infty(\real)$ and $\epsilon \in 
\real^*$.
Without going further into  the  details of the decomposition \eq{457}, 
 we  simply  note that  the variable $t$ is now the fiber coordinate on   
the principle bundle $P$ and that  under the
transformations  \eq{444}  the coefficients of the  metric  \eq{445}, 
which are now functions of $r$  alone,  transform according to
$$
\align
 A(r) &\to  \epsilon^2 A(f(r)),  \qquad               B(r) \to \epsilon[ 
f'B(f(r)) + 2g'A(f(r))]
\\
C(r)      &\to (f')^2C(f(r)) + f'g'B(f(r)) +(g')^2 A(f(r)) \qquad  D(r)\to 
D(f(r)) .
\endalign
$$
Consequently, the sections of $\tilkappaG(E)$  can be  written  as
$$
 \tils(r) = [\tilde g (r) , \tilde \omega(r), \tilde  h(r)],
$$
where
$$
	\tilde           g(r)  = [ C(r) -  \frac{B(r)^2}{4 A(r)}]\, dr^2, 
               \quad 
	\tilde \omega(r)  = \frac{B(r)}{2A(r)} dr \otimes \vect{t}, 
               \quad \text{and}\quad 
                \tilde          h(r)  =   A(r) dt^2  + D(r)\, d\Omega^2 .
$$
Here  $\tilde  g(r)$ is a metric on $\tilM$, $\tilde \omega(r)$ is a 
connection on $P$ pulled back to $\tilde M$, and  $ \tilde h(r)$ is a map  
from $\tilde M$ into the $G$ invariant
metrics on $ \real \times S^2$.

The detailed expression for the reduced  operator  $\tilDelta$ for the 
stationary,  rotationally invariant   metrics 
can be found  in  any introductory text  on general relativity.  Here  we
simply point  out that   by computing    the action of  $G$ on 
$\text{Sym}^2(TM)$,
we can  deduce  that the  reduced operator will  have the form
$$
	\tilDelta = \tilDelta^{tt} \vect{t} \otimes \vect{t} +  \tilDelta^{rt} ( 
\vect{r} \otimes\vect{t}  +  \vect{t}\otimes \vect{r}) 
           +  \tilDelta^{rr} \vect{r}\otimes \vect{r}   + 
\tilDelta^{\Omega}( \vect{\phi }\otimes \vect{\phi} +  \frac{1}{\sin^2\phi}
                 \vect{\theta}\otimes \vect{\theta}),
$$
where $\tilDelta^{tt}$,   $\tilDelta^{rt}$, $\tilDelta^{rr}$  and  
$\tilDelta^\Omega$ are smooth functions  on the 2-jets
of  the  bundle  $(r, A, B, C, D) \to(r)$.  
In other words, of the  ten components in  the   field equations, the  
dynamic reduction diagram automatically
implies that 6 of these components vanish.   Moreover,  the  reduced 
operator  $\tilDelta$ is constrained  by the 
reduced    Bianchi identities. Since  $d t$ and $dr$ provide a  basis for 
the invariant one forms on $M$, we  know that  
the reduction   of $\Div_g S$  is  a linear combination  of   $dt $ and 
$dr$,
$$ 
\widetilde{\Div_{g }\ S}  =  \tilde S_t \,dt  + \tilde S_r \, dr.
$$
 By
direct computation,  one finds that   the $dt $ and $dr$ components of  
the reduced Bianchi identities are
$$
\gather
\frac{1}{2\gamma}  \frac{d\hfill}{dr} [ \gamma (  2A \tilDelta^{rt} + B 
\tilDelta^{rr})]=0 
\\
\squash{5}{and}{10}
\\
\frac{1}{2}\bigl(\frac{1}{\gamma} \frac{d\hfill}{dr} [ 
\gamma(2C\tilDelta^{rr} +B\tilDelta^{rt})] 
-\dot A \tilDelta^{tt}  - \dot C \tilDelta^{rr} -\dot B \tilDelta^{rt}  -2 
\dot D \tilDelta^{\Omega}\bigr) =0
\endgather
$$
where $\dsize \gamma = D\sqrt{\frac{1}{4}B^2 - AC}$.
It follows from the first  of these identities and  the  transformation 
properties  of  $A$, $B$, $\tilDelta^{rt}$ and $\tilDelta^{rr}$ 
under the  residual scaling $t\to \epsilon t$ that
$$
2A \tilDelta^{rt} + B \tilDelta^{rr}=0.
$$
This same identity can be derived by   first  observing that the principle 
of symmetric criticality holds for the action $G$   and  then  by    
 applying  Noether's  second theorem to the reduced Lagrangian with  
symmetry  $\tilbigG_{\text{eff}}$, 

 Consequently of the four  ODE  arising  in the  stationary, spherically 
symmetric reduction of the  field equations
one need only solve the   two equations
$$
\tilDelta^{tt} = 0 \quad\text{and}\quad  \tilDelta^{rr}= 0.
$$ 
The remaining two  equations
$$
\tilDelta^{rt} = 0 \quad\text{and}\quad \tilDelta^{\Omega} = 0
$$
will  automatically be satisfied (assuming $\dot D \neq 0$, $A\neq 0$). We 
stress that  
these  conclusions   actually hold true  for  the  stationary, 
rotationally invariant reductions of  any generally covariant  metric 
field equations derivable from  a variational principle. \endEx

\Ex{\State 404 .\ \smc Static, Spherically Symmetric Reductions}   A  
metric   is {\it static} and spherically  symmetric  if,  in addition to  
being   invariant under  time  translations  and  rotations, it is  
invariant  under   time reflection. The  symmetry group
$G$   now includes the transformations $t \to t+c$ and $t \to -t$ and 
therefore the isotropy subgroup $G_{x_0}$ of the point  $x_0 = 
(t_0,\bfx_0)$
now includes the reflection    $t \to  2t_0 -t $.  The   fibers of the 
kinematic bundle  are  now  constrained by  \eq{412}  along with  
$$
 bgb^{\text{t}}= g, \quad\text{where}\quad  b= \text{diag}[-1,1,1,1].
$$
This  forces  $B = 0$ in \eq{414} so that the   fibers of the kinematic   
bundle  are  now   3 dimensional and the 
general invariant section is
$$
	ds^2 = A(r) dt^2   + C(r) dr^2  + D(r) d\,\Omega^2.
$$
   The   automorphism  group  for
this  bundle  is   now  $ r \to f(r)$ and $t\to \epsilon t$ and   the 
$A(\tilM) $  summand  in \eq{457}  does not appear. 
This example shows that  while  discrete
symmetries  will  never  result in a  reduction of the  dimension of  the 
orbit space $\tilM$, that is,  the number of independent variables,  
discrete symmetries  can    reduce the  fiber dimension of the  kinematic  
bundle, that is,  the number of    dependent variables. 

\Ex{\State407 . \ \smc Plane  Waves}  As our  next  example  from  general 
relativity, we consider a  class of plane wave metrics 
 \cite{bondi-pirani-robinson:1959a}. We take $M= \real^4$ with 
coordinates  $(u,v,x,y)$ and 
let $P(u)$  and $Q(u)$ be arbitrary  smooth functions satisfying  
$P'(u)>0$ and  $ Q'(u)> 0 $.
 The  symmetry group  on $M$ is  the five-parameter   transformation group 
$$
\align
u' &= u,
\qquad
v' =v + \varepsilon_1 + \varepsilon_4 x +\varepsilon_5y + 
1/2\bigl(\varepsilon_2\varepsilon_4 + \varepsilon_3\varepsilon_5 + 
                 \varepsilon_4^2P(u) +\varepsilon^2_5Q(u)\bigr),
\Tag458
\\
\vspace{2\jot}
x' &=x+ \varepsilon_2 +  \varepsilon_4 P(u),
\qquad
y' = y + \varepsilon_3 + \varepsilon_5 Q(u),
\endalign
$$
with infinitesimal generators $\dsize  V_1 = \vect{v}$,  $\dsize V_2= 
\vect{x}$, $\dsize V_3= \vect{y}$, 
$$
               V_4= x \vect{v} + P(u) \vect{x} \quad\text{and}\quad    
V_5= y \vect{v} + Q(u) \vect{y}. 
$$ 

The  only non-vanishing brackets are
$$
	[\,V_2,V_4\,] = V_1\quad\text{and}\quad [\,V_3, V_5\,]  =V_1
$$
so that,  regardless of the choice of functions  $P$ and $Q$, the 
abstract  Lie algebras  or groups are the same although the 
actions are  generically different for different choices of $P$ and $Q$.  
The  coordinate function $u$ is the only invariant and  the  orbits of 
this action are 3-dimensional. Therefore,
at each point   the isotropy  subgroup is two dimensional and it   is 
easily  seen that,  at $\bfx_0 = (u_0,v_0,x_0,y_0)$,
the infinitesimal   isotropy  $\Gamma_{\bfx_0}$ is generated by
$$
	Z_1 =  V_4  -x_0 V_1 - P(u_0) V_2 
               \quad\text{and} \quad
               Z_2 =  V_5 - y_0  V_1 -Q(u_0) V_3.
$$ 
 At   $\bfx\in M$ the   metric components  $g = [\,g_{ij}\,]$   of a $G$ 
invariant  metric    satisfy the isotropy conditions
$$
	ga_1 + a_1^{t}g = 0 \quad\text{and}\quad  ga_2 + a_2^{t} g = 0,     
\Tag417
$$
where 
$$
a_1 = 
\bmatrix 0                           &     0     &      0      &   0   \\  
                0                           &     0     &       1     &   
0  \\
               P'(u)          &    0     &       0     &   0   \\
               0                             &    0     &       0     &   
0   
\endbmatrix    
\quad\text{and}\quad
a_2  = 
\bmatrix 0                           &     0     &       0      &   0   
\\  
                0                           &     0     &       0     &    
1  \\
                0                           &    0     &        0     &   
0   \\
                Q'(u)          &    0      &        0     &       0       
\endbmatrix.
$$  
We  find that the solutions to \eq{417} are 
$$
	g_1= \bmatrix   1  & 0 & 0 & 0 \\  0 & 0 &0 &0 \\   0& 0 &0 &0 \\  0& 
0&0&0 \endbmatrix
               \qquad\text{and}\qquad
               g_2=  \bmatrix   0   & -1 & 0 & 0 \\  -1 & 0 &0 &0 \\  0& 0 
&  \dfrac{1}{P'(u)}  &0 \\  0 & 0 & 0& \dfrac{1}{Q'(u)} 
                            \endbmatrix.
$$
Thus  the  kinematic reduction diagram  is 
$$
\CD
(u,A ,B)        @ <\q_{\kappaG}<<    (x^i, A,B)     @>\iota>>   (x^i,  
g_{ij})  \\
@VVV                                                    
@VVV                                      
@VVV                                \\
(u)                  @<\q_M <<                   (x^i)              @>\id 
>>          (x^i), 
\endCD
$$
and the  inclusion  map  $\iota$ sends $(A,B)$  to  $ds^2 = Adu^2 + B 
\gamma$, where
$$ 
\gamma = -2 du \,dv    + \frac{dx^2}{P'(u)}  +\frac{dy^2}{Q'(u)}.
$$
The most general  $G$ invariant  metric  is
$$
	ds^2 = A(u) du^2 + B(u) \gamma.
\Tag450
$$
From  the  form  of the most general    $G$ invariant  symmetric type 
$\binom{2}{0}$ tensor, we are assured that  the reduced field equations
take the form  
$$
	\tilDelta =   
               \tilDelta^{vv}\vect{v} \otimes \vect{v}  +  
               \tilDelta^\gamma (  -\vect{u} \otimes  \vect{v} - 
\vect{v}\otimes \vect{u}   + P'(u)   \vect{x} \otimes \vect{x}  +
                Q'(u) \vect{y}\otimes  \vect{y}). 
$$
Every  $G$  invariant     one-form   is  a multiple of  $du$ so that  
there is  only one non-trivial component to the  contracted 
Bianchi identities and, indeed,  by direct computation  we find that
$$
	\Div_{\tilg}  \tilDelta^\flat =    \frac{d\hfill}{d u} \bigl( B \tilde 
\Delta^\gamma \bigr) \,d\,u. 
$$
Since  this must vanish identically,  we conclude that  the 
$\tilDelta^\gamma$ component of  the  reduced  field equations
is of the form
$$
	\tilDelta^\gamma =  \frac{c}{B},
$$
where $c$ is a constant.  Either the  constant $c$  is non-zero, in which 
case the  reduced equations are inconsistent and there  are  no 
$G$ invariant  solutions,  or   else $c=0$ and the  reduced equations 
consist of  just  the single equation $\tilDelta^{vv}=0$.  For  generally 
covariant
metric theories  the  case $c\neq 0$ can only arise   when the field  
equations contain  a  cosmological term  \cite{torre:1999a}.
 
It is easy to check that while the isotropy algebras  $\Gamma_{\bfx_0}$ 
are all two-dimensional abelian subalgebras, on disjoint orbits 
none   are  conjugate under the  adjoint  action of $G$.   Hence  the  
group action \eq{458} is not simple and consequently the kinematic bundle
for this action need  not decompose according to  \eq{457}.  Indeed,  
the   tensor  $\gamma$ cannot be identified with 
any  $G$ invariant  quadratic  form on the  orbits $G/G_{x_0}$. \endEx

\Ex{\State414 .  \  \smc Symplectic Reduction and Group Invariant 
Solutions}  It is important to  recognize the   fundamental   differences 
between   symplectic reduction and   Lie  symmetry  reduction  for  group 
invariant solutions of a Hamiltonian system with symmetry. Let  $M$ be an 
even dimensional manifold  with symplectic form 
$\omega$ and   let $H\:M \to \real$ be  the  Hamiltonian  for a dynamical 
system  on $M$. For the  purposes of this example, it suffices
to consider reduction by a  one dimensional  group of     Hamiltonian
symmetries generated by a  vector field  $V$   with  associated momentum 
map $J$, 
$$
V\hook  \omega  = d\, J.
\Tag435
$$

 In symplectic  reduction  the reduced  space  $\widehat M$ is 
obtained by  {\bf[i]} restricting to the submanifold of $M$ 
defined by
$$
J = \mu \equiv constant,
$$
and then  {\bf[ii]} quotienting this submanifold by the action of the 
transformation  group generated by $V$.   Both $\omega$ and $H$   descend 
to   $\widehat M$   and  the reduced  equations are  the 
associated    Hamiltonian system on  $\widehat M$.  Since   $\dim\widehat 
M =\dim M - 2$, the reduction in the  number of dependent variables is 
2.   The solution to the  original Hamiltonian   equations are obtained
from that of the reduced   Hamiltonian equations by  quadratures.  

To   compare with symmetry reduction for group invariant solutions, we 
transcribe   Hamilton's equations into the operator-theoretic setting used 
to construct  the kinematic and  dynamic reduction diagrams. Let 
$E=M\times \real \to \real$
be extended phase space  so that the differential operator  
characterizing  the  canonical equations  is  the one-form  valued
operator  on  $J^1(E)$   defined by 
$$
	\Delta = X \hook \omega  - d\, H. 
$$
Here  $X$ is the  total derivative  operator  given,  in standard 
canonical coordinates $(u^i,p_i)$ on $M$,  by 
$$
	X =  \frac{d\hfill}{dt} =  \vect{t} + \dot u^i  \vect{u^i} +  \dot  p_i  
\vect{p_i}. 
$$
It  is not difficult to show  that if  $V$ is any vector field on $M$, 
then  the prolongation of  $V$ to $J^1(E)$
satisfies  $[X,\pr^1 V] =0$ and therefore $V$ is a  symmetry of   the 
operator $\Delta$ whenever   $V$ is a symmetry of
$\omega$ and $H$.  

Since $V$ is a vertical vector field   on $E$ it is  ``all
isotropy'' and the  kinematic  bundle  is the fixed point set  for the 
flow of $V$, 
$$
\kappaGamma(E) =\{(t, u^i, p_i) \, | \, V (u^i, p_i) = 0\,\}.
$$
The   dimension of  $\kappaGamma(E)$  therefore  depends upon  the choice 
of $V$   and  is generally  less than   the dimension of $E$  by
more than  2 (the decrease in the  dimension in the case of symplectic 
reduction). In short, it is 
not possible to  identify  the  fibers of the kinematic bundle  with the  
reduced phase space $\widehat M$.  Moreover,  from \eq{435}, it follows 
that {\it points in  $\kappaGamma(E)$    always correspond to points   on 
the singular level sets of  the momentum map and, typically, to points 
where the 
level sets  fail to  be a  manifold}. Thus the invariant  solutions are    
problematic from the
viewpoint of   symplectic reduction and  are  subject  to  special  
treatment. See, for example,  \cite{arms-gotay-jennings:1990a} and  
\cite{gotay-bos:1986a}. 
Finally,  there  is no  guarantee that the  reduced equations for the  
group invariant solutions possess
any natural  inherited  Hamiltonian  formulation.

We illustrate these general  observations with  some specific examples. 
First,  if   $V$  is  a translation symmetry    
of  a mechanical  system,  then $J$ is a linear   function and symplectic 
reduction
yields  all  the solutions to Hamilton's equations  with  a given fixed 
value  for the first integral  $J$.  Since the vector field $V$ never 
vanishes,
the kinematic bundle is  empty  and there   are  {\it no}  group invariant 
solutions.

 Second, for   the  classical   3-dimensional central force problem
$$
\ddot u = - f(\rho) u,\quad \ddot v = -f(\rho) v, \quad \ddot 
w = -f(\rho) w,
$$
where $\rho = \sqrt{u^{2} + v^{2} + w^{2}} $,  the 
extended  phase space $E$ is ${\real}\times \real^6 \to \real$  with 
coordinates 
$$
(t, u,v,w,p_{u},p_{v},p_{w}) \to (t),
$$ 
 the symplectic structure on  phase space is 
$
\omega = du \wedge  dp_{u} +  dv \wedge dp_{v} +  dw \wedge dp_{w}
$
and  the  Hamiltonian  is
$
H = {1\over 2} (p_{u}^{2} + p_{v}^{2}
+ p_{w}^{2}) + \phi(\rho),
$
where
$
\phi'(\rho) = \rho f(\rho).$
The vector field
$$
V = -u{\partial\over\partial v} + v{\partial\over\partial u} - p_{u} 
{\partial\over \partial p_{v}}  + p_{v} {\partial\over\partial 
p_{u}}
$$
is a Hamiltonian symmetry.

The kinematic bundle  for the $V$ invariant sections of $E$ is 
$$
\CD
(t, w, p_w) @< \id <<           (t,w,p_w) @> \iota  >>             (t, 
u,v,w, p_u,p_v,p_w) \\ 
@V\pi VV                           @VV \pi V                            @ 
VV \pi V    \\
(t)               @< \id<<            (t)          @>\id 
>>                 (t) \ ,
\endCD
\Tag440
$$
where $\iota(t,w,p_w) = (t,0,0,w,0,0,p_w)$,
the invariant sections are of the form
$$
t \to (0,0,w(t), 0,0, p_w(t)), 
$$
and the reduced differential operator for the  $V$  invariant solutions  is
$$
\tilDelta = (\dot w - p_{w})\, dp_w  - (\dot p_{w}  + wf(|w|) )\, dw.
$$

Let us compare this state of affairs with that obtained by 
symplectic reduction based upon  the Hamiltonian vector field $V$.  The 
momentum  map  associated to this
symmetry is  the angular momentum
$$
	J = -up_{v} + vp_{u}.
$$
The level sets  $J=\mu$ are  manifolds except  for $\mu=0$.  The level set 
$J=0$ is the  product  of a plane   and  a cone     whose 
vertex is  precisely   the fiber of the kinematic bundle.   
To implement the  symplectic  reduction, we  
introduce canonical cylindrical coordinates 
$(r,\theta,w,p_{r},p_{\theta},p_{w})$, 
where
$$
u=r\cos\theta,\quad v=r\sin\theta,
$$
$$
 p_{u}=p_{r}\cos\theta  - { p_{\theta}\over r}\sin\theta \,,
\quad
p_{v}=p_{r}\sin\theta  + {p_{\theta}\over r}\cos\theta \, .
$$
{\it Note that this change of coordinates  fails  precisely at   points of 
the kinematic bundle. } 
In terms of these phase space coordinates,
the symplectic structure is still in canonical form
$
\omega =  dr \wedge dp_{r}  +  d\theta \wedge dp_{\theta}+ 
dw \wedge  dp_{w}.
$,
the Hamiltonian is
$\dsize 
H = {1\over 2}(p_{r}^{2} + {\mu^2\over r^{2}}p_{\theta}^{2} + p_{w}^{2})
+ \phi(\sqrt{r^{2}+w^{2}}),
$
and the momentum map is
$
J = -p_{\theta}.
$
 We can  therefore  describe the symplectic 
reduction  of  $E$ by the diagram 
$$
\CD
(t, r,w, p_r, p_w) @<<<   (t,r,\theta, w, p_r, p_w) @>\iota>> (t,r, 
\theta, w, p_r, p_\theta, p_w) \\
@VVV                                 @VVV                           @VVV \\
(t)                             @<<<  (t)                        
@>>>                    (t) .
\endCD
$$
The 
reduced symplectic structure is then $\hat \omega=  dr \wedge dp_{r}\wedge 
dr+ dw \wedge dp_{w}$,
the reduced 
Hamiltonian is 
$\dsize 
\widehat H = {1\over 2}(p_{r}^{2} + {\mu^2\over r^{2}} + p_{w}^{2}) + 
\phi(\sqrt{r^{2}+w^{2}})
$,
and the reduced equations of motion are
$$
\dot r = p_{r},\quad \dot p_{r} = 
-rf(\sqrt{r^{2}+w^{2}}) +{\mu^2\over r^{3}},\quad \dot w = 
p_{w},\quad \dot p_{w} = -wf(\sqrt{r^{2}+w^{2}}).
$$
Given a choice of $\mu$ and  solutions to these 
reduced equations, we get a solution to the full equations via
$
\theta = -\mu t + const.
$
\endEx
\equationnumber=1
\statementnumber=0
\sectionnumber=7

\
\subheading{7.  Appendix }  We summarize a  few    technical  points  
concerning group actions on fiber bundles
and  the  construction of the  kinematic  and dynamic reduction diagrams. 
For details, see  \cite{anderson-fels:1999a}.

\medskip
\noindent
{\bf A. \smc Transversality and Regularity.} 
 Let $G$ be a  finite dimensional Lie group acting  projectably on a 
bundle $\pi\:E\to M$.  We say that 
$G$ acts   {\deffont transversally}   on  $E$ if,   for each fixed  $p\in 
E$  and each fixed $g\in G$, the  equation
$$
\pi(g\cdot p) = \pi(p) \quad\text{implies that}\quad  g\cdot p = p.
\Tag501
$$
Thus each orbit of $G$  intersects each fiber  of $E$   exactly once. For 
transverse group actions the orbits  of $G$ in $E$ are
diffeomorphic  to the orbits of $G$ in $M$ under the  projection map 
$\pi\:E \to M$.  Projectable, transverse actions always
satisfy  the infinitesimal 
transversality condition \equationlabel{2}{104}{} but the  converse is 
easily seen to be false.

Let us say  that the  action   of $G$   on $M$ is {\deffont regular}   if  
the  quotient space $\tilM = M/G$ is a  smooth manifold  and the  quotient 
map
$\qM\:M \to \tilM$  defines $M$ as a bundle over  $\tilM$.  The   
construction of the  orbit manifold $\tilM$   is  discussed  in 
various texts, for example,
 \cite{abraham-marsden:1978a},  \cite{anderson-fels-torre:2000a},  
\cite{olver:1993a},   \cite{palais:1957a}.  
The assumption that the action of $G$ on $M$ is  regular is a standard  
assumption in Lie  symmetry reduction. For simplicity   we  suppose 
that  $\tilM$ is a  manifold without  boundary but, as   Example 
\statementlabel{6}{420} shows,  this  assumption can  be relaxed  
in  applications.

The fundamental properties  of  transverse group actions   are  described 
in the following theorem which is proved in
 \cite{anderson-fels:1999a}.

\proclaim{Theorem \State502   {\smc (The Regularity Theorem for Transverse 
Group Actions)}  }   
Let  $G$ be a Lie group which acts projectably  and transversally on  the 
bundle $\pi\:E\to M$.   
Suppose that   $G$ acts regularly on $M$. \smallskip
\noindent
{\bf[i]}  Then $G$ acts regularly on $E$ and $\tilE = E/G $ is a bundle 
over $\tilM$.
\smallskip
\noindent
{\bf[ii]} If  the orbit manifold $\tilM$ is Hausdorff,  then the orbit 
manifold $\tilE$ is also Hausdorff.
\smallskip
\noindent
{\bf[iii]} The bundle $E$  can be identified with the  pullback of the  
bundle $\tilpi\:\tilE\to \tilM$  via the quotient map
$ \qM\:M \to \tilM$. 
\smallskip
\noindent
{\bf[iv]}  Let $\tilU$ be an open set  in $\tilM$ and let $U = \q_{\sssize 
M}^{-1}(\tilU)$. There  is a  one-to-one correspondence between the smooth 
$G$ invariant sections of $E$ over $U$ and the sections of $\tilE$ over 
$\tilU$.
\endproclaim

\bigskip

\noindent
{\bf B.  \smc  Transversality and the   Kinematic Bundle.} Lemma 
\statementlabel{3}{206} implies that the  action of $G$ on $E$
always  restricts to  a transverse action on  the set $\kappaG(E)$.  In 
fact, it is not difficult to characterize  $\kappaG(E)$ as the
largest  subset of $E$   on which $G$ acts transversally or, 
alternatively, as the  smallest set through   which all locally defined  
invariant sections
factor. {\it  For Lie symmetry reduction  without transversality  the  
assumption that 
$\kappaG(E)$ is an  imbedded subbundle  of $E$ now  replaces the   
infinitesimal transversality condition  \equationlabel{2}{104}{}
as the  underlying hypothesis for the action of $G$  on $E$ }(together, of 
course,  with the regularity   of  the action of $G$  on $M$).
In particular, the assumption that  the  dimension of $\kappaGpt{x}(E)$  
is constant as $x$ varies over $M$  is clearly a necessary condition if  
one hopes to parameterize the  space  of $G$  invariant  local  sections 
of $E$ in terms of a  fixed number of arbitrary  functions.   
There are a  variety of  general results which one can apply to  check  
whether $\kappaG(E)$ 
is subbundle of $E$.   To begin with, if  $x,y\in M$  lie on the same $G$ 
orbit, that is, if   $y=g\cdot x$ for some $g\in G$, then it is not 
difficult to prove that
$$
	\kappaGpt{y}(E) = g \cdot \kappaGpt{x}(E) .
$$
By virtue of this observation it suffices to  check  that  the 
restrictions of  $\kappaG(E)$ to  the  cross-sections of  the action of 
$G$ on $M$
are subbundles.   For Lie group actions $G$  which admit slices on $M$,  
it is not difficult to establish  (see  \cite{anderson-fels-torre:2000a})  
that the kinematic bundles
for the induced actions on tensor bundles of $M$   always exist. For  
compact groups acting  by  isometries on hermitian  vector bundles the 
existence 
of the kinematic bundle is  established in  
 \cite{bruning-heintze:1979a}.

Granted that   $\kappaG(E) \to M$ is a bundle,  Theorem 
\statementlabel{3}{202}  now  follows from  Theorem  \st{502}.  
  Theorem \st{502}  also shows that  there is  considerable redundancy
in the  hypothesis of 
Theorem \statementlabel{3}{202}.

 We emphasize that the  action of $G$ on  $E$ itself need not be 
regular  in order  to construct  a   smooth kinematic reduction diagram. 
This is well illustrated by  Example 19 in Lawson  \cite{lawson:1980a}
(p.  23).

\bigskip
\noindent
{\bf C. \smc The Bundle of Invariant Jets.} The  following theorem 
summarizes the  key properties of  the bundle $\Inv^k(E) \to M$.

\proclaim{Theorem \State301  } Let $G$ be a projectable group action on 
$\pi\:E\to M$ and suppose that  $E$
admits a smooth kinematic reduction diagram  \equationlabel{3}{206}{}.  

\smallskip
\noindent
{\bf[i]} Then $\Inv^k(E)$ is a $G$ invariant   embedded submanifold of 
$J^k(E)$.

\smallskip
\noindent
{\bf[ii]}  The action of  $G$ on $\Inv^k(E)$ is transverse and regular.

\smallskip
\noindent 
{\bf[iii]} The  quotient  manifold $\Inv^k(E)/G$ is diffeomorphic to  
$J^k(\tilkappaG(E))$ and the diagram 
$$
\CD
	J^k(\tilkappaG(E) )   @<\q_{\Inv}<<       \Inv^k(E)      @>\iota>>       
J^k(E)  \\ 
       	@VVV                       @VVV                           
@VVV       \\
            \tilM    @<\qM<<       M      @>\id>>       M
\endCD
$$
commutes. 
\endproclaim

This  theorem implies that   the same   hypothesis   on the action of $G$  
on the bundle  $\pi\:E\to M$   needed to
 insure that  the kinematic reduction diagram is a diagram of smooth 
manifolds and  maps   also     insures  that   the  bottom
row of the dynamical  reduction diagram \equationlabel{4}{305}{} exists.  
Therefore  to guarantee  the   smoothness  of the entire dynamic 
reduction  diagram
one need only assume, in addition, that  $\Cal D_{\sssize \Inv}$ is a  
subbundle  of $\Cal D$.

\bigskip 
\noindent
{\bf D. \smc The Automorphism   Group of the Kinematic Bundle. }  For 
computations of  the  automorphism group of the kinematic bundle
 it is often
advantageous to use the fact that $\tilbigG{}^*$ fixes every $G$ 
invariant  section of $E$, that $\tilbigG$   preserves the
space  of  $G$  invariant sections and that, conversely, under very mild 
assumptions, these properties characterize these  groups.

\proclaim{Theorem \State610 }   Assume that there is a $G$ invariant  
section through each point of $\kappaG(E)$. 
Then the 
group $\tilbigG{}^*$ coincides with the  subgroup of $\bigG$ which fixes 
every  
invariant section of $E$,
$$ 
	\tilbigG{}^* =
         \{\, a \in \bigG \,|\, a\cdot s  = s\quad\text{for all  $G$ 
invariant sections $s\:M\to E$}\,\} 
$$
and 
 the   group $\tilbigG$    coincides with the subgroup   of $\bigG$ which  
preserves the
set of $G$ invariant sections of $E$,
$$
	\tilbigG
                = \{\, a \in \bigG \,|\, \text{ $a\cdot s$ is $G$  
invariant for all  $G$ invariant sections $s\:M\to E$}\,\}.
$$
\endproclaim

\newpage

\InitializeRef

\heading References \endheading

\end

\smallskip
\noindent
Finally,  we should also mention the  problem of  classifying  all  
inequivalent  group  invariant solutions to a given system of differential
equations. In the past this problem has  been addressed  by  first 
classifying all the subgroups of the full symmetry group of the
system of differential equations  (see, for example, Winternitz  \cite{}, 
\cite{}).  In actuality one is really   interested in following  problem 
\smallskip
\noindent
{\bf[x]} Given a  Lie group action  $\bigG$ on  $E$, classify  all the  
subgroups  $G\subset \bigG$  which  define inequivariant   kinematic
reduction diagrams.
\smallskip
\noindent

For our next   examples  we  look  at two    well-known reductions of  the 
harmonic map equation. Let $M$ and $N$  be two  Riemannian manifolds with  
metrics $g$ and $h$  respectively.  Let $E=M\times  N  \to M$ and   let us 
identify    the sections of $E$ with  maps $s:M\to N$.
Let $\Cal D \to J^2(E)$ be the   bundle  with fiber $T_uN$ at the point 
$(x^i, u^\alpha,  u^\alpha_i, u^\alpha_{ij} )$  
and define  the harmonic map operator 
$\Delta:J^2(E) \to \D$ by 
$$
\Delta =  - g^{ij}\biggl(  u^\alpha_{ij} -  \Gamma^l_{ij} u^\alpha_l + 
\Gamma^\alpha_{\beta\gamma}u^\beta_i 
               	u^\gamma_j\biggr) \,	\vect{u^\alpha}
$$
where $\Gamma^l_{ij}$ and  $\Gamma^\alpha_{\beta\gamma}$ are the  
Christoffel symbols for the metrics $g$ and $h$.
For  symmetry reductions of the  harmonic map equation one is typically  
interested in   solutions which are
invariant  with respect  to a subgroup of  $\text{Iso}(g) \times 
\text{Iso}(h)$, the product of the isometry
groups of the metric $g$ and $h$.  Specifically,  let $G$ be  a subgroup 
of $\text{Iso}(g)$, let $\rho\:G \to \text{Iso}(h)$ be a  group
homomorphism  and  define the action of $G$ on $E$ by
$$
g\cdot (x,u) =  (g\cdot x  \rho(g) \cdot u)
$$
Then a  section of $E$ is  $G$ invariant if the corresponding map 
$s\:M\to N$ is $G$  equivariant   with respect to the map $\rho$, that is,
$$
s(g\cdot x) = \rho(g) \cdot s(x)\quad \text{ for all}\quad g \in G.
$$

When the manifolds $M$ and $N$ are the standard spheres $S^m\subset 
\real^{n+1}$ and $S^m\subset \real^{m+1}$
 there are a number of  simplications
one  can use to   evaluate the harmonic map operator.  First, suppose that 
the target manifold  $N = S^{m}$.  Let  $\iota_m\:S^m\to \real^{m+1}$ be 
the standard inclusion map and let 
$$
	\phi= \iota_m\circ s\:M \to \real^{m+1}.
$$    Then  it is not difficult to  show that
$$
\bigl((\iota_m)_* \bigr)\Delta\bigl(j^2(s)(x)\bigr) = 
\Delta'\bigl(j^2(\phi)(x)\bigr), 
$$
 where 
$$
	\Delta'=  \bigl(\Delta^M \phi^\alpha   - \lambda \phi^\alpha\bigr) 
\,\vect{\phi^\alpha}
	\quad\text{and}\quad
	\lambda=  |d\phi|^2 =   \delta_{\alpha\beta} g^{ij} \phi^\alpha_i 
\phi^\beta_j 
$$
 Here $\Delta^M$ is the Laplacian  for  functions  $f\:S^n \to \real$ 
acting on  the individual  components of  the map $\phi^\alpha$,
$$
\Delta^M \phi^\alpha = - g^{ij}(  \phi^\alpha_{ij} -  \Gamma^l_{ij} 
\phi^\alpha_l )
$$ 
Thus $s\:M \to S^m$ is  harmonic if and only the map $\phi$  solves 
$\Delta'\bigl(j^2(\phi)(x)\bigr) = 0$.
 This  formula avoids having to introduce  coordinates
on $S^m$ and computing
the  Christoffel symbols $\Gamma$ in these coordinates.

We note that   for each section $\phi$,  $\Delta'(j^2(\phi)(x))$  is 
tangent to the sphere $S^m$ at the point $\phi(x)$ and therefore 
we  can   view  $\Delta'$  as a section of the  bundle  $\Cal D'_{\perp} 
\to J^2(S^n \times \real^{n+1})$,  the fibers  of  which are 
tangent vectors in  $R^{m+1}$ which are  orthogonal to the    radial 
direction  $(\phi^\alpha)$.....  
$$
\Cal D'_{\perp} = 
\Tag429
$$This constraint plays an  critical  role in  computing the  reductions 
of the   harmonic map operator. 

Secondly, if $M = S^n$ and 
 if   $\Phi\:U\to \real^{m+1}$ be a map which extends $\phi$ to a map on an
open set $U\subset \real^{n+1}$, where  $S^n \subset  U$, then  $\phi= 
\Phi\circ i_n$ and 
$$
\Delta^{S^n} \phi^\alpha=\Delta^{S^n}( \Phi^\alpha\circ\iota_n) = \bigl( 
\Delta^{\real^n}\Phi^\alpha + R^2(\Phi^\alpha)
+n R(\Phi^\alpha )\bigr)\circ \iota_n,
$$
where  $R$ is the  radial vector field 
$$
	R=   \frac{ x^i}{r} \vect{x^i},\quad\text{and}\quad r = \sqrt{(x^1)^2 + 
(x^2)^2 + \cdots+ (x^{n+1})^2}.
$$
With  this formula one need not introduce coordinates on the domain sphere 
$S^n$ nor compute the Christoffel symbols
$\Gamma^l_{ij}$.

#############################################################

\Ex{\State405 }  The  kinematic  bundle for the action of $G=\LieSO(n+1)$  
on the bundle  of   Lorentz  metrics, or more generally, on   tensor  
bundles  over $M=R\times R^{n+1}-\{\,0\,\}$ is  be easily computed  by a   
judicious application  of   the fundemental  theory of invariant theory 
for the orthogonal group.  This theorem  states that  if 
$T\: V^k  \to  \real $  a  multi-linear map, where $V= \real^n$, which is  
$\LieSO(n)$ invariant in the sense that
$$
	 T( R\cdot X_1, R\cdot  X_2,\ldots, R\cdot  X_s) = 
               (R\cdot T)(X_1,X_2\ldots, X_s) \quad\text{for all $R\in 
\LieSO(n)$}
$$ 
then $T$ is a sum of products of inner product  $<X_i, X_j>$ and the 
determinant  $\det(X_1,X_2,\ldots, X_n)$.  The components of 
$T$ are therefore sums of  products of  the identity matrix $\delta_{ij}$ 
and the  permutation symbol $\varepsilon_{i_1i_2\cdot i_n}$.
The second fundemental theorem of  classical invariant theory for 
$\LieSO(n)$ asserts that   the only relationship between these  tensors is 
$$
	\det(X_1,X_2,\ldots, X_n)^2  = \det (  <X_i, X_j>)
$$ 
so that products of pairs of $\epsilon$ can always  be replaced by   
products of metrics.

For example, the most  general rank  $4$  tensor is
$$
	T_{ijhk}  = a \delta_{ij}\delta_{hk} + b \delta_{ik}g_{jh} + c 
\delta_{ih}\delta{jk} + d \varepsilon_{ijhk}
$$

Now  let $T_s(M)$  be the  bundle   of rank $s$ covariant tensors  on 
$M$.  (Since  the manifold $M$ admits a  $G$ invariant  metric  the 
general case 
of  tensor bundles of arbitrary  valence can be reduced to this case by  
lowering indices.) 
The isotropy subgroup of the  point  $x_0 =(t_0,\bfx_0) \in M$ is
the  othogonal  group $G_x= \LieSO(n-1)_{\bfx0}$ which fixes the vector 
$\bfx_0$ and therefore 
$$
\kappaGpt{x_0}(T_s(M)) = \{\, T\in [T_s(M)]_{x_0} \,| \,  R\cdot T = T  
\quad\text{for all $R \in \LieSO(n-1)_{\bfx0}$} \,\}.
$$
We  claim  that $\kappaGpt{x_0}$ is the  tensor   algebra generated by 
the   one forms
 $\alpha = dt$ and 
$$
\beta = dr  =  \frac{1}{r} (x^1 dx^1  + x^2 dx^2 + \cdots   + x^n dx^n),
$$
evaluated at $x_0$, and the symmetric two form
$$
\gamma= (dx^1)^2 + (dx^2)^2 +\cdots + (dx^{n+1})^2
$$
In  particular,   the  kinematic bundle for the  symmetric rank 2  
covariant tensors  has   four dimensional  fiber  spanned by 
$\alpha^2$,  $\alpha \beta$, $ \beta^2$ and $\gamma$ and the general form 
of the most general invariant  is the same
is the four-dimensional case.
 
The  kinematic  bundle for  the  skew-symmetry  rank 2 tensors   has  one 
dimensional  fiber   $dt \wedge dr$.

To  prove this  result   we cannot apply the  fundamental  theorem  
directly since $T$is a mult-linear map on an underlying  tangent space of 
dimension $n+2$  and is invariant  under a subgroup of $\LieSO(n+2)$ but  
this problem
is easily circumvented.  Write
$$
	T =  \sum_{a,b} \alpha^a\otimes \beta ^b\otimes T_{ab}
$$
where  $T_{ab}$ is of type $s - a -b$ and  
$$
              T_{ab}(\vect{t}, \vect(r), X , \dots )  = 0.
$$
It now follows that 
$$
	T_{ab}(X_1,X_2, \ldots, X_k)  =   T_{ab}(p(X_1), 
p(X_2),                                 	\ldots,  p(X_k)),
\Tag420
$$
 where 
$$
	p(X) = X - \alpha(X) \vect{x}  -\beta(X) \vect{r}
$$ 
is the  projection of  vector $X\in T_{x_0}(M)$ onto the  $n$ dimensional 
subspace   $V$ 
orthogonal   to $\vect(t)$ and $ \bfx0$ with  respect to 
the  metric $ds^2= dt^2 + \gamma$    Since  the projection map $p$ 
commutes with  rotations   \LieSO( in $\real^n$ about the   $\bfx_0$ 
axis,    
we can  now apply the  fundemental theorem to the right hand side  of  
\eq{420}, viewed as a multi-linear  map on $V$ to  deduce that  $T_ab$ is  
a sum of  products of  inner products 
$$
<p(X_i),p(X_j)> = \gamma(X_i, X_j) - \beta(X_i) \beta(X_j)
$$ 
and the  result  follows.
\endEx  

**********************************************************
\Ex{\State414  .\ \smc Sympletic Reduction and Group Invariant Solutions}  
It is important to  recognize the   fundemental   differences between  the 
symplectic reduction and    symmetry  reduction  for  group invariant 
solutions of a Hamiltonian system with symmetry. Let  $M$ be a even 
dimensional manifold  with symplectic form 
$\omega$ and   let $H\:M \to \real$ be  the    Hamiltonian  for a dynamic 
system  on $M$. For the  purposes of this example, it suffices
to consider reduction by a single    Hamiltonian
symmetry $V$  with  associated momentum map $J$, 
$$
V\hook  \omega  = d\, J.
\Tag435
$$

 In symplectic  reduction  the reduced  space  $\hat M$ is 
obtained by  {\bf[i]} restricting to the submanifold of $M$ 
defined by
$$
J = \mu \equiv constant,
$$
and then  {\bf[ii]} quotienting this space by the action of the 
transformation  group generated by $V$.   Both $\omega$ and $H$   descend 
to   $\widehat M$   and  the reduced  equations are  the 
associated    Hamiltonian system on  $\widehat M$.  Since   $\dim\widehat 
M =\dim M - 2$, the reduction in the  number of dependent variables is 
2.   The solution to the  original Hamiltonian   equations are obtained
from that of the reduced   Hamiltonian equations by  quadratures.  

To   compare with Lie symmetry reduction, we transcribe   Hamilton's 
equations into the bundle-operator setting used to construct  the 
kinematic and  dynamic reduction diagrams. Let $E=M\times \real \to \real$
be extended phase space  so that the differential operator  
characterizing  the  canonical equations  is  the one-form  valued
operator  on  $J^1(E)$   defined by 
$$
	\Delta = X \hook \omega  - d\, H. 
$$
Here  $X$ is the  total derivative  operator  given,  in standard 
canonical coordinates $(u^i,p_i)$ on $M$,  by 
$$
	X =  \frac{d\hfill}{dt} =  \vect{t} + \dot u^i  \vect{u^i} +  \dot  p_i  
\vect{p_i}. 
$$
It  is not difficult to show  that if  $V$ is any vector field on $M$, 
then  the prolongation of  $V$ to $J^1(E)$
satisfies  $[X,\pr^1 V] =0$ and therefore $V$ is a  symmetry of   the 
operator $\Delta$ whenever   $V$ is a symmetry of
$\omega$ and $H$.  

Since $V$ is a vertical vector field   on $E$, it is,  in some sense ``all
isotropy'', and the  kinematic  bundle  is the fixed pint set  for the 
flow of $V$, 
$$
\kappaGamma(E) =\{(t, u^i, p_i) \, | \, V (u^i, p_i) = 0\,\}.
$$
The   dimension of  $\kappaGamma(E)$  therefore  depends upon  the choice 
of $V$   and  is generally  less than   the dimension of $E$  by
more that  2,  the reduction  in  dimension in the case of symplectic 
reduction. In short, it is simply 
not possible to  generally  identify  the  fibers of the kinematic bundle  
with the  reduced phase space $\widehat M$.  Moreover,  from \eq{435}, it 
follows that {\it points in  $\kappaGamma(E)$    always correspond to 
points   on the singular level sets of  the momentum map and, it appears, 
to points where the 
level sets  fail to  be a  manifold}. Thus the solutions derived by Lie 
symmetry  reduction are    problematic from the
viewpoint of   symplectic reduction and  are  subject  to  special  
treatment. See, for example, \cite{arms-gotay-jennings:1990a} and 
\cite{gotay-bos:1986a}. 
Finally,  there  is no  {\it a priori} guarantee that the  reduced 
equations for the  group invariant solutions possess
any natural  inherited  Hamiltonian  formulation.

In particular,  in  the  case   $V$  is  a translation symmetry    of  a 
mechanical  system,  $J$ is a linear   first integral and symplectic 
reduction
yields  all  the solutions to Hamilton's equations  with  a given fixed 
value  for $J$.  Since the vector field $V$ never vanishes,
the kinematic bundle is  empty  and there   are  {\it no}  group invariant 
solutions.

 For   the  classical   3-dimensional central force problem
$$
\ddot u = - f(\rho) u,\quad \ddot v = -f(\rho) v, \quad \ddot 
w = -f(\rho) w,
$$
where $\rho = \sqrt{u^{2} + v^{2} + w^{2}} $,  the 
extended  phase space $E$ is ${\real}\times \real^6 \to \real$  with 
coordinates 
$$
(t, u,v,w,p_{u},p_{v},p_{w}) \to (t),
$$ 
 the symplectic structure on  phase space is 
$
\omega = dp_{u}\wedge du + dp_{v}\wedge dv + dp_{w}\wedge dw.
$
and  the  Hamiltonian  is
$
H = {1\over 2} (p_{u}^{2} + p_{v}^{2}
+ p_{w}^{2}) + V(r),
$
where
$
V'(\rho) = \rho f(\rho).$
This  system  enjoys rotational symmetry  as defined, for example,  by  
the vector field
$$
V = u{\partial\over\partial v} - v{\partial\over\partial u} + p_{u} 
{\partial\over \partial p_{v}} -p_{v} {\partial\over\partial 
p_{u}}. 
$$

The kinematic bundle  for the $V$ invariant sections of $E$ is 
$$
\CD
(t, w, p_w) @<<<           (t,w,p_w) @>> \iota  >             (t, u,v,w, 
p_u,p_v,p_w) \\ 
@VVV                           @VVV                    @        VVV    \\
(t)               @<<<            (t)          @>>>                 (t) \ ,
\endCD
\Tag440
$$
where $\iota(t,w,p_w) = (t,0,0,w,0,0,p_w)$,
the invariant sections are of the form
$$
t \to (0,0,w(t), 0,0, p_w(t)), 
$$
and the reduced differential operator for the  $V$  invariant solutions  is
$$
\tilDelta = (\dot w - p_{w})\, dp_w  + (\dot p_{w}  + wf(|w|) )\, dw.
$$

Let us compare this state of affairs with that obtained by 
symplectic reduction based upon  the Hamiltonian vector field $V$.  The 
moment map  associated to this
symmetry is  the angular momentum
$$
	J = up_{v} - vp_{u},
$$
The level sets  $J=\mu$ are  manifolds except  for $\mu=0$ and except at  
those points in  phase space which
characterize the  fibers of the kinematic bundle.   
To implement the  symplectic  reduction, we  
introduce canonical cylindrical coordinates 
$(r,\theta,w,p_{r},p_{\theta},p_{w})$, 
where
$$
u=r\cos\theta,\quad v=r\sin\theta,
$$
$$
 p_{u}=\cos\theta 
p_{r} - {1\over r}\sin\theta p_{\theta},\quad
p_{v}=\sin\theta 
p_{r} + {1\over r}\cos\theta p_{\theta}.
$$
{\it Note that this change of coordinates  fails  precisely at   points of 
the kinematic bundle. } 
In terms of these phase space coordinates,
the symplectic structure is still in canonical form
$
\omega = dp_{r}\wedge dr + dp_{\theta}\wedge d\theta + 
dp_{w}\wedge dw
$,
the Hamiltonian is
$
H = {1\over 2}(p_{r}^{2} + {1\over r^{2}}p_{\theta}^{2} + p_{w}^{2})
+ V(\sqrt{r^{2}+w^{2}}),
$
and the moment map is
$
J = p_{\theta}.
$
 We can  therefore  describe the symplectic 
reduction  of  $E$ by the diagram 
$$
\CD
(t, r,w, p_r, p_w) @<<<   (t,r,\theta, w, p_r, p_w) @>\iota>> (t,r, 
\theta, w, p_r, p_\theta, p_w) \\
@VVV                                 @VVV                           @VVV \\
(t)                             @<<<  (t)                        
@>>>                    (t) .
\endCD
$$
The 
reduced symplectic structure is then $\hat \omega= dp_{r}\wedge 
dr+dp_{w}\wedge dw$,
the reduced 
Hamiltonian is 
$
\widehat H = {1\over 2}(p_{r}^{2} + {\mu\over r^{2}} + p_{w}^{2}) + 
V(\sqrt{r^{2}+w^{2}})
$,
and the reduced equations of motion are
$$
\dot r = p_{r},\quad \dot p_{r} = 
-rf(\sqrt{r^{2}+w^{2}}) +{\mu\over r^{3}},\quad \dot w = 
p_{w},\quad \dot p_{w} = -wf(\sqrt{r^{2}+w^{2}}).
$$
Given a choice of $\mu$ and  solutions to these 
reduced equations, we get a solution to the full equations via
$
\theta = \mu t + const.
$
\endEx

***************************************************************

\bigskip 
\subheading{Appendix}  We use the  formula for the  Laplacian in terms of 
the Hodge star operator to calculate the Laplacian. This 
turns out  to be a  reasonable  efficient  procedure.  We  first compute 
the  Laplacian of the  $P_A =  A(t) \, \bfx$    to be 
\def\d{d\,}

$$
	*\d P_A 
            =   A'(*\d t)  \bfx +A(t)( *\d  \bfx)
$$
and therefore
$$ 
	\d (* \d P_A)
 =        \bigl[A'' (\d t \wedge  *\d t) + A'(d *\d t)\bigr]  \, \bfx
                + A'  \bigl[\d \bfx   \wedge   (*\d t) +    \d t \wedge 
*\d \bfx)\bigr]  
$$
We now  compute
$$
\d t  = \frac{ -s \,\d r + r \, \d s}{r^2 +s^2} 
$$
and
$$
\align
\d (* \d t)
& 	=  \d \biggl( \frac{ -s \, *\d r + r \, *\d s}{r^2 +s^2}\biggr) =  
                   -\frac{s}{r^2 +s^2} \d (* \d r) + \frac{r}{r^2+s^2} 
\d(*\d s) 
\\
\vspace{2\jot}
&           = \frac{1}{r^2 +s^2} \biggl(  -(n-2) \frac{s}{r} + \frac{r}{s} 
\biggr)\, \nu .
\endalign
$$
Since
$$
\d t \wedge (*\d t) = \nu , \quad 
\d \bfx \wedge (* \d t) =  \d t \wedge * \d \bfx = -\frac{s 
\bfx}{r(r^2+s^2)}\, \nu, 
$$
we   arrive at (on $S^n$)
$$
	\Delta^{\real^{n+1}} P_A = * \d ( * \d P_A) = 
	\bigl[ A'' +  \biggl( -)\frac{\sin(t)}{\cos(t)} + \frac{\cos(t)}{\sin(t)} 
\biggr) A'  \bigr]\, \bfx
$$

Likewise  to find the  Laplacian  of $P_B= B(t) \bfy $ we compute
$$
	\d(* \d P_B) 
          =    \bigl[ B'' +  \biggl( -n\frac{\sin(t)}{\cos(t)} + 
\frac{\cos(t)}{\sin(t)}   \biggr) B' \bigr]\, \bfy
$$

We   can also use this  formalism to compute
$$
\align
	|d P_A| *  
&=          \text{tr}(\d P_A \wedge *\d P_A)
\\
\vspace{2\jot}
&=         \text{tr} \bigl( \bigl[A'\d t  +  A\,\d \bfx\bigr] \wedge  
\bigl[A' (*\d t)   +  A\,*  \d \bfx)    \bigr]   \bigr)
\\ 
\vspace{2\jot}
& =     (A')^2  + nA^2  - 2 \sin(t) A A'  .
\endalign
$$

**************************

For the vaccum Einstein equations, we have that $c=0$ and, with the change 
of  variables
$$
	B=  e^{q(u)}\quad Q = e^{M(u)}\quad P = e^{N(u)},
$$ 
the reduced equations are
$$
\ddot q  - \frac12 (\dot q)^2 = - \frac{ \dot M^2  + \dot N^2}{4}  
-\frac{\ddot M + \ddot N}{2}
$$

One very  interesting aspect of  this symmetry reduction  is that  there  
is  a larger  group which  fixes  the  metric \eq{450}, namely the
homotety
$$
V* =  x\vect{x} + y \vect{y} +  \mu(u) \vect{v} +  g_{ij} \vect{g_{ij}} 
$$

*********************

The rigorous justification of  the Lie method   consists of proving  that 
since   \eq{102} is a symmetry algebra
for the differential equations  \eq{101} the parameter variable  $\hatx^i$ 
which may appear explicitly in the
differential equations  \eq{101}  and  in the  solution ansatz \eq{106}  
can always be eliminated in the 
reduced equations  \eq{107}.

************************************************

\subheading{5. Concluding Remarks}  The kinematic  and  dynamic reduction 
diagrams  give a precise  description of the
reduced equations for the  group invariant solutions of a system of 
differential equations  and provide a basis for the  systematic 
study of the reduced equations.  In particular, one would like to 
understand     what     different  geometric 
properties of the  original equations  can  be inherited by the reduced 
equations and  what properties of the action of the symmetry  assure
this inheritance.  We mention the following  topics  to illustrate this 
general problem and to briefly describe some partial results.
We  intend  to report on  these  in greater detail   elsewhere.

\smallskip
{\bf[ii]}  There is surprising    little in the literature  concerning   
the formal  solvability of the   reduced equations. 
If the  original system of equations are of Cauchy-Kovalevskaya  type  and 
if  no characteristic directions are tangent to the orbits
of $G$ acting on $M$, then it is not difficult to show (using Finzi's 
Theorem \cite{olver:1993a}, pXX ) that the reduced equations are of 
Cauchy-Kovalevskaya type.  More generally, we  anticipate theorems
relating the involutivity  and Cartan characters  of the  original 
equations    to the involutivity and Cartan characters of the reduced
equations.  In particular it is important  to carefully   study the 
properties of  the symbol of the reduced equations and to relate the
Spencer cohomology of the reduced symbol to  that of the original symbols. 
See, for example,  Guillemin \cite{guillemin:1968a}.  For equations
such as the Einstein equations and the Yang-Mills equations whose symbols 
are totally degenerate  one would like to characterize
those reductions for which such degeneracy persists.  This question is 
impinges upon  the nature of residual gauge group (see {\bf[i]}) for these
equations and to the reduction of the Noether identities.
\medskip
\noindent
{\bf[iii]} Palais'  principle of symmetric  criticality  
\cite{palais:1979a}, \cite{palais:1985a} examines the   problem of 
characterizing those group actions for which
the  reduction of a system of Euler-Lagrange equations  are always 
variational.   Two obstructions
naturally arise  which reflect  characteristics of the base reduction on 
the left-hand side of the kinematic reduction diagram 
\equationlabel{3}{206}{}
and  of the  fiber reduction  in the  right-hand side.  The obstruction 
arising from reduction in  the independent variables was studied
in Anderson and Fels \cite{anderson-fels:1997a}.  The  obstruction coming 
from the fiber reduction is described in  elementary representation 
theoretic
data  for the action  of the isotropy group on the space of  $\pi$ 
vertical vectors on $E$ and is  the  finite dimensional
analog of  the  obstruction described in \cite{palais:1979a}, Theorem.  
Closely related to the  principle of symmetric criticality  is the problem 
of  reduction of  invariant  conservation laws and Hamiltonian structures.

\subheading{5. Concluding Remarks}  The kinematic  and  dynamic reduction 
diagrams  give a precise  description of the
reduced equations for the  group invariant solutions of a system of 
differential equations  and provide a basis for the  systematic 
study of the reduced equations.  In particular, one would like to 
understand     what     different  geometric 
properties of the  original equations  can  be inherited by the reduced 
equations and  what properties of the action of the symmetry  assure
this inheritance.  We mention the following  topics  to illustrate this 
general problem and to briefly describe some partial results.
We  intend  to report on  these  in greater detail   elsewhere.

\smallskip{\bf[iii]} Palais'  principle of symmetric  criticality  
\cite{palais:1979a}, \cite{palais:1985a} examines the   problem of 
characterizing those group actions for which
the  reduction of a system of Euler-Lagrange equations  are always 
variational.   Two obstructions
naturally arise  which reflect  characteristics of the base reduction on 
the left-hand side of the kinematic reduction diagram 
\equationlabel{2}{206}{}
and  of the  fiber reduction  in the  right-hand side.  The obstruction 
arising from reduction in  the independent variables was studied
in Anderson and Fels \cite{anderson-fels:1997a}.  The  obstruction coming 
from the fiber reduction is described in  elementary representation 
theoretic
data  for the action  of the isotropy group on the space of  $\pi$ 
vertical vectors on $E$ and is  the  finite dimensional
analog of  the  obstruction described in \cite{palais:1979a}, Theorem.  
Closely related to the  principle of symmetric criticality  is the problem 
of  reduction of  invariant  conservation laws and Hamiltonian structures.

***************************************************************************
\advance\refnumb by 1 \ref\no\the\refnumb
\referencetag        harnad-schnider-vinet:1979b
\by                        J. Harnad, S. Schnider, and L.  Vinet
\paper                  The Yang-Mills system in compactified Minkowski 
space; invariance conditions and  $\text{\bf SU}(2)$
                              invariant solutions
\jour                     J . Mathematical Physics                      
\vol                      20
\issue                    5
\yr                        1979
\pages                  931--942                  
\endref\smallskip

\advance\refnumb by 1 \ref\no\the\refnumb
\referencetag        harnad-shnider-tafel:1980  
\by                        J. Harnad, S. Shnider and J. Tafel
\paper                  Group actions on principal bundles and dimensional 
reduction
\jour                     Lett.  Mathematical Physics                     
\vol                      4
\yr                        1980  
\pages                  107--113                 
\endref\smallskip

\advance\refnumb by 1 \ref\no\the\refnumb
\referencetag        harnad-schnider-vinet:1980a
\by                        J. Harnad, S. Schnider, and  L.  Vinet
\paper                   Group actions on princpal bundles and invariance  
conditions for gauge fields
\jour                     J . Mathematical Physics                      
\vol                      21
\issue                    12
\yr                        1980
\pages                  2719--2724               
\endref\smallskip

\advance\refnumb by 1 \ref\no\the\refnumb
\referencetag        harnad-vinet:1978a
\by                        J. Harnad and L. Vinet
\paper                  On the  $\text{\bf U}{2}$ invariant solutions to 
Yang-Mills  equations in compactified Minkowski space
\jour                     Phys. Lett. B                    
\vol                      76
\issue                    5 
\yr                        1978
\pages                  589--592                   
\endref\smallskip
\Refs
\nofrills{}


\NoBlackBoxes

\advance\refnumb by 1 \ref\no\the\refnumb
\referencetag       abraham-marsden:1978a
\by 	              R. Abraham and J. Marsden 
\book	              Foundations of Mechanics 
\bookinfo           2nd ed.
\publ 	              Benjamin-Cummings
\publaddr           Reading, Mass
\yr 	             1978
\endref\smallskip

\advance\refnumb by 1 \ref\no\the\refnumb
\referencetag 	anderson-fels:1997a
\by 		I. M. Anderson and M. E. Fels
\paper            	Symmetry Reduction of Variational Bicomplexes 
                  	and the principle of symmetry criticality     
\yr 		1997
\vol 		112
\jour  		Amer. J. Math.
\pages 		609--670
\endref\smallskip


\advance\refnumb by 1 \ref\no\the\refnumb
\referencetag 	anderson-fels:1999a
\by 		I. M. Anderson and M. E. Fels
\paper            	Transverse group actions on bundles   
\paperinfo            In  preparation 
\endref\smallskip



\advance\refnumb by 1 \ref\no\the\refnumb
\referencetag       anderson-fels-torre:2000a
\by 	              I.  M. Anderson, Mark  E. Fels, Charles G. Torre
\book	              Symmetry Reduction of Differential Equations  
\bookinfo           in  preparation 
\endref\smallskip

\advance\refnumb by 1 \ref\no\the\refnumb
\referencetag        arms-gotay-jennings:1990a 
\by                        J. A. Arms,  M. J. Gotay,  G. Jennings
\paper                  Geometric and Algebraic  Reduction  for Singular 
Momentum Maps
\jour                     Adv. in Math                    
\vol                       79
\yr                         43--103
\pages                   1990              
\endref\smallskip




\advance\refnumb by 1 \ref\no\the\refnumb
\referencetag       beckers-harnad-perroud-winternitz:1978a   
\by                       J. Beckers, J. Harnad, M. Perrod, and P. 
Winternitz 
\paper                 Tensor  fields invariant  under  subgroups of the 
conformal group of space-time 
\jour                    J. Mathematical 
Physics                             
\vol                     19(10) 
\yr                       1978
\pages                 2126--2153                        
\endref\smallskip



\advance\refnumb by 1 \ref\no\the\refnumb
\referencetag       beckers-harnad-jasselette:1979a   
\by                       J. Beckers, J. Harnad and P. Jasselette     
\paper                 Spinor fields invariant  under space-time  
transformations
\jour                    J. Mathematical 
Physics                             
\vol                     21(10) 
\yr                       1979
\pages                 2491--2499                         
\endref\smallskip

\advance\refnumb by 1 \ref\no\the\refnumb
\referencetag        bleecker:1979a    
\by                        D. D. Bleecker
\paper                  Critical mappings of Riemannian manifolds
\jour                     Trans. Amer. Math. Soc.                     
\vol                       254
\yr                        1979
\pages                  319--338                  
\endref\smallskip

\advance\refnumb by 1 \ref\no\the\refnumb
\referencetag       bleecker:1979b 
\by                       D. D. Bleecker
\paper                 Critical Riemannian manifolds
 \jour                   J. Differential Geom.                        
\vol                     14
\yr                        1979 
\pages                  599--608                   
\endref\smallskip



\advance\refnumb by 1 \ref\no\the\refnumb
\referencetag        bluman-kumei:1989a
\by 		G. W. Bluman and S. Kumei
\book		Symmetries and  Differential Equations
\bookinfo            Applied Mathematical Sciences, 81
\publ                    Springer-Verlag
\publaddr            New York-Derlin 
\yr                        1989 
\endref


\advance\refnumb by 1 \ref\no\the\refnumb
\referencetag         bruning-heintze:1979a
\by       	                J. Br\'uning and E. Heintze
\paper    	 Representations of compact lie groups and  elliptic operators
\jour     		 Inventiones Math.
\vol      		50
\yr       		1979
\pages    	169--203
\endref\smallskip




\advance\refnumb by 1 \ref\no\the\refnumb
\referencetag        bondi-pirani-robinson:1959a
\by       	               H. Bondi, F.Pirani,  I. Robinson 
\paper    	Gravitational waves  in general relativity III. Exact plane 
waves
\jour     		 Proc.  Roy. Soc. London  A
\vol      		251
\yr       		1959
\pages    	 519--533
\endref\smallskip




\advance\refnumb by 1 \ref\no\the\refnumb
\referencetag    coquereaux-jadczyk:1988a       
\by                    R. Coquereaux and A. Jadczyk      
\book               Riemannian Geometry, Fiber Bundles, Kaluza-Klein 
Theories and all that    
\bookinfo         Lecture Notes in Physics
\vol                   16
\publ                 World Scientific 
\publaddr         Singapore
\yr                     1988 
\endref\smallskip


\advance\refnumb by 1 \ref\no\the\refnumb
\referencetag     	david-kamran-levi-winternitz:1986a
\by 		D. David, N. Kamran, D. Levi and P. Winternitz
\paper 		Symmetry reduction for the Kadomtsev-Petviashvili equation using 
a loop algebra
\yr 		1986
\vol 		27
\jour 		J. Mathematical Physics
\pages 		1225--1237
\endref\smallskip



\advance\refnumb by 1 \ref\no\the\refnumb
\referencetag       eells-ratto:1993a
\by                       J. Eells and A. Ratto                   
\book                  Harmonic Maps and Minimal Immersions with 
Symmetries 
\bookinfo            Annals of Mathematical Studies 
\vol                     130
\publ                   Princeton Univ. Press
\publaddr           Princeton
\yr                      1993
\endref\smallskip


\advance\refnumb by 1 \ref\no\the\refnumb
\referencetag       fels-olver:1997a
\by                       M. E. Fels and P. J.  Olver
\paper                 On relative invariants
\jour                    Math. Ann.       
\yr                       1997
\vol                     308
\pages                 609--670
\endref\smallskip


\advance\refnumb by 1 \ref\no\the\refnumb
\referencetag       fels:1999a
\by                       M. E. Fels 
\paper                  Symmetry  reductions of the Euler equations
\paperinfo            In preparation                   
\endref\smallskip



\advance\refnumb by 1 \ref\no\the\refnumb
\referencetag        fushchich-shtelen-slavutsky:1976a
\by		W. I.  Fushchich, W. M. Shtelen, S. L. Slavutsky
\paper 	 	 Reduction and  exact solutions of the Navier-Stokes equations
\jour 		Topology
\yr 		1976
\vol 		15 
\pages 		165--188
\endref\smallskip



\advance\refnumb by 1 \ref\no\the\refnumb
\referencetag       gaeta-morando:1997a
\by                       G.  Gaeta and  P. Morando
\paper                  Michel theory of symmetry breaking and gauge 
theories
\jour                     Annals of Physics                      
\vol                       260 
\yr                         1997 
\pages                   149--170                  
\endref\smallskip



\advance\refnumb by 1 \ref\no\the\refnumb
\referencetag        gotay-bos:1986a
\by                        M. J. Gotay and L. Bos
\paper                   Singular  angular momentum mappings
\jour                      J.  Differential  Geom.                     
\vol                       24
\yr                         1986
\pages                   181--203                  
\endref\smallskip




\advance\refnumb by 1 \ref\no\the\refnumb
\referencetag        grundland-winternitz-zakrzewski:1996a
\by 		A. M. Grundland,  P. Winternitz, W. J. Zakrewski
\paper 		On the solutions of the ${C}{\text{P}}\sp 1$ model
		in $(2+1)$ dimensions
\jour     	 	J.  Math. Phys.
\vol                     37
\issue                   3             
\yr 	              1996
\pages                 1501--1520
\endref\smallskip


\advance\refnumb by 1 \ref\no\the\refnumb
\referencetag       harnad-schnider-vinet:1979a
\by                       J. Harnad, S. Schnider and L. Vinet
\paper                  Solution to  Yang-Mills equations on  $\barM^4$ 
under subgroups of $\text{\bf O}(4,\,2)$  
\inbook               Complex manifold techniques in the theoretical 
physics (Proc. Workshop, Lawrence, Kan. 1978)                    	
		Research Notes in Math.  
\vol                      32                       
\eds                      
\yr                        1979
\publ                   Pitamn
\publaddr            Boston
\pages                  219-230
\endref\smallskip



\advance\refnumb by 1 \ref\no\the\refnumb
\referencetag 	ibragimov:1995a
\by       		N. H. Ibragimov
\book     	CRC Handbook of Lie Group Analysis of Differential Equations, 
          		Volume 1
\bookinfo 	Symmetries, Exact Solutions and Conservation Laws.
\yr       		1995
\publ     		CRC Press
\publaddr  	Boca Raton, Florida
\endref\smallskip 




\advance\refnumb by 1 \ref\no\the\refnumb
\referencetag      jackiw-rebbi:1976a
\by                     R. Jackiw and C. Rebbi
\paper                Conformal properties of a Yang-Mills pseudoparticle
\jour                   Phys. Rev.  D
\vol                    14  
\yr                     1976 
\pages                517--523
\endref\smallskip 




\advance\refnumb by 1 \ref\no\the\refnumb
\referencetag       kovalyov-legare-gagnon:1993a
\by                       M. Kovalyov,  M. L\'egar\'e, and L. Gagnon
\paper                 Reductions  by isometries of the self-dual 
Yang-Mills equations in 
	              four-dimensional Euclidean space
\jour     	              J. Mathematical Physics
\vol      	              34(7)
\yr       	              1993
\pages                 3245--3267
\endref\smallskip 

\advance\refnumb by 1 \ref\no\the\refnumb
\referencetag      lawson:1980a
\by                     H. B. Lawson   
\book                 Lectures on Minimal Submanifolds
\bookinfo            Mathematics Lecture Series
\vol                    9
\publ                  Publish or Perish
\publaddr           Berkeley
\yr                      1980
\endref\smallskip




\advance\refnumb by 1 \ref\no\the\refnumb
\referencetag        legare-harnad:1984a
\by                        M. L\'egar\'e  and J. Harnad
\paper                   $SO(4)$  reduction of the Yang-Mills equations  
for the classical gauge group 
\jour                      J.  Mathematical Physics                   
\vol                       25
\issue                     5
\yr                         1984
\pages                   1542--1547                 
\endref\smallskip


\advance\refnumb by 1 \ref\no\the\refnumb
\referencetag       legare:1995a 
\by                       M. L\`egar\'e
\paper                  Invariant spinors and reduced Dirac equations 
under subgroups of the 
                              Euclidean  group in four-dimensional 
Euclidean space
\jour                     J. Mathematical Physics                     
\vol                      36
\issue                    6
\yr                        1995  
\pages                  2777--1791                   
\endref\smallskip



\advance\refnumb by 1 \ref\no\the\refnumb
\referencetag     olver:1993a
\by 		P. J. Olver
\book 		Applications of Lie Groups to Differential Equations
\bookinfo          (Second Ed.)
\publ 		Springer
\publaddr 	New York
\yr 		1986
\endref\smallskip



\advance\refnumb by 1 \ref\no\the\refnumb
\referencetag     ovsiannikov:1982a
\by 		L. V. Ovsiannikov
\book 		Group Analysis of Differential Equations
\publ 		Academic Press
\publaddr 	New York
\yr 		1982
\endref\smallskip



\advance\refnumb by 1 \ref\no\the\refnumb
\referencetag       palais:1957a 
\by                      R. S. Palais
\book                 A   Global Formulation of the Lie theory of 
Transformation Groups  
\bookinfo          Memoirs of the Amer. Math Soc.
\vol                    22
\publ                  Amer. Math. Soc.
\publaddr          Providence, R.I.
\yr                      1957
\endref\smallskip



\advance\refnumb by 1 \ref\no\the\refnumb
\referencetag 	  palais:1979a
\by       		  R. S. Palais
\paper    	  The  principle of symmetric criticality 
\jour     	                 Comm. Math. Phys. 
\vol      		  69 
\yr       		  1979
\pages    	  19--30
\endref\smallskip 


\advance\refnumb by 1 \ref\no\the\refnumb
\referencetag        palais:1985a      
\by                        R. S. Palais
\paper                  Applications of the symmetric criticality 
principle in 		mathematical
                              physics and differential geometry
\inbook                Proc. U.S.-- China Symp. on Differential Geometry 		
and    Differential Equations II                       
\yr                        1985 
\endref\smallskip



\advance\refnumb by 1 \ref\no\the\refnumb
\referencetag       rogers-shadwick:1989a   
\by                       C. Rogers and W. Shadwick   
\book                  Nonlinear boundary value problems in science and 
engineering  
\bookinfo           Mathematics in Science and Enginering
\vol                     183
\eds                     W. F. Ames
\publ                   Academic Press
\publaddr            Boston
\yr                       1989
\endref\smallskip



\advance\refnumb by 1 \ref\no\the\refnumb
\referencetag       smith:1975a
\by                       R.T. Smith 
\paper                  Harmonic mapings of  spheres
\jour                    Amer. J. of Math  
\vol                      97     
\yr                        1975
\pages                   364--385
\endref\smallskip




\advance\refnumb by 1 \ref\no\the\refnumb
\referencetag          stephani:1989a    
\by                          H. Stephani            
\book                     Differential Equations and their Solutions using 
symmetries
\ed                         M. MacCallum                 
\publ                     Cambridge University Press      
\publaddr             Cambridge
\yr                         1989        
\endref\smallskip

\advance\refnumb by 1\ref\no\the\refnumb
\referencetag        torre:1999a
\by                        C. G. Torre
\paper                   Gravitational waves: Just plane symmetry
\paperinfo             preprint gr-qc/9907089
\endref
\smallskip

\advance\refnumb by 1 \ref\no\the\refnumb
\referencetag    	toth:1990a
\by                    	G. T\'oth    
\book                 	Harmonic and Minimal Immersions through 
representation  theory
\bookinfo             Prespectives in Math.
\publ                    Academic  Press	 
\publaddr             Boston
\yr                     	1990
\endref\smallskip

\advance\refnumb by 1 \ref\no\the\refnumb
\referencetag        urakawa:1993a
\by                        H. Urakawa
\paper                  Equivariant  harmonic maps between  compact 
Riemannian manifolds of cohomogenity 1
\jour                     Michigan Math. J.                     
\vol                      40
\yr                        1993
\pages                  27--50                   
\endref\smallskip



\advance\refnumb by 1 \ref\no\the\refnumb
\referencetag        vorobev:1991a
\by                        E. M. Vorob'ev
\paper                  Reduction  of  quotient equations for differential 
equations with symmetries
\jour                     Acta Appl. Math.                     
\vol                      23
\yr                        1991
\pages                  1991               
\endref\smallskip





\advance\refnumb by 1 \ref\no\the\refnumb
\referencetag      winternitz:1990a
\by                     P. Winternitz
\paper                Group theory and exact solutions of
                         partially integrable equations
\inbook              Partially Integrable Evolution Equations
\eds                   R. Conte and N. Boccara
\publ                  Kluwer Academic Publishers                       
\yr                     1990
\pages                515 -- 567
\endref\smallskip


\advance\refnumb by 1 \ref\no\the\refnumb
\referencetag 	winternitz-grundland-tuszynski:1987a
\by 		P. Winternitz, A. M. Grundland, J. A. Tuszy\'nski
\paper 		Exact solutions of the multidimensional classical 
		$\phi^6$ -- field equations obtained by symmetry reduction
\yr 		1987
\vol 		28
\jour 		J. Mathematical Phys.
\pages 		2194--2212
\endref\smallskip



\endRefs
